\documentclass[fleqn,usenatbib]{mnras}

\usepackage[T1]{fontenc}
\usepackage{ae,aecompl}
\usepackage[normalem]{ulem}

\usepackage{graphicx}	
\usepackage{amsmath}	

\usepackage{amssymb}	

\usepackage{txfonts} 

\usepackage{hyperref}
\usepackage{caption}
\usepackage{multirow}

\usepackage{epstopdf}
\usepackage{journals}

\usepackage[none]{hyphenat}
\graphicspath{{images/}{images/particle}}
\DeclareGraphicsExtensions{.pdf,.png,.jpg, .jpeg}
\usepackage{natbib}
\usepackage{color}

\usepackage{newtxtext,newtxmath}
\usepackage{tikz,xcolor,hyperref}
\definecolor{lime}{HTML}{A6CE39}
\DeclareRobustCommand{\orcidicon}{%
	\begin{tikzpicture}
	\draw[lime, fill=lime] (0,0) 
	circle [radius=0.16] 
	node[white] {{\fontfamily{qag}\selectfont \tiny ID}};
	\draw[white, fill=white] (-0.0625,0.095) 
	circle [radius=0.007];
	\end{tikzpicture}
	\hspace{-2mm}
}

\foreach \x in {A, ..., Z}{%
	\expandafter\xdef\csname orcid\x\endcsname{\noexpand\href{https://orcid.org/\csname orcidauthor\x\endcsname}{\noexpand\orcidicon}}
}

\usepackage{tikz}
\usetikzlibrary{matrix}
\usetikzlibrary{calc}
\input{pgfgraphicsmacros.tex}
\input{reviewmacros.tex}

\usepackage{threeparttable}

\title[Barless flocculent galaxies: a dynamic puzzle]
{Barless flocculent galaxies: a dynamic puzzle}

\author[Daria Zakharova, Natalia Ya. Sotnikova, Anton A. Smirnov, Sergey S. Savchenko]{Daria Zakharova$^{1,2,3}\orcidA{}$\thanks{E-mail: dzakharovaa@gmail.com}, Natalia Ya. Sotnikova$^{3}$, Anton A. Smirnov$^{4}$, \newauthor Sergey S. Savchenko$^{3,4,5}$\\
$^{1}$Dipartimento di Fisica e Astronomia "Galileo Galilei", Universita' degli studi di Padova, Vicolo dell'Osservatorio, 3, I-35122, Padova, Italy\\
$^{2}$INAF - Osservatorio astronomico di Padova, Vicolo dell'Osservatorio, 5, I-35122, Padova, Italy\\
$^{3}$St. Petersburg State University,
Universitetskij pr.~28, 198504 St. Petersburg, Stary Peterhof, Russia\\
$^{4}$Central (Pulkovo) Astronomical Observatory of RAS, Pulkovskoye Chaussee 65/1, 196140 St. Petersburg, Russia\\
$^{5}$Special Astrophysical Observatory, Russian Academy of Sciences, 369167 Nizhnij Arkhyz, Russia 
}

\date{2020}

\date{Accepted XXX. Received YYY; in original form ZZZ}

\pubyear{2022}

\begin{document}

\label{firstpage}
\pagerange{\pageref{firstpage}--\pageref{lastpage}}
\maketitle

\begin{abstract}
We draw attention to the bright galaxies that do not show a bar in their structure but have a flocculent spiral structure. Using the THINGS' and HERACLES' kinematic data for four barless galaxies (NGC~2841, NGC~3512, NGC~5055, NGC~7331) we built their mass models including dark halos. We concluded that the fraction of the dark matter does not exceed 50\% within the optical radii of the galaxies. This is too little to explain the lack of a bar in these galaxies. In an attempt to understand the featureless structure of these galaxies we constructed several $N$-body models with an initially reduced content of dark matter. We concluded that, in addition to the low mass of the dark halo, the decisive factor that leads to a barless disc is the start from an initially unstable state. An isolated dynamically cold disc (with the Toomre parameter $Q < 0.5$) settled into rotational equilibrium passes trough the short stage of violent instability with fragmentation and formation of stellar clumps. After that, it evolves passively and ends up with a featureless structure.  We assume that the barless flocculent galaxies studied in the present work may be descendants of galaxies at high redshifts with rotation curves which are consistent with the high mass fraction of baryons relative to the total dark matter halo.
\end{abstract}

\begin{keywords}
galaxies: bar -- galaxies: kinematics and dynamics -- galaxies: structure
\end{keywords}

\section{Introduction}
\label{sec:intro}
\par
According to observational statistics, up to 60$\%$ of disc galaxies possess bars (e.g., \citealp{Eskridge+2000,Menendez-Delmestre+2007,Marinova_Jogee2007,Barazza+2008,Aguerri+2009}). These statistics mean that a bulk of galaxies do not show bars in their structure. Meanwhile, in numerous $N$-body simulations, the bar is an almost inevitable result of the evolution of the galaxy model. Rather specific  model parameters are needed to suppress bar formation, at least on a timescale of 8 billion years or more. This can be achieved if a stellar disc is very hot dynamically (e.g., \citealp{Athanassoula_Sellwood1986}), or a model has a very compact initial classical bulge \citep{Fujii+2018,Saha+2018,Kataria_Das2018}, or the disc is embedded in a very massive dark halo (e.g., \citealp{Ostriker_Peebles1973}). Some other possibilities are discussed, for example, in \citet{Sellwood+2019}.
\par
The lack of the bar in some bright galaxies seems mysterious in terms of stellar dynamics. \citet{Sellwood+2019}, not for the first time, raises the question of the absence of a bar in the  M~33 galaxy. The authors provide an $N$-body model of M~33, taking into account its modern observable characteristics (photometric parameters and a rotation curve). In their study, it turned out that for all considered physical constraints, the model is not stable with respect to the bar formation. In only two cases, the authors managed to prevent the formation of a bar in the model. In the first case, it deals with an increase in the random motion of stars (an increase of the Toomre parameter $Q$). However, this solution was rejected as the developing multi-arm spirals in such a model contradict the bi-symmetric spiral pattern, which is observed in the IR range. In the second case, the mass-to-light ratio of the stars was reduced to $M/L_{K}=0.23$ in solar units, which contradicts the basic stellar population models with $M/L_{K}=0.52$. For the basic model, the mass of a dark halo within four radial scales of the stellar disc can be estimated as three disc masses. Bar formation can be suppressed only by doubling it. 
\par
We draw attention to bright galaxies that do not show a bar in their structure. Such galaxies have an even lower relative mass of a dark halo than M~33 and look much more puzzling from a dynamic point of view.
These galaxies are included in the THINGS survey \citep{Walter+2008} and some of them (e.g. NGC~2841) have rotation curves falling to the periphery, which distinguishes them from most galaxies that have an extended plateau in the rotation curve. We focused on the galaxies NGC~2841, NGC~3521, NGC~5055, and NGC~7331. The analysis of their isophotes in the S$^4$G images (PA and ellipticity profiles) does not show the presence of an oval-like distortion or, at best, demonstrates only faint traces of such a distortion \citep{Salo_etal2015}. In SDSS images galaxies demonstrate only flocculent spirals. The velocity difference between the peak of the rotation curve and the last point can reach 30~km/s--40~km/s, although NGC~7331 rather has a plateau in the rotation curve. The peak is located at about two radial scales of the stellar disc and is associated with the disc rotation curve. Even gently falling rotation curve indirectly indicates a reduced dark matter content in these galaxies, both within the optical radius of the galaxy and within the virial radius of the halo. It should be noted that we consider only massive flocculent galaxies without bar which shows the changing in circular velocity between peak and periphery at least 10\%.
\par
The rotation curves of the THINGS galaxies were extracted from spectral data cubes and decomposed in several previous works \citep{deBlok+2008,Katz+2014,Frank+2016,Saburova+2016,DiTeodoro_Peek2021,ManceraPina+2022}. Such a large number of decompositions of galaxies according to the same kinematic data is explained by the introduction of additional factors into consideration (the influence of the gas component, especially the molecular gas, the influence of the adiabatic compression of the halo by baryonic matter, additional models of the dark halo, radial motions etc.). In some cases, the decomposition results are close, in some cases they diverge greatly, which is explained by different restrictions that are imposed on the models. We also redid the kinematic analysis using $^{\mathrm{3D}}${\tt BBarolo} \citep{DiTeodoro_Fraternali2015}. We focused only on galaxies with rotation curves that show signs of falling towards the periphery or, judging by the decomposition given in the literature, show a rotation curve of the stellar disc at the optical radius of the galaxy that lies above the rotation curve associated with the dark halo. One feature of our analysis is that we used S$^4$G photometric models \citep{Salo_etal2015} and modern $M/L$ calibrations \citep{Querejeta+2015} based directly on 3.6~$\mu m$ photometry. We obtained and emphasize it that the relative mass of the dark matter within four radial disc scales is very small (0.5-1.0 of the disc mass). Such a small mass of a dark matter and the lack of bars in these galaxies is strange from the dynamic point of view, and this needs to be explained. In this paper, we provide one of the possible explanations, based on the analysis of several $N$-body models evolving from different initial conditions.
\par
In Section~\ref{sec:sample}, we list the selected galaxies and give their main parameters. In Section~\ref{sec:cubes}, we describe how we use kinematic data (the THINGS' data by \citealp{Walter+2008} for the periphery of galaxies and HERACLES' data by \citealp{Leroy+2009} for central regions). Here, we provide the extracted density profiles of the gas and rotation curves obtained from the velocity field using the $^{\mathrm{3D}}${\tt BBarolo} package \citep{DiTeodoro_Fraternali2015}. 
In Section~\ref{sec:mass_models}, we build the mass model of all baryonic components. In Section~\ref{sec:dark_halo}, we describe the dark matter mass models. 
In Section~\ref{sec:decomposition}, we decompose the extracted rotation curves taking into account S$^4$G photometric models \citep{Salo_etal2015}, modern $M/L$ calibrations \citep{Querejeta+2015},
the contribution of the atomic and molecular gas \citep{Leroy+2009},
and several models for the dark halo. In Section~\ref{sec:scaling}, we testify scale relations for dark halo parameters. In Section~\ref{sec:discussion}, we discuss baryon-dominated galaxies in the cosmological context. In Section~\ref{sec:nbody}, we provide dynamic models for the galaxies considered. We give our conclusions in Section~\ref{sec:conclusions}.
\section{Sample and data}

\subsection{Investigated objects}
\label{sec:sample}
\begin{figure}
\centering
\includegraphics[width=0.23\textwidth]{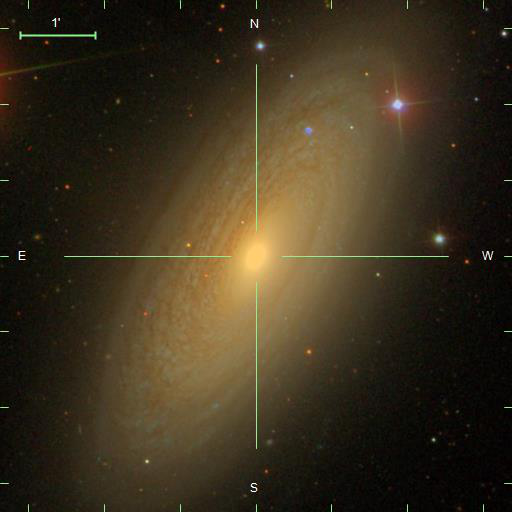}
\includegraphics[width=0.23\textwidth]{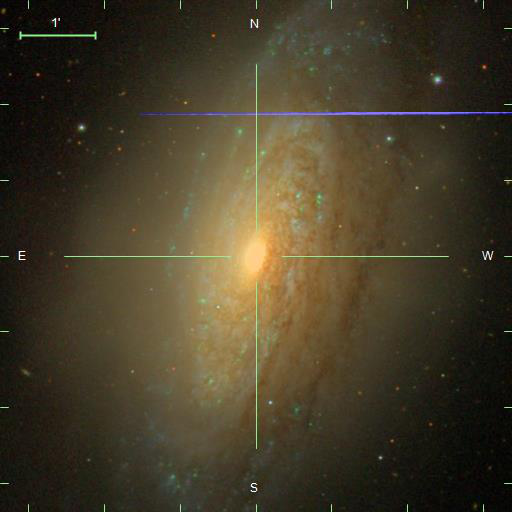}
\includegraphics[width=0.23\textwidth]{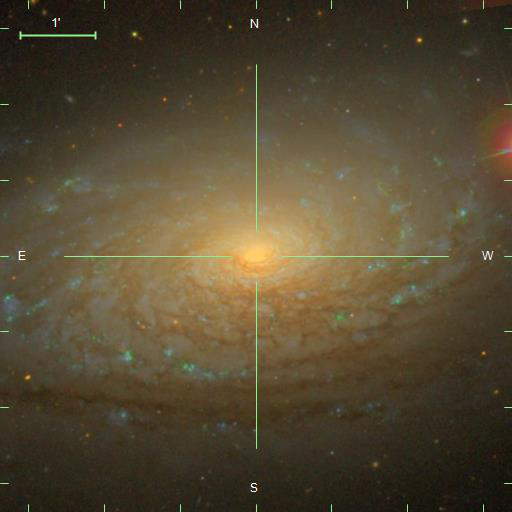}
\includegraphics[width=0.23\textwidth]{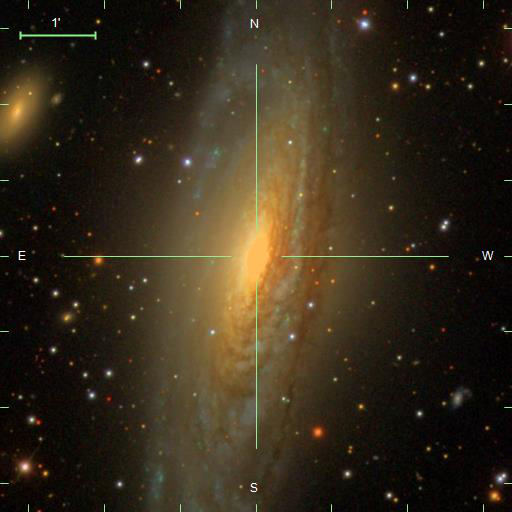}
\caption{SDSS (gri) images of galaxies studied. From left to right, from top to bottom --- NGC~2841, NGC~3521, NGC~5055, and NGC~7331.}
\label{fig:flocc}
\end{figure}

We paid attention to 4 bright galaxies (\mbox{$M_B < - 19.5$~mag}) from the THINGS survey, which show a decrease in the rotation speed  from the maximum value to the periphery by no less than 30~km/s. These are NGC~2841, NGC~3521, NGC~5055, and NGC~7331. Fig.~\ref{fig:flocc} shows SDSS ($gri$) images of these galaxies. They demonstrate flocculent spirals (Arm Class 3 according to \citealp{Elmegreen_Elmegreen1987}) and do not have a bar or even any noticeable oval-like distortion in the central regions\footnote{According to data collected on the S$^4$G portal  (\href{https://www.oulu.fi/astronomy/S4G\_PIPELINE4/MAIN/}{https://www.oulu.fi/astronomy/S4G\_PIPELINE4/MAIN/}) the PA profiles for NGC~2841, NGC~3521, and NGC~5055 do not demonstrate any twisting. The ellipticity profiles change a little beyond the area of the bulge.}. As will be shown below, these are galaxies in which mass of baryonic matter within the optical radius is greater than the mass of the dark matter.
\par
The list of investigated galaxies is given in Table~\ref{tab:general}. Here one can find data on the morphological types, inclinations, and distances. Information is also given on the photometric model for each galaxy from the S$^4$G survey \citep{Salo_etal2015}. NGC~7331 galaxy is not included in the S$^4$G survey, and the photometric decomposition for it into stellar sub-structures was done in this work (see Section~\ref{sec:mass_models}). The last column of the table describes the observed morphology.

\begin{table}
\centering
\caption{General description of the morphology of galaxies}
\begin{tabular}{llllll}
\hline
Name & type & $i$ & $D$ & model & \multicolumn{1}{l}{morphology} \\
\multicolumn{1}{l}{} & 
\multicolumn{1}{l}{(t)} & 
\multicolumn{1}{l}{($^\circ$)} & 
\multicolumn{1}{l}{(Mpc)} & 
\multicolumn{1}{c}{} & 
\multicolumn{1}{l}{(Arm Class)} \\
\hline
\hline
NGC~2841       & SAa    & \textcolor{black}{71} & \multicolumn{1}{|r|}{14.1\phantom{0}} & bd & 3\\
\hline
NGC~3521       & SAab   & \textcolor{black}{70} & \multicolumn{1}{|r|}{{13.2}\phantom{0}} & bdd & 3, LSB\\
\hline
NGC~5055       & SAbc   & \textcolor{black}{55} &  \multicolumn{1}{|r|}{{\phantom{1}8.99}} & ndd & 3\\
(M~63) & & & & & \\
\hline
NGC~7331       & SAb    & \textcolor{black}{77} & 14.7 & nbdd & 3\\
\hline 
\multicolumn{6}{p{0.45\textwidth}}
{\textit{Description:} type (t): http://cdsportal.u-strasbg.fr/;
inclination ($i$): 
\citet{DiTeodoro_Peek2021}; derived with kinematic modeling;
distance ($D$): the Extragalactic Distance Database, http://edd.ifa.hawaii.edu, \citet{EDD_2017}; model: b -- bulge, d -- disc, n -- nuclear core \citep{Salo_etal2015}, the model for NGC~7331 was built in the present work; morphology: Arm Class according to \citet{Elmegreen_Elmegreen1987}, 3 --- fragmented arms uniformly distributed around the galactic centre.} 
\end{tabular}
\label{tab:general}
\end{table}

\subsection{Data cubes}
\label{sec:cubes}
$^{\mathrm{3D}}${\tt BBarolo} (‘3D-Based Analysis of Rotating Objects from Line Observations’) package \citep{DiTeodoro_Fraternali2015} was used to obtain the rotation curves of galaxies from the data cubes of the THINGS and HERACLES surveys. $^{\mathrm{3D}}${\tt BBarolo} is a package for determining kinematic data of galaxies from observations of emission lines ($\mathrm{HI,\ CO,\ C^{+}}$). It implements the most common method of inclined rings \citep{Rogstad+1974}: the galaxy is divided into several concentric rings with different radii, tilts, and position angles so that inside each ring the circular velocity $V_ {\mathrm{rot}}(R)$ is constant and it depends only on the distance from the centre $R$.
\par
$^{\mathrm{3D}}${\tt BBarolo} can work with any observational data associated with emission lines with very different spatial resolutions. The code creates a model and compares it with observational data, minimizing the difference between the real and model data cubes.
\par
The results of our 3D modeling applied to the THINGS' galaxies NGC~2841, NGC~3521, NGC~5055 and NGC~7331 are presented in the~Appendix~\ref{sec:appendix}. They include the comparison between the data and the best-fit model, which was double-checked, as well as the best-fit model parameters derived by $^{\mathrm{3D}}${\tt BBarolo}: the rotation curves, surface density profiles of gas, inclinations, position angles and radial velocities. In general, our model rotation curves are in good agreement with the results presented in~\citet{DiTeodoro_Peek2021}.

\subsection{Hydrogen surface density profiles}

\subsubsection{$\mathrm{HI}$ distribution}
To obtain the surface density of neutral gas, we used the result of the $^{\mathrm{3D}}${\tt BBarolo} $\mathrm{3DFIT}$ procedure. At the normalization stage, the package builds a surface density distribution of $\mathrm{HI}$ and produces the profile in $\mathrm{M_{\odot}}$/pc$^2$ units. The resulting profile was then adjusted to take into account the presence of helium and other metals by multiplication of the data generated by the package $^{\mathrm{3D}}${\tt BBarolo} by a factor of 1.36.
\par
Fig.~\ref{fig:comp-with-leroy2} ({\it left} plot) shows the surface density profile of neutral hydrogen for the NGC~5055 galaxy obtained by $^{\mathrm{3D}}${\tt BBarolo}. For comparison, the profile presented in ~\citet{Leroy+2008} is given in the same plot. One can see a good agreement in general. The discrepancy near the maximum is due to the use of different masks when extracting data from 3D cubes.

\subsubsection{$\mathrm{H}_2$ distribution}
The observed luminosity of $\mathrm{CO}$ 
$I_{\mathrm{CO}} \ \mathrm {[K \ km/s]}$ was converted into the surface density of molecular hydrogen $\Sigma_{\mathrm{H_2}} \ [M_{\odot}$/pc$^2]$ using the calibration \citep{Leroy+2013}:

\begin{equation}
\displaystyle \Sigma_{\mathrm{H_2}} = 6.3  I_{\mathrm{CO}} \left( \frac{0.7}{R_{21}} \right) \left( \frac{\alpha^{1-0}_{\mathrm{CO}}}{4.35}\right) ,
\label{eq:H2}
\end{equation}
where $R_{21}=0.7$ is the $\mathrm{CO_{2 \rightarrow 1}}$-to-$\mathrm{CO_{1 \rightarrow 0}}$ line ratio, and $ \alpha^{1-0}_{\mathrm{CO}}=4.35 M_{\odot}$/pc$^2$ (K km/s)$^{-1}$ is the conversion factor for the Milky Way for the transition CO $J=1 \to 0$. Although there are improved estimates of $\alpha^{1-0}_{\mathrm{CO}}$ for galaxies from the HERACLES \citep{Sandstrom+2013}, it was shown that the change of this coefficient does not affect the decomposition of rotation curves \citep{Frank+2016}, therefore, the value for the Milky Way was used in the present work. The expression~\eqref{eq:H2} already includes the coefficient 1.36 to account for the helium and other heavy elements. 
\par
The corresponding values of $I_{\mathrm{CO}}$ can also be obtained using $^{\mathrm{3D}}${\tt BBarolo}. The package constructs the distribution of the intensity over the rings in units of [K km/s].
\par
\begin{figure} 
\centering
\includegraphics[width=0.45\textwidth]{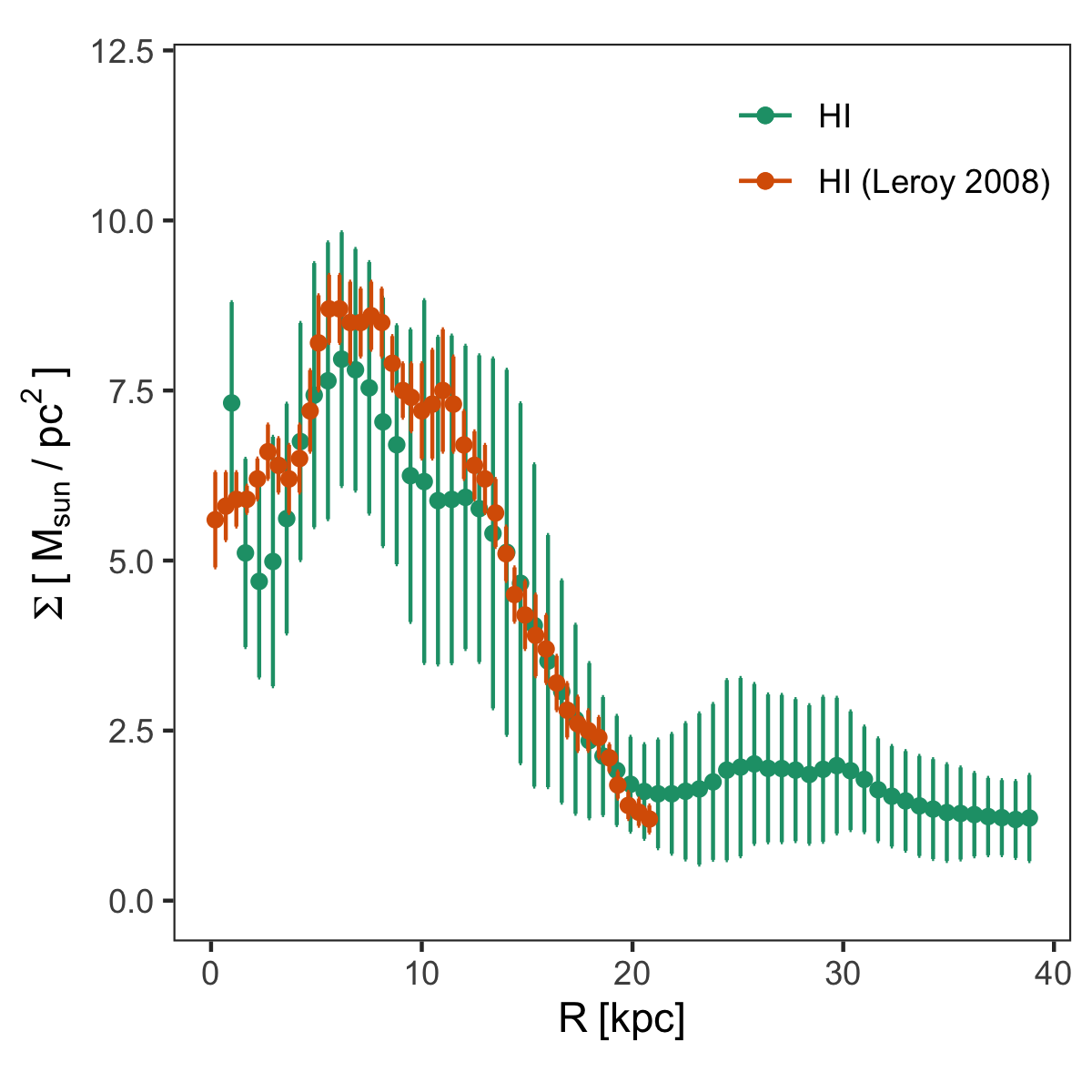}
\includegraphics[width=0.45\textwidth]{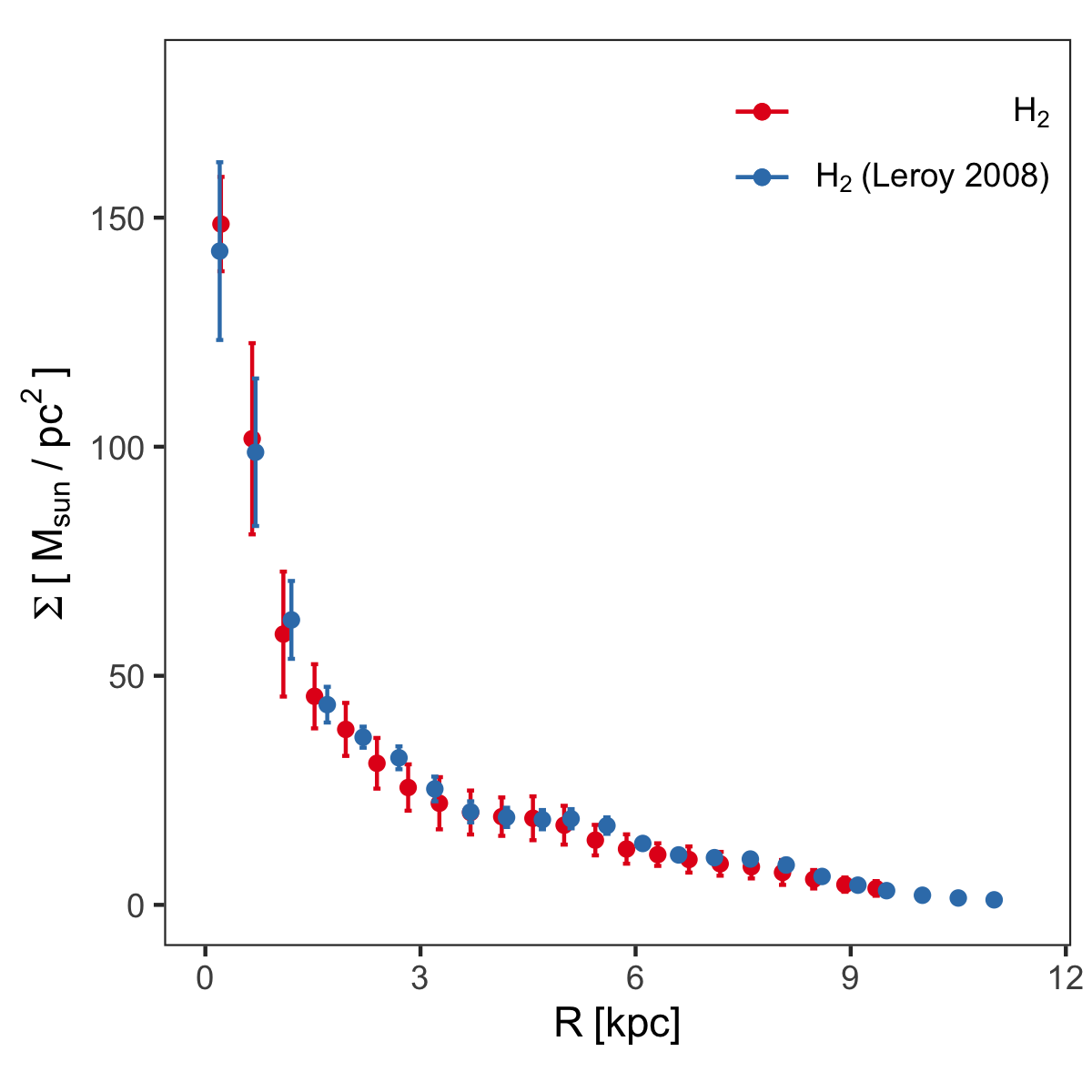}
\caption{Surface density profiles of gas for galaxy NGC~5055. For comparison, the profiles from \citet{Leroy+2008} are shown. {\it Left}: neutral gas; {\it right}: molecular gas.}
\label{fig:comp-with-leroy2}
\end{figure}

    
As an example, Fig.~\ref{fig:comp-with-leroy2} ({\it right} panel) shows the surface density profile of the molecular gas for galaxy NGC~5055.

\section{Photometric and mass models for stellar subsystems}
\label{sec:mass_models}
Mass models of stellar sub-structures were built on the basis of multi-component photometric decomposition of the images from the S$^4$G in the band 3.6~$\mu$m \citep{Salo_etal2015}. The decomposition results are collected on the \href{https://www.oulu.fi/astronomy/S4G\_PIPELINE4/MAIN/}{S$^4$G portal},
as well as in the tables of the \href{http://vizier.cfa.harvard.edu/viz-bin/VizieR?-source=J/ApJS/219/4}{VizieR database}\footnote{\href{http://vizier.cfa.harvard.edu/viz-bin/VizieR?-source=J/ApJS/219/4}{http://vizier.cfa.harvard.edu/viz-bin/VizieR?-source=J/ApJS/219/4}}.
\par
For the disc, the exponentional law for the sutface brightness is adopted (``expdisc'')
$$
I_\mathrm{d}(R) = I_0 \, \exp(-R/h) \, ,
$$
where $I_0$ is the central surface brightness of the disc observed face-on and $h$ denotes the exponential scale length. The bulge component is described with a S{\' e}rsic profile (``sersic'')
$$
\displaystyle
I_\mathrm{b}(R) = I_\mathrm{e} \, \exp \left(-k \left[ \left(R/r_\mathrm{e} \right)^{1/n_\mathrm{b}} - 1 \right] \right) \, ,
$$
where $I_\mathrm{e}$ is the surface brightness at the effective radius
$r_\mathrm{e}$ (isophotal radius encompassing half of the total flux of
the component). The S{\' e}rsic-index $n_\mathrm{b}$ describes the shape
of the radial profile, which becomes steeper with increasing $n_\mathrm{b}$. For example, $n_\mathrm{b}=1$ corresponds to an
exponential profile and $n_\mathrm{b}=4$ to a de Vaucouleurs profile.
The factor $k$ is a normalization constant determined by $n_\mathrm{b}$. The galaxy NGC~5055 contains a central component (a small bulge) with angular size so small that it cannot be resolved in the S$^4$G images
($r_\mathrm{e} < \mathrm{FWHM} = 2\overset{''}{.}1$ of S$^4$G images). This component is fit with a PSF-convolved point source and denoted in Table~\ref{tab:general} as `n' (nuclear core). In this case the free parameter is the total magnitude $m_\mathrm{b}$. In \citet{Salo_etal2015} the free parameters are $m_\mathrm{d} = - 2.5 \log_{10} (2 \pi \, I_0 \,   h^2)$ (instead of $I_0$) and the integrated magnitude $m_\mathrm{b}$ (instead of $I_\mathrm{e}$).
\par
For the galaxies studied, the decomposition parameters are listed in Table~\ref{tab:phot_decomposition}. The stellar magnitudes are given in the AB system ($M_\odot=6.02$).

\begin{table*}
\centering
\footnotesize
\caption{Two-Component decomposition parameters \citep{Salo_etal2015}}
\begin{tabular}{|r|r|r|r|r|r|r|r|r|r|r|}
\hline
\multicolumn{2}{|c|}{} & \multicolumn{5}{c|}{Bulge} & \multicolumn{4}{c|}{Disc} \\
\cline{3-11}
\multicolumn{1}{|c|}{NGC} & \multicolumn{1}{|c|}{$R_\mathrm{eff}$} & \multicolumn{1}{c|}{$f_\mathrm{b}$} & \multicolumn{1}{c|}{$m_\mathrm{b}$} & \multicolumn{1}{c|}{$r_\mathrm{e}$} & \multicolumn{1}{c|}{$n_\mathrm{b}$} & \multicolumn{1}{c|}{$f_\mathrm{d}$} & \multicolumn{1}{c|}{$m_\mathrm{d}$} & \multicolumn{1}{c|}{$q_\mathrm{d}$} & \multicolumn{1}{c|}{$h$} & 
\multicolumn{1}{c|}{$\mu_0$}
\\
& 
\multicolumn{1}{c|}{$('')$} & 
\multicolumn{1}{c|}{} & 
\multicolumn{1}{c|}{(mag)} & 
\multicolumn{1}{c|}{$('')$} & 
\multicolumn{1}{c|}{} & 
\multicolumn{1}{c|}{} & 
\multicolumn{1}{c|}{(mag)} & 
\multicolumn{1}{c|}{} & 
\multicolumn{1}{c|}{$('')$} & 
\multicolumn{1}{c|}{}
\\
\hline
2841 & 114.03 & \multicolumn{1}{|l|}{0.166} & \multicolumn{1}{|l|}{10.50} & 10.92 & 2.25 &
         0.834 & 8.75 & 0.457 & 56.28 & 19.50\\
\hline
3521 & 76.76 & \multicolumn{1}{|l|}{0.104} & \multicolumn{1}{|l|}{10.57} & 8.31 & 2.87 &
         0.674 & 8.54 & 0.514 &  39.05 & 18.49\\
&&&&&&   0.222 & 9.73 & 0.453 & 108.98 & 21.91\\
\hline
5055   & 114.94 & \multicolumn{1}{|l|}{0.016$^\dag$} & 12.46$^\dag$ & - & - &
         0.170 & 9.88 & 0.556 &  14.99 & 17.75\\
&&&&&&   0.814 & 8.18 & 0.556 &  73.57 & 19.51\\
\hline
{7331} & {50.0\phantom{0}} & \multicolumn{1}{|l|}{{0.003$^\ddag$}} & \multicolumn{1}{|l|}{{14.64$^\ddag$}} & {-} & {-} &
         {0.573} & {8.86} & {0.642} & {29.55} & {18.25}\\
     & & \multicolumn{1}{|l|}{{0.096}} & \multicolumn{1}{|l|}{{10.80}} & {6.5\phantom{0}} & {2.74} & {0.328} & {9.46} & {0.571} & {107\phantom{.00}} & {21.85}\\
\hline
\hline
\multicolumn{11}{p{0.67\textwidth}}
{{\textit{Description:} ($R_\mathrm{eff}$) --- the effective radius from one-component decomposition, ($f_\mathrm{b}$) --- ``sersic'' fraction of the total model flux, ($m_\mathrm{b}$) --- ``sersic'' total 3.6 $\mu$m AB magnitude, ($r_\mathrm{e}$) --- ``sersic'' effective radius, ($n_\mathrm{b}$) --- ``sersic'' parameter, ($f_\mathrm{d}$) --- ``expdisc'' fraction of the total model flux, ($m_\mathrm{d}$) --- ``expdisc'' total 3.6 $\mu$m AB magnitude, ($q_\mathrm{d}$) --- ``expdisc'' axis ratio, ($h$) --- ``expdisc'' exponential scale length, ($\mu_0$) --- ``expdisc'' central surface face-on brightness. \textit{Note}: for the unresolved nuclear cores, symbols $\dag$ and $\ddag$ denote the PSF fraction of the total model flux and the PSF total 3.6 $\mu$m AB magnitude. \textit{Source:} \citep{Salo_etal2015}. For NGC~7331 --- parameters of decomposition were determined in the present paper.} }%
\end{tabular}
\label{tab:phot_decomposition}
\end{table*}

Photometric decomposition gives the surface brightness profile $I(R)$ of all stellar components. To go to surface densities, one needs to know the ratio $\gamma_{3.6}= M/L$. It was calculated using the calibration formula from \cite{Querejeta+2015}. Unlike previous studies, the $M/L$ ratio was calculated without intermediate recalibration using 2MASS data in the $K_\mathrm{s}$-band. To calculate the colour index [3.6~$\mu$m] -- [4.5~$\mu$m], we used integral magnitudes in the corresponding bands from \citet{Sheth+2010}. The corresponding values of $\gamma_{3.6}$ can be found in Table~\ref{tab:M_L}. They were calculated via the formula~\eqref{eq:ml} and calibrated for a ``diet'' Salpeter IMF, equivalent to a standard Salpeter (\citeyear{Salpeter1955}) IMF, but with a reduced number of very low mass stars. The values $\gamma_{3.6}-0.15\mathrm{dex}$ correspond to the Kroupa (\citeyear{Kroupa2001}) IFM. The calibration accuracy is $\pm0.1\mathrm{dex}$.

\begin{equation}
\mathrm{lg}(M/L) = -0.339(\pm0.057)\times([3.6~\mu\mathrm{m}]-[4.5~\mu\mathrm{m}]) - 0.336(\pm0.002) 
\label{eq:ml}
\end{equation}

\begin{table}
\centering
\footnotesize
\caption{Mass-luminosity ratio}
\begin{tabular}{|l|c|c|c|c|c|}
\hline
NGC     & 3.6$\mu$m & 4.5$\mu$m & [3.6] -- [4.5]$\mu$m 
& $\gamma_{3.6}$ & $\gamma_{3.6}$ - 0.15dex \\
\hline
2841 & 8.713 & 9.204 & -0.491 & 0.68 & 0.48 \\
\hline
3521 & 8.202 & 8.656 & -0.454 & 0.66 & 0.47 \\ 
\hline
5055 & 8.018 & 8.469 & -0.451 & 0.66 & 0.46 \\ 
\hline
{7331} & {8.250} & {8.725} & {-0.475} & {0.67} & {0.47} \\
\hline
\end{tabular}
\label{tab:M_L}
\end{table}

\begin{figure}
\centering
\includegraphics[width=0.46\textwidth]{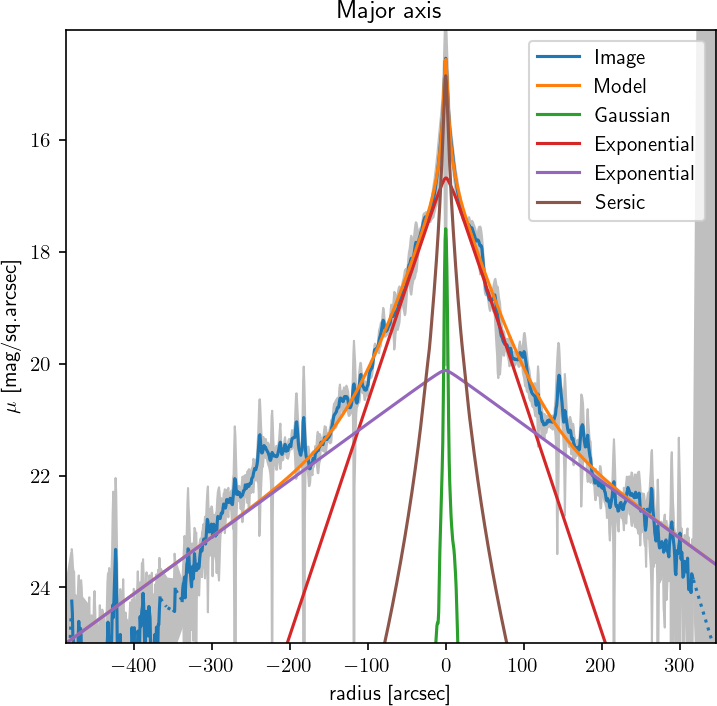}
\caption{The surface brightness profile in the band 3.6~$\mu$m and its decomposition for the galaxy NGC~7331. 
}
\label{fig:N7331_mu}
\end{figure}

NGC~7331 is not included in the S$^4$G survey. For this galaxy the surface brightness profile was extracted from the Spitzer telescope archive.  
{For this galaxy we constructed our own photometric model. It consists from an unresolved nuclear (PSF), a small bulge and two exponential discs (see Table~\ref{tab:phot_decomposition}).} Fig.~\ref{fig:N7331_mu} shows the surface brightness profile {of the galaxy NGC~7331 in the band 3.6~$\mu$m and the profiles of all components.} Stellar magnitudes are given in the AB system. The ratio $M/L$ was calculated using the calibration formula~\eqref{eq:ml}. 

\section{Dark halo models}
\label{sec:dark_halo}
Three models were used to describe the dark matter halo: the pseudo-isothermal (ISO) halo model, the Navarro-Frank-White (NFW) model with a density peak at the centre \citep{NFW1996,NFW1997}, and the Burkert \citeyearpar{Burkert1995} model. The density profile in the Burkert model coincides with the NFW density profile at large distances from the centre, but in the centre the density does not have a peak, but reaches a final value.
\par
\citet{deBlok+2008}, who used the ISO model and the NFW model, came to the conclusion that for bright galaxies with $M_B<- 19$ both dark halo models are equally well suited for describing rotation curves. Several galaxies are distinguished, for which the formal solution for the NFW model gives the model parameters that are incompatible with cosmological simulations (either very small values of the halo concentration parameter, or too large). For dwarf galaxies with $M_B>-19$, the ISO halo model is more suitable. In addition, \citet{Saburova+2016} showed that the Burkert dark halo model (with a smoothed density profile at the centre compared to the NFW model) gives a nondegenerate solution for almost all galaxies, in contrast to the ISO and NFW models. The authors stress that models with NFW profile fail more often than madels with other profiles when the baryonic surface density is fixed.

\subsection{The pseudo-isothermal (ISO) model}
\label{sec:iso}
\par
The ISO dark halo model has a radial density profile:
\begin{equation}
\rho_\mathrm{iso} = \frac{\rho_0}{1+(r/R_\mathrm{c})^2} \, ,
\label{eq:iso}
\end{equation}
where $\rho_0$ is the central density, and $R_\mathrm{c}$ is the radius of the core, where the density is almost constant.
\par
The contribution to the rotation curve is calculated by the formula
\begin{equation}
V^2_\mathrm{h,iso}(R) = V^2_{\infty} \left[1 - 
\frac{R_\mathrm{c}}{R} \arctan\left(\frac{R}
{R_\mathrm{c}} \right)  \right] \, .
\label{eq:v_iso}
\end{equation}
At $R \to \infty$, the circular speed reaches a constant value $V_ {\infty} = \sqrt{4 \pi G \rho_0 R_\mathrm{c}^2}$. The model has two fit parameters $V_{\infty}$ and $R_\mathrm{c}$.

\subsection{The Navarro–Frenk–White (NFW) model}
\label{sec:NFW}
The Navarro-Frank-White model (NFW) \citep{NFW1996, NFW1997} was proposed on the basis of numerical cosmological simulations of the dark halo formation. Its density profile is
\begin{equation}
\rho_\mathrm{nfw} = \frac{\rho_\mathrm{s}}{r/r_\mathrm{s} \left(1 + r/r_\mathrm{s} \right)^2} \, ,
\label{eq:nfw}
\end{equation}
where the density $\rho_\mathrm{s}$ and the scale length $r_\mathrm{s}$ are parameters of the model. It is believed that this profile is universal, its parameters practically do not depend on the halo mass, rotation and formation epoch, but depend on the chosen cosmological model. At small distances from the center, the density diverges (there is a density peak). 
\par
Typically, two other parameters (instead $\rho_\mathrm{s}$ and $r_\mathrm{s}$) are used to describe the NFW halo profile.
The scale length $r_\mathrm{s}$ can be expressed trough $r_{200}$, the radius of the sphere, inside which the average density of dark matter is 200 times greater than the critical density of the Universe. This can be written via the condition
$$
M_{200} \equiv 200 \rho_\mathrm{cr} \frac{4 \pi}{3} r_{200}^3 \, ,
$$
where $M_{200}$ is the total mass of dark matter inside a sphere of radius $r_{200}$. The radius $r_{200}$ is close to the radius of the sphere, inside which the dark halo is virialized, i.e. is in dynamic equilibrium. Because of this, $M_{200}$ is often referred to as the virial mass. The radius $r_{200}$ for a given cosmology can be expressed through the circular velocity at this radius $r_{200}$ \citep{Mo+1998}:
$$
r_{200} = \frac{V_{200}}{10 H(z)},
$$
here $H(z)$ is the Hubble constant at redshift $z$. The coefficient ``10'' is not an exact number, but depends on the accepted ``cosmology''. For the standard LCDM cosmology, it is about ten. Then the first independent (fitting) parameter of the model can be considered as $V_{200}$.
\par
Another parameter can be the concentration parameter
$$
c \equiv \frac{r_{200}}{r_\mathrm{s}} \, .
$$
The concentration parameter $c$ is not completely independent and is rather strongly related to the history of merges, i.e. generally speaking it depends on the halo mass and ``cosmology''. But within the same ``cosmology'' there is a scatter of this parameter, and in the first approximation its dependence on the virial mass (or $V_{200}$) can be neglected, considering $c$ to be a free (fitting) parameter.
\par
If we denote $R_\mathrm{h}=R/r_{200}$, then the rotation curve of this model has the following form:
\begin{equation}
V^2_\mathrm{h,nfw}(R_\mathrm{h}) = V^2_{200}\, \frac{f_\mathrm{nfw}(R_\mathrm{h} c)}{R_\mathrm{h} f_\mathrm{nfw}(c)}\, ,
\label{eq:v_nfw}
\end{equation}
where $\displaystyle f_\mathrm{nfw}(x) = \lg(1+x) - \frac{x}{1+x}$
The rotation curve of such a model has two fit parameters, the velocity $V_{200} $ and the concentration parameter $c$.

\subsection{The Burkert model}
\label{sec:Burkert}
The density profile for the Burkert \citeyearpar{Burkert1995} model is
\begin{equation}
\rho_\mathrm{burk} = \frac{\rho_\mathrm{s}}{(r + r_\mathrm{s}) \left(r^2 + r_\mathrm{s}^2 \right)} \, .
\label{eq:burk}
\end{equation}
At large distances from the centre, this profile behaves in the same way as the density profile~\eqref{eq:nfw} for the NFW model. At $r=r_\mathrm{s}$ both profiles give the same density value $\rho_\mathrm{s}/4$. For the same values of the core radius $r_\mathrm{s}$ and the density $\rho_\mathrm{s}$, the mass of the NFW halo inside the core radius is approximately 1.5 times greater than the mass in the Burkert model. Inside a sphere of radius $3r_\mathrm{s}$ both masses are approximately equal, and then the cumulative mass of the NFW model is about 10\% less than the cumulative mass of the Burkert model. Thus, the NFW model has a slightly higher concentration of matter towards the center. The maximum of the rotation curve in the NFW model is reached approximately at $2r_\mathrm{s}$, and in the Burkert model at approximately $3r_\mathrm{s}$.
\par
It is convenient to describe the rotation curve of this model with parameters similar to those for describing the rotation curve of the NFW model. By analogy, one can take the parameter $V_{B200}$ --- the circular velocity at the virial radius $r_{B200}$, and the concentration parameter $c_B=r_{B200}/r_\mathrm{s}$. Then the circular speed at the distance $R_\mathrm{h}=R/r_{B200}$ can be calculated by the formula
\begin{equation}
V^2_\mathrm{h,burk}(R_\mathrm{h}) = V^2_{B200}\, 
\frac{f_\mathrm{burk}(R_\mathrm{h} c_B)}{R_\mathrm{h}f_\mathrm{burk}(c_B)} \, ,
\label{eq:v_burk}
\end{equation}
where $\displaystyle f_\mathrm{burk}(x) = \lg\frac{1 + x}{\sqrt{1 + x^2}}  - 
\arctan (x)$. Similar to the NFW model the rotation curve of the Burkert model has two fit parameters, the velocity $V_{B200} $ and the concentration parameter $c_B$.

\subsection{Adiabatic contraction of the NFW dark halo}
\label{sec:AC}
The compression of dark matter halos by baryonic component distorts the initial NFW density profiles predicted by cosmological simulations. We use the algorithm by \citet{Blumenthal+1986} to account for the compession. 
This scheme assumes a spherically symmetric halo and particles in circular orbits. If the halo responds adiabatically to the slow assembly of the disc and remains spherical as it contracts, the angular momentum of the individual dark matter particles is conserved, e.g. $rV_\mathrm{c}(r)=\mathrm{const}$. For circular orbits, if, in addition, the distribution of the baryons has spherical symmetry, we have $V^2_\mathrm{c}= G M(r)/r$, where $M(r)$ is the enclosed mass within radius $r$. This
reduces further to $rM(r)=\mathrm{const}$. \citet{Blumenthal+1986} formulate the last condition as following
\begin{equation}
r_\mathrm{i} M(r_\mathrm{i}) = r_\mathrm{f} M_\mathrm{f}(r_\mathrm{f}) \, ,
\label{adiab}
\end{equation}
where $r_\mathrm{i}$ and $r_\mathrm{f}$ are the initial and final radii of a spherical shell of the dark halo. $M(r)$ gives the initial distribution of the matter (dark and baryonic) with the NFW density profile NFW, $M_\mathrm{f}(r)$ is the final mass profile.
\par
When the bulge is also presented $M_\mathrm{f}(r)$ and $M(r)$ are connected as \citep{Dutton+2007}:
\begin{equation}
M_\mathrm{f}(r_\mathrm{f}) = M_\mathrm{d}(r_\mathrm{f}) + M_\mathrm{b}(r_\mathrm{f}) + 
(1 - m_\mathrm{d} - m_\mathrm{b}) M(r_\mathrm{i}) \, ,
\label{mass_f}
\end{equation}
where $m_\mathrm{d} \equiv M_\mathrm{d}/M(r_{200})$ and $m_\mathrm{b} \equiv M_\mathrm{b}/M(r_{200})$.
\par
Then the dark halo mass profile $M_\mathrm{h}(r)$ after adiabatic contraction will be
\begin{equation}
M_\mathrm{h}(r) = M_\mathrm{f}(r) - M_\mathrm{d}(r) - M_\mathrm{b}(r) \, .
\label{mass_h}
\end{equation}
\par
In fact, it is necessary to find the relationship between the initial radius
$r_\mathrm{i}$ of the dark halo spherical shell and its final value $r_\mathrm{f}$. This dependency, based on Eqs.~\eqref{adiab} and \eqref{mass_f},
can be written as
\begin{equation}
r_\mathrm{i} = r_\mathrm{f} \left(1 - m_\mathrm{d} - m_\mathrm{b}+ 
\frac{M_\mathrm{d}(r_\mathrm{f})+M_\mathrm{b}(r_\mathrm{f})}{M(r_\mathrm{i})} \right) \, ,
\label{rad}
\end{equation}
where $M(r_\mathrm{i})$ is the mass distribution in the NFW model. Eq.~\eqref{rad} is solved iteratively and $r_\mathrm{i}$ is found as the function of $r_\mathrm{f}$. After this, the mass distribution in the compressed halo is defined as
\begin{equation}
M_\mathrm{h}(r_\mathrm{f}) =  
M(r_\mathrm{i}(r_\mathrm{f})) \, .
\nonumber
\end{equation}
\par
Blumental scheme overpredicts the compression of the halo and, for, example \citet{Katz+2014} used another formalism, by \citet{Young_1980}, and applied it to a sample of galaxies from the THINGS. The Young's method conserves the radial action in addition to the angular momentum and leads to a less strong compression of the halo, so that the initial values of the concentration parameter may still turn out to be too large. That is why we used the Blumental scheme for the galaxy NGC~2841.

\section{Results of decomposition}
\label{sec:decomposition}
When decomposing the rotation curves, we considered the contribution of four subsystems: the bulge, the stellar disc, the gas disc, and the dark halo. For stellar components (bulge and disc) we have only photometric models. Models of mass distribution are obtained by taking into account factors $(M/L)_\mathrm{b}$ and $(M/L)_\mathrm{d}$, respectively. Then, the contribution of each subsystem is accounted for by the formula
\begin{equation}
V^2(R) = (M/L)_\mathrm{b} \, V^2_\mathrm{b}(R) + (M/L)_\mathrm{d} \, V^2_\mathrm{d}(R) + 
V^2_\mathrm{g}(R) + V^2_\mathrm{h}(R) \, ,
\end{equation}
where $V_\mathrm{d}$, $V_\mathrm{b}$, $V_\mathrm{g}$ and $V_\mathrm{h}$ are the contributions to the circular speed by the bulge, disc,  $\mathrm{HI}$ and $\mathrm{H}2$ discs and the dark matter halo, respectively, $(M/L)_\mathrm{d}$ ($(M/L)_\mathrm{b}$) are the disc (bulge) mass-to-light ratio.
\par
Bulge (typical for galaxies of early morphological types) makes the main contribution to the rotation curve in the central region of the galaxy. The rotation curve, associated only with the bulge, rapidly decreases towards the periphery. For the bulge, the Sersic model \citep{Sersic1968} of the surface brightness profile is adopted. The contribution of the bulge to the rotation curve of the galaxy was calculated according to \citet{Noordermeer2008}. The contribution of the exponential stellar disc to the rotation curve was calculated using the formula by \citet{Casertano1983}. The thickness of the disc was taken into account. We adopted $z_0 = 0.2 h$, where $z_0$ is the vertical scale and $h$ is the exponential scale of the disc. The contribution of the gas to the rotation curve is calculated by the formula for an infinitely thin disc \citep{Freeman1970} with a radial surface density profile $\Sigma_\mathrm{g}(R)$. The formulas for the dark halo contribution are described in Section~\ref{sec:dark_halo}. 
\par
The decomposition of the rotation curves refers to the so-called problems with degeneracy, when different sets of free parameters can give equally good (in the sense of the smallness of the $\chi^2$ value) solutions. There is always arbitrariness.
\par
There are four free parameters: $(M/L)_\mathrm{d}$, $(M/L)_\mathrm{b}$ and two parameters that describe the halo model. Preference is given to models with a fixed value of $(M/L)_\mathrm{d}$, which corresponds to the Kroupa IMF. The value of $(M/L)_\mathrm{b}$, as a rule, varied freely. Solutions with unrealistically large ($>1.3$) and unrealistically small values ($<0.1$) of $(M/L)_\mathrm{b}$ were discarded. Sometimes it was necessary to vary $(M/L)_\mathrm{d}$ within the calibration error. Below are only solutions for which iterations converge and which give the physical values of the parameters.
\par
Decomposition of the rotation curve is the so-called multidimensional nonlinear least-squares fitting. To find the best-fit parameters we used the weighted nonlinear least-squares fitting solver from the gsl (GNU Scientific Library) library, which is based on the Levenberg–Marquardt algorithm. The algorithm uses a generalized trust region to keep each step of fitting under control. The gsl solver finds a solution even if it starts very far off the final minimum. We tested this by randomly varying the initial guess. Additionally, we made 20,000 random realizations of the observed rotation curve, varying it within the error and assuming that the error is Gaussian distributed. The solutions converge to our best $\chi^2$ solution for the original observational rotation curve (see Appendix~\ref{sec:appendix2}). When fitting, we discarded the most central points on the rotation curve, since galaxies are observed at rather high inclination and there is a beam-smearing problem for them. Because of this, difficulties could arise in finding a solution for $M/L_\mathrm{b}$ if $M/L_\mathrm{d}$ was fixed. In this case, the solver tends to attribute the rotation curve at the center to the disc only, giving negative or very small values for the bulge. We rejected these solutions for NGC~3521 and NGC~7331 and fixed $(M/L)_\mathrm{b}=1.4 (M/L)_\mathrm{d}$. Such a fixation is close to that done in \citet{ManceraPina+2022}, although there the factor before $(M/L)_\mathrm{d}$ was a completely free parameter in the rotation curve decomposition\footnote{\citet{ManceraPina+2022} used a more sophisticated technique in such cases. They imposed a prior $(M/L)_\mathrm{b}=1.4(M/L)_\mathrm{d}$, varying the factor 1.4 with a small Gaussian scatter.}. For galaxies NGC~2841 and NGC~5055, solutions for the fixed and free ratios $M/L_\mathrm{d}$, although they led to a redistribution of mass between the bulge and disc and slightly different dark halo parameters, did not generally change the conclusion that these galaxies have a reduced content of dark matter within four radial scale of a disc.
\par
Tables~\ref{tab:ISO_RCdec}-\ref{tab:Burkert_RCdec} summarize fit parameters for three types of halos. $M_\mathrm{h}/M_\mathrm{baryon}$ gives the halo-to-baryonic mass ratio within four exponential scales of the heaviest disc while $f_\mathrm{DM}(R_\mathrm{eff})$ is the relative dark matter mass within the effective radius of the galaxy.

\begin{table*}
\caption{ISO dark halo parameters}
\centering
\begin{tabular}{|l|l|c|c|r|l|c|c|c|}
\hline
\hline\multicolumn{9}{|c|}{ISO halo}\\ 
\hline
Name & \multicolumn{2}{|c|}{$(M/L)_\mathrm{d}$} & $(M/L)_\mathrm{b}$ &
\multicolumn{1}{|c|}{$R_\mathrm{c}$} & \multicolumn{1}{|c|}{$V_{\infty}$} & $\chi$ &
$M_\mathrm{h}/M_\mathrm{baryon}$ & $f_\mathrm{DM}(R_\mathrm{eff})$\\ 
& & & (free) & \multicolumn{1}{|c|}{(kpc)} & \multicolumn{1}{|c|}{ (km/s)} & & &\\
\hline
NGC~2841 & fixed & 0.68 & \textcolor{black}{$1.09 \pm 0.19$} & \textcolor{black}{$2.7 \pm \phantom{1}0.5$} & \textcolor{black}{$261.8 \pm \phantom{1}2.8$} & \textcolor{black}{0.23} & \textcolor{black}{1.34} & \textcolor{black}{0.41}\\
NGC~2841 & free & \textcolor{black}{$0.72 \pm 0.26$} & \textcolor{black}{$1.15 \pm 0.41$} & \textcolor{black}{$3.2 \pm \phantom{1}3.5$} & \textcolor{black}{$262.4 \pm \phantom{1}5.7$} & \textcolor{black}{0.23} & \textcolor{black}{1.21} & \textcolor{black}{0.41}\\
\hline
\hline
NGC~3521 & fixed & \multicolumn{1}{|c|}{-} & \multicolumn{1}{|c|}{-} & \multicolumn{1}{|c|}{-} & \multicolumn{1}{|c|}{-} & \multicolumn{1}{|c|}{-} & \multicolumn{1}{|c|}{-} & \multicolumn{1}{|c|}{-}\\
NGC~3521 & free & \textcolor{black}{$0.25 \pm 0.01$} & \textcolor{black}{$0.35 \pm 0.01$} & \textcolor{black}{$4.6 \pm 1.0$} & \textcolor{black}{$176.9 \pm \phantom{1}5.6$} & \textcolor{black}{1.07} & \textcolor{black}{0.50} & \textcolor{black}{0.16}\\
\hline
\hline
NGC~5055 & fixed & 0.46 & \textcolor{black}{$0.38 \pm 0.04$} & \textcolor{black}{$6.3 \pm \phantom{1}0.4$} & \textcolor{black}{$175.8 \pm \phantom{1}3.3$} & \textcolor{black}{1.22} & \textcolor{black}{0.63} & \textcolor{black}{0.13}\\
NGC~5055 & free & \textcolor{black}{$0.40 \pm 0.02$} & \textcolor{black}{$0.47 \pm 0.05$} & \textcolor{black}{$4.0 \pm \phantom{1}0.6$} & \textcolor{black}{$168.3 \pm \phantom{1}2.6$} & \textcolor{black}{0.86} & \textcolor{black}{0.89} & \textcolor{black}{0.22}\\
\hline
\hline
NGC~7331 & fixed & \multicolumn{1}{|c|}{-} & \multicolumn{1}{|c|}{-} & \multicolumn{1}{|c|}{-} & \multicolumn{1}{|c|}{-} & \multicolumn{1}{|c|}{-} & \multicolumn{1}{|c|}{-} & \multicolumn{1}{|c|}{-}\\
NGC~7331 & free & \textcolor{black}{$0.31 \pm 0.53$} & \textcolor{black}{$0.43 \pm 0.74$} & \textcolor{black}{$2.2 \pm 13.7$} & \textcolor{black}{$198.0 \pm 10.3$} & \textcolor{black}{0.84} & \textcolor{black}{0.71} & \textcolor{black}{0.41}\\
\hline
\hline
\end{tabular}
\label{tab:ISO_RCdec}
\end{table*}

\begin{table*}
\caption{NFW dark halo parameters}
\centering
\begin{tabular}{|l|l|c|c|r|l|c|c|c|c|c|c|c|c|}
\hline
\hline
\multicolumn{9}{|c|}{NFW halo}\\ 
\hline
Name & \multicolumn{2}{|c|}{$(M/L)_\mathrm{d}$} & $(M/L)_\mathrm{b}$ & \multicolumn{1}{|c|}{$c$} & \multicolumn{1}{|c|}{$V_{200}$} & $\chi$ & $M_\mathrm{h}/M_\mathrm{baryon}$ & $f_\mathrm{DM}(R_\mathrm{eff})$\\ 
& & & (free) & & \multicolumn{1}{|c|}{(km/s)} & & &\\
\hline
NGC~2841 & fixed & 0.68 & \textcolor{black}{$1.15 \pm 0.10$} & \textcolor{black}{$14.4 \pm 0.8$} & \textcolor{black}{$188.3 \pm \phantom{1}3.2$} & \textcolor{black}{0.26} & \textcolor{black}{1.32} & \textcolor{black}{0.39}\\
NGC~2841$^\#$ & fixed & 0.68 & \textcolor{black}{$0.59 \pm 0.11$} & \textcolor{black}{$5.0 \pm 0.3$} & \textcolor{black}{$246.4 \pm \phantom{1}8.3$} & \textcolor{black}{0.29} & \textcolor{black}{1.66} & \textcolor{black}{0.51}\\
NGC~2841 & free & \textcolor{black}{$0.83 \pm 0.10$} & \textcolor{black}{$1.07 \pm 0.12$} & \textcolor{black}{$11.2 \pm 2.0$} & \textcolor{black}{$198.3 \pm \phantom{1}8.8$} & \textcolor{black}{0.23} & \textcolor{black}{1.00} & \textcolor{black}{0.32}\\
\hline
\hline
NGC~3521 & fixed & \multicolumn{1}{|c|}{-} & \multicolumn{1}{|c|}{-} & \multicolumn{1}{|c|}{-} & \multicolumn{1}{|c|}{-} & \multicolumn{1}{|c|}{-} & \multicolumn{1}{|c|}{-} & \multicolumn{1}{|c|}{-}\\
NGC~3521 & free & \textcolor{black}{$0.22 \pm 0.01$} & \textcolor{black}{$0.31 \pm 0.01$} & \textcolor{black}{$11.7 \pm 2.3$} & \textcolor{black}{$129.0 \pm \phantom{1}6.8$} & \textcolor{black}{1.04} & \textcolor{black}{0.73} & \textcolor{black}{0.27}\\
\hline
\hline
NGC~5055 & fixed & 0.46 & \textcolor{black}{$0.32 \pm 0.06$} & \textcolor{black}{$4.3 \pm 0.5$} & \textcolor{black}{$169.8 \pm 11.3$} & \textcolor{black}{2.22} & \textcolor{black}{0.58} & \textcolor{black}{0.15}\\
NGC~5055 & free & \textcolor{black}{$0.34 \pm 0.02$} & \textcolor{black}{$0.47 \pm 0.05$} & \textcolor{black}{$11.5 \pm 1.3$} & \textcolor{black}{$125.2 \pm \phantom{1}3.5$} & \textcolor{black}{0.85} & \textcolor{black}{1.21} & \textcolor{black}{0.34}\\
\hline
\hline
NGC~7331 & fixed & \multicolumn{1}{|c|}{-} & \multicolumn{1}{|c|}{-} & \multicolumn{1}{|c|}{-} & \multicolumn{1}{|c|}{-} & \multicolumn{1}{|c|}{-} & \multicolumn{1}{|c|}{-} & \multicolumn{1}{|c|}{-}\\
NGC~7331 & free & \textcolor{black}{$0.32 \pm 0.03$} & \textcolor{black}{$0.45 \pm 0.04$} & \textcolor{black}{$14.0 \pm 4.6$} & \textcolor{black}{$141.1 \pm 13.2$} & \textcolor{black}{0.94} & \textcolor{black}{0.66} & \textcolor{black}{0.39}\\
\hline
\hline
\multicolumn{7}{p{0.57\textwidth}}
{\textit{Note}: Symbol $\#$ denotes the model with adiabatic contraction. See Section~\ref{sec:AC}.}
\end{tabular}
\label{tab:NFW_RCdec}
\end{table*}

\begin{table*}
\caption{Burkert dark halo parameters}
\centering
\begin{tabular}{|l|l|c|c|r|l|c|c|c|c|c|c|c|c|}
\hline
\hline
\multicolumn{9}{|c|}{Burkert halo}\\ 
\hline
Name & \multicolumn{2}{|c|}{$(M/L)_\mathrm{d}$} & $(M/L)_\mathrm{b}$ & \multicolumn{1}{|c|}{$c_B$} & \multicolumn{1}{|c|}{$V_{B200}$} & $\chi$ & $M_\mathrm{h}/M_\mathrm{baryon}$ & $f_\mathrm{DM}(R_\mathrm{eff})$\\ 
& & & (free) & & \multicolumn{1}{|c|}{(km/s)} & & &\\
\hline
NGC~2841 & fixed & 0.68 & \textcolor{black}{$1.35 \pm 0.10$} & \textcolor{black}{$8.0 \pm 0.6$} & \textcolor{black}{$151.6 \pm \phantom{1}0.5$} & \textcolor{black}{0.48} & \textcolor{black}{1.23} & \textcolor{black}{0.34}\\
NGC~2841 & free & \textcolor{black}{$0.95 \pm 0.09$} & \textcolor{black}{$1.07 \pm 0.12$} & \textcolor{black}{$6.0 \pm 1.1$}
& \textcolor{black}{$165.4 \pm \phantom{1}0.9$} & \textcolor{black}{0.25} & \textcolor{black}{0.76} & \textcolor{black}{0.24}\\
\hline
\hline
NGC~3521 & fixed & \multicolumn{1}{|c|}{-} & \multicolumn{1}{|c|}{-} & \multicolumn{1}{|c|}{-} & \multicolumn{1}{|c|}{-} & \multicolumn{1}{|c|}{-} & \multicolumn{1}{|c|}{-} & \multicolumn{1}{|c|}{-}\\
NGC~3521 & free & \textcolor{black}{$0.23 \pm 0.01$} & \textcolor{black}{$0.31 \pm 0.01$} & \textcolor{black}{$6.6 \pm 1.4$} & \textcolor{black}{$105.4 \pm \phantom{1}0.9$} & \textcolor{black}{1.09} & \textcolor{black}{0.63} & \textcolor{black}{0.22}\\
\hline
\hline
NGC~5055 & fixed & 0.46 & \textcolor{black}{$0.32 \pm 0.06$} & \textcolor{black}{$4.3 \pm 0.3$} & \textcolor{black}{$116.4 \pm \phantom{1}1.4$}  & \textcolor{black}{2.16} & \textcolor{black}{0.60} & \textcolor{black}{0.16}\\
NGC~5055 & free & \textcolor{black}{$0.38 \pm 0.02$} & \textcolor{black}{$0.45 \pm 0.05$} & \textcolor{black}{$6.2 \pm 0.7$}
& \textcolor{black}{$104.2 \pm \phantom{1}0.5$} & \textcolor{black}{0.88} & \textcolor{black}{1.02} & \textcolor{black}{0.26}\\
\hline
\hline
NGC~7331 & fixed & \multicolumn{1}{|c|}{-} & \multicolumn{1}{|c|}{-} & \multicolumn{1}{|c|}{-} & \multicolumn{1}{|c|}{-} & \multicolumn{1}{|c|}{-} & \multicolumn{1}{|c|}{-} & \multicolumn{1}{|c|}{-}\\
NGC~7331 & free & \textcolor{black}{$0.37 \pm 0.02$} & \textcolor{black}{$0.52 \pm 0.03$} & \textcolor{black}{$5.5 \pm 1.6$} & \textcolor{black}{$137.1 \pm \phantom{1}4.2$} & \textcolor{black}{1.09} & \textcolor{black}{0.39} & \textcolor{black}{0.27}\\
\hline
\hline
\end{tabular}
\label{tab:Burkert_RCdec}
\end{table*}

\subsection{NGC~2841}
NGC~2841 --- a spiral galaxy without a bar in the constellation Ursa Major.
\par
\begin{figure}
\centering
\includegraphics[width=0.23\textwidth]{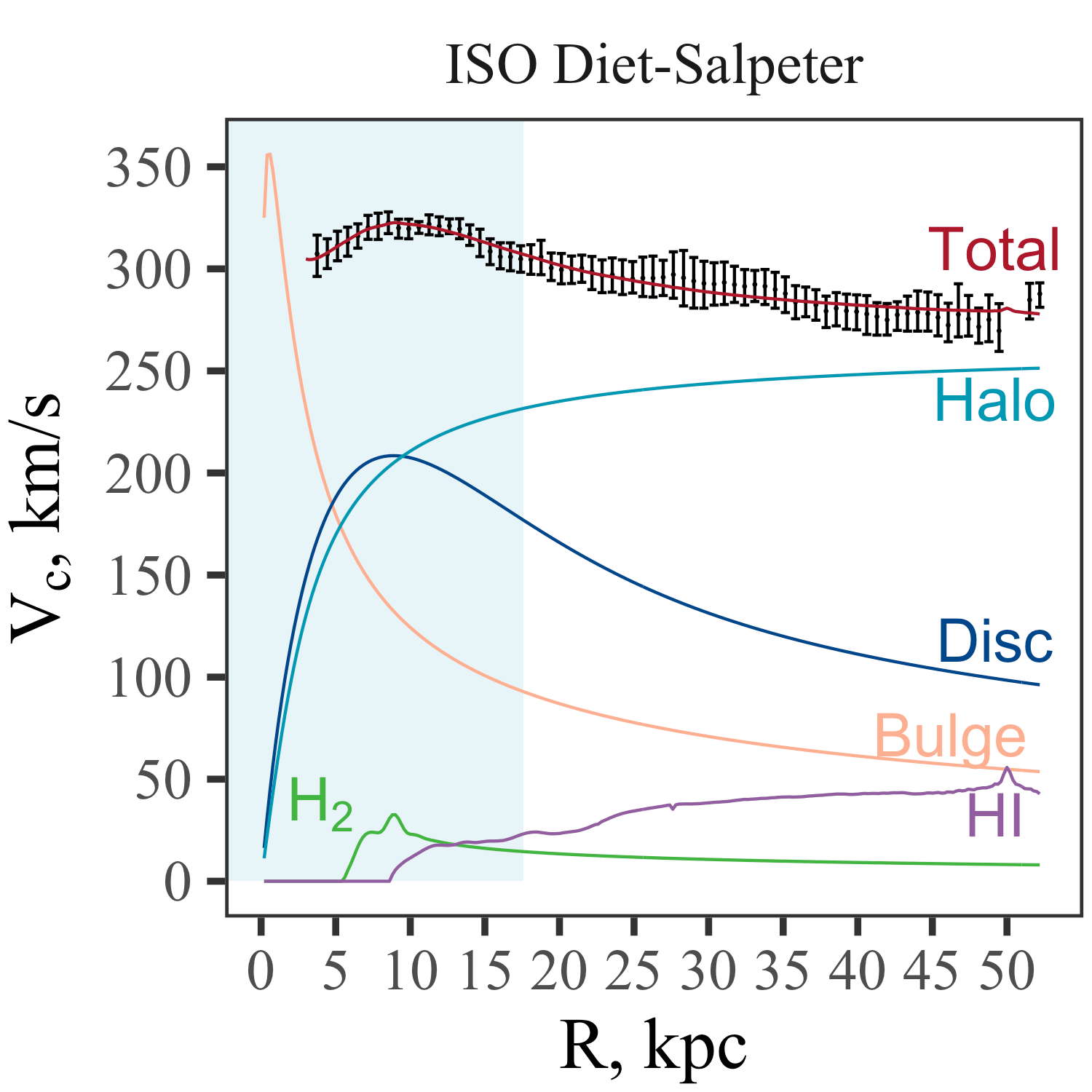}
\includegraphics[width=0.23\textwidth]{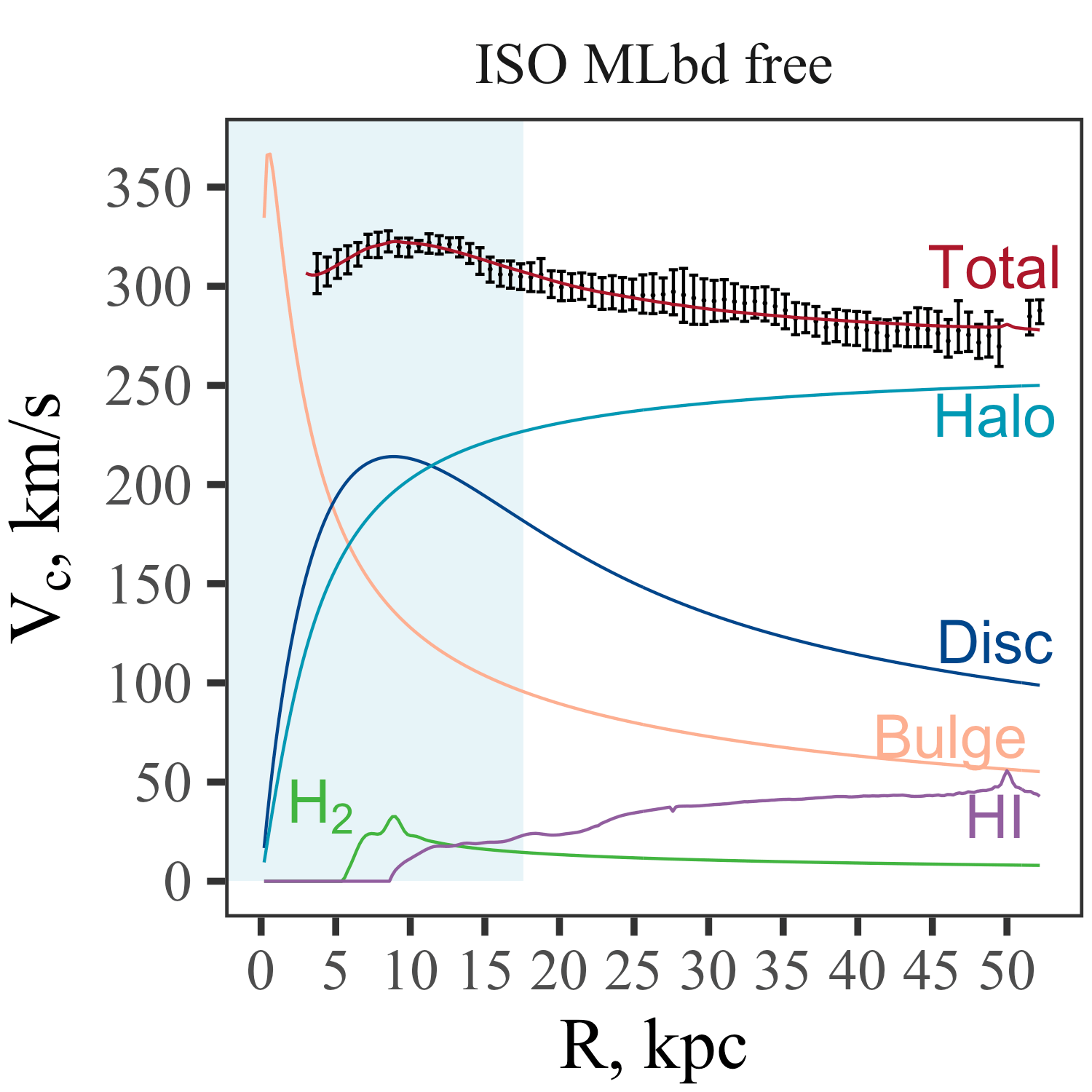}
\caption{Decomposition of the rotation curve~(black points) of NGC~2841 for the ISO halo. {\it Left} --- for a fixed value $(M/L)_\mathrm{d}=0.68$. {\it Right} --- with a free parameter $(M/L)_\mathrm{d}$. The colored zone corresponds to $R=4h$.}
\label{fig:N2841_ISO}
\end{figure}
\begin{figure}
\centering
\includegraphics[width=0.23\textwidth]{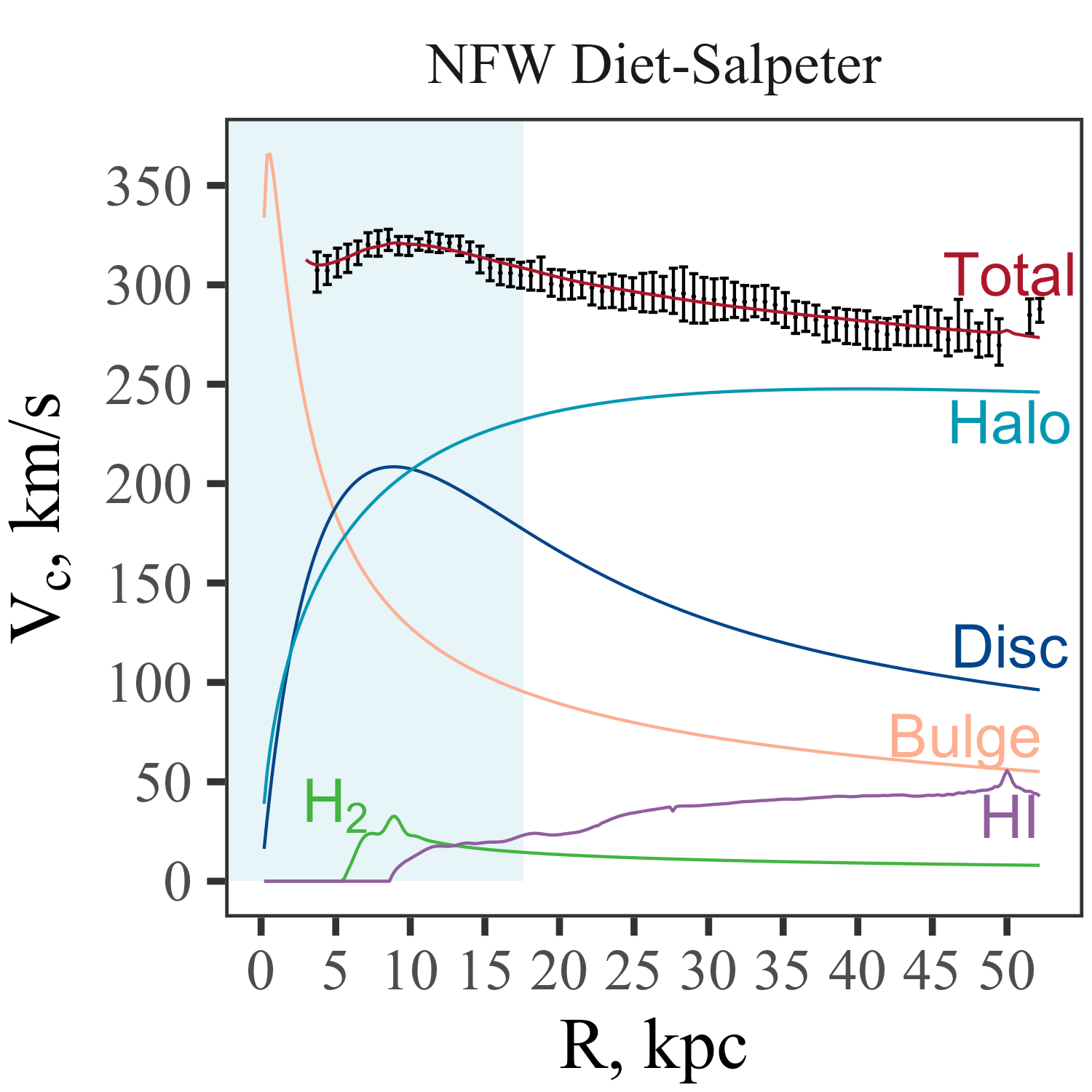}
\includegraphics[width=0.23\textwidth]{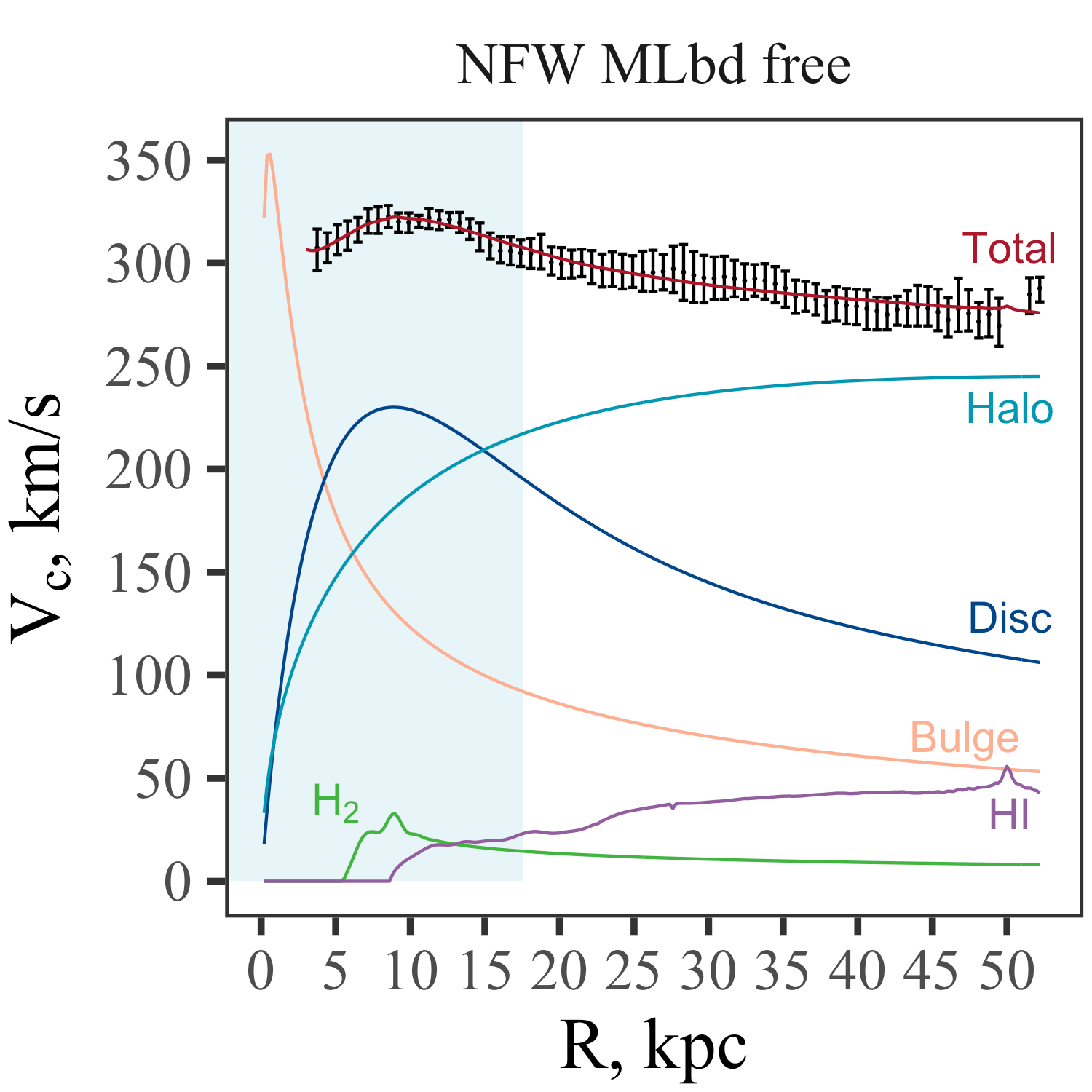}
\caption{Decomposition of the rotation curve of NGC~2841 for the NFW halo. Symbols, colored zone and $(M/L)_\mathrm{d}$ as in Fig.~\ref{fig:N2841_ISO}.}
\label{fig:N2841_NFW}
\end{figure}
\begin{figure}
\centering
\includegraphics[width=0.23\textwidth]{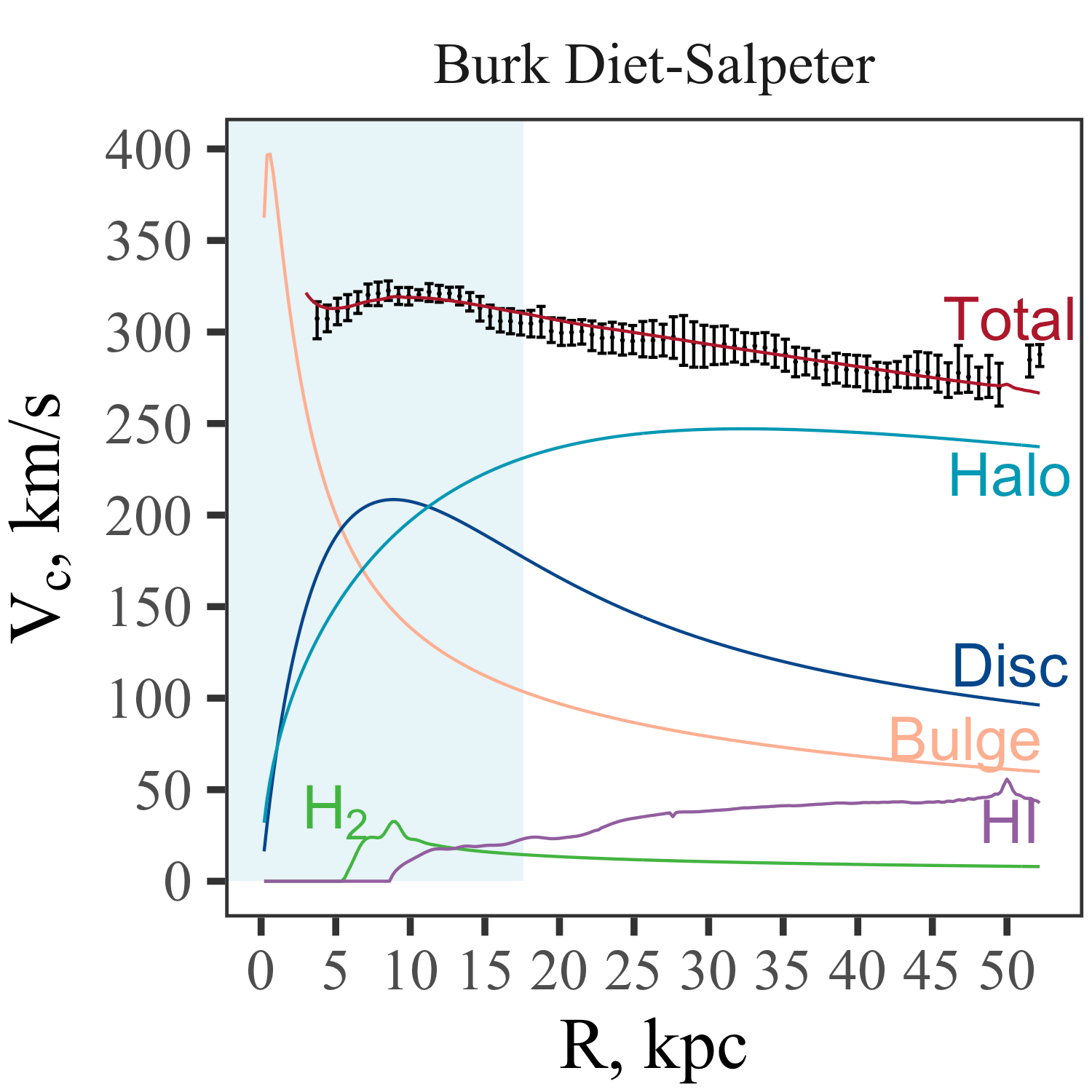}
\includegraphics[width=0.23\textwidth]{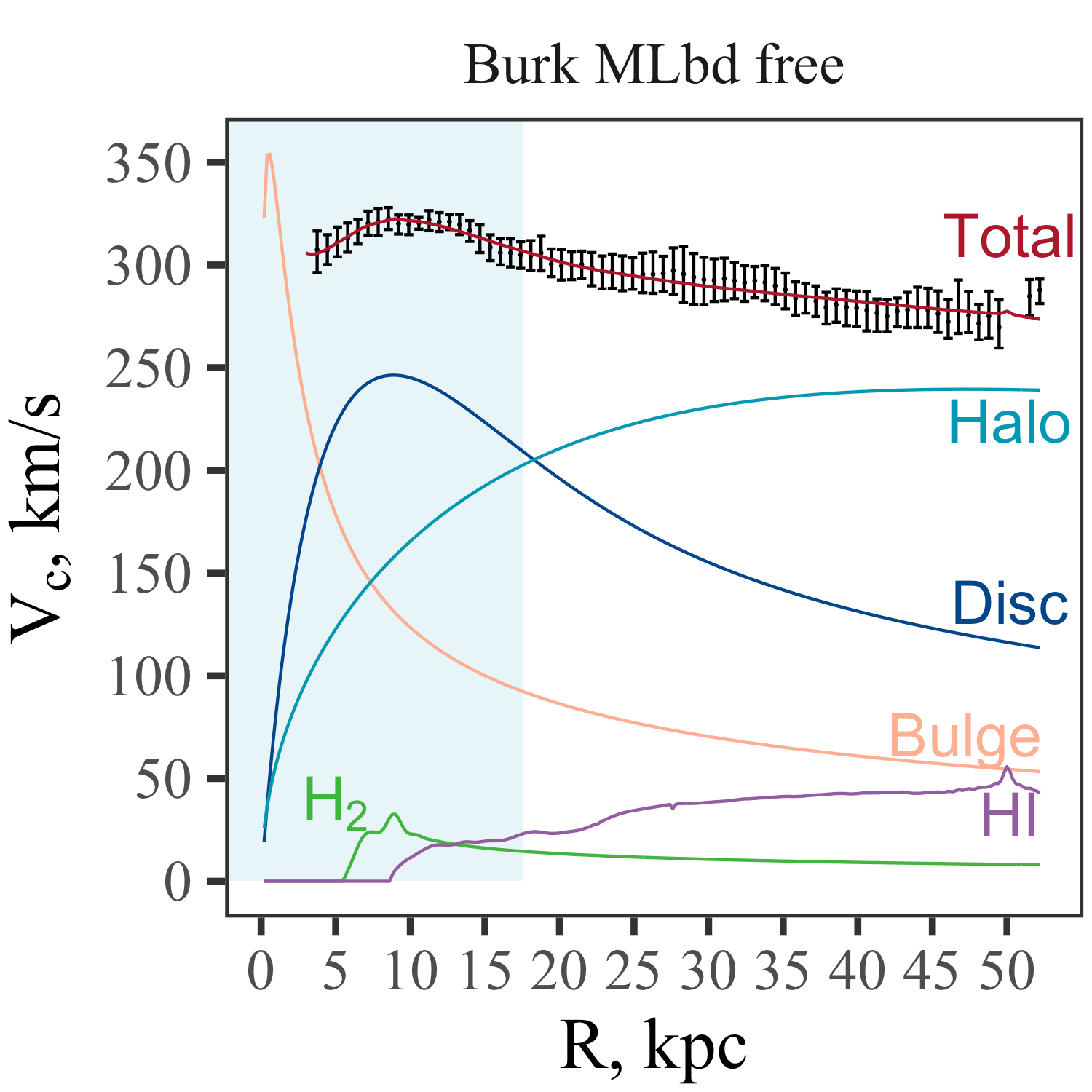}

\caption{Decomposition of the rotation curve of NGC~2841 for the Burkert halo. Symbols, colored zone and $(M/L)_\mathrm{d}$ as in Fig.~\ref{fig:N2841_ISO}.}
\label{fig:N2841_Burk}
\end{figure}

The rotation curve of this galaxy is traced up to 50~kpc and shows a gentle but clear fall by $\Delta V=40$~km/s. 
If the value of $(M/L)_\mathrm{d}$ is fixed at 0.48, which corresponds to the Kroupa IMF, then all the obtained solutions will give high values of $\chi$. The best solution with a minimal value of $\chi$ can be reached if $(M/L)_\mathrm{d}$ and $(M/L)_\mathrm{b}$ are free parameters. In this case, the values of $(M/L)_\mathrm{d}$ are obtained at the upper boundary of this value for the ``diet'' Salpeter IMF taking into account calibration errors ($(M/L)_\mathrm{d} \approx $1.0), regardless of the dark halo model. The parameters of the NFW halo correspond to the cosmological models and are very close to the parameters given by \citet{ManceraPina+2022} for the free ratio $(M/L)_\mathrm{d}$. 
Although in this work the ratio $(M/L)_\mathrm{b}$ was set as $(M/L)_\mathrm{b}=f\,(M/L)_\mathrm{d}$ with $f$ in a small range of values around 1.4, and in our case this ratio was a completely free parameter, without any restrictions, we obtained $(M/L)_\mathrm{b}=1.3(M/L)_\mathrm{d}$ for this model.
The resulting mass of the disc approximately equal to the mass of the dark halo within four exponential scales of the disc (for all types of halo). This is seen in Fig.~\ref{fig:N2841_ISO}--\ref{fig:N2841_NFW} ({\it right} plots) and in Fig.~\ref{fig:N2841_Burk}: at $R=4h$, the circular speed associated with the disc alone is greater than the circular speed associated with the halo alone. The halo-to-baryonic mass ratio within four exponential scales of the disc varies from 0.8 to 1.2 for the free parameter $(M/L)_\mathrm{d}$ (see Tables~\ref{tab:ISO_RCdec}-\ref{tab:Burkert_RCdec}). 
\par
Given the tendency to get the best solution with a higher value of $(M/L)_\mathrm{d}$, we examined the solutions when fixing the ratio $M/L=0.68$, which corresponds to the ``diet'' Salpeter IMF. 
For all models, a stellar disc does not show itself as baryon-dominated, but the total mass of baryonic matter (disc $+$ bulge $+$ gas) turns out to be only slightly less than the mass of the dark matter within four exponential scales of the disc (Figs.~\ref{fig:N2841_ISO}- \ref{fig:N2841_NFW}, {\it left} plots; Tables~\ref{tab:ISO_RCdec}-\ref{tab:Burkert_RCdec}). Moreover, as in the case of the free value of $(M/L)_\mathrm{d}$, the obtained dark halo parameters remain physical, and for the NFW model, they are consistent with cosmological simulations (see Section~\ref{sec:scaling} and Fig~\ref{fig:cosm-NFW}), although the concentration parameter $c$ is somewhat overestimated. Halo parameters are compiled in Tables~\ref{tab:ISO_RCdec}-\ref{tab:Burkert_RCdec}.

\subsection{NGC~3521}
NGC~3521 is a spiral galaxy in the constellation Leo.
\par
The rotation curve of the galaxy NGC~3521 shows a general decrease in the velocity of about 40~km/s (from 235~km/s at ~155~arcsec to 200~km/s at ~600~arcsec). 
This drop is created mainly by a sharp jump at about $R\approx320$~arcsec. 
The gap may be related to a very perturbed and asymmetric kinematics (e.g. \citealp{Casertano_vanGorkom1991}). This is confirmed by other authors \citep{deBlok+2008,DiTeodoro_Peek2021} and by our major-axis position-velocity diagram. \citet{Casertano_vanGorkom1991} suggested that this galaxy experienced some interaction. However, the galaxy has the regular morphology.
Due to perturbed kinematics the rotation curve of the cold gas might not trace well the gravitational potential (and thus the mass distribution) of this galaxy. However, we decided to leave this galaxy for further analysis, since in the central regions the rotation curve follows the rotation curve of a more massive disc surprisingly well. 
At the same time, in the very outer parts the approaching and receding sides of the rotation curve are symmetric. 
\par
Even for $(M/L)_\mathrm{d}=0.47$ the contribution of the disc component is overestimated. When this ratio is set to be free it decreases substantially and the modelled rotation curve better describes the observed one. 
Due to the high inclination, the central part of the rotation curve, where the bulge should dominate and give a sharp gradient, is not very reliable even for the HERACLES' data (for the effects of the inclination in the central regions, see, for example, \citealp{Zasov_Khoperskov2003,Stepanova_Volkov2013} and \citealp{Frank+2016}, their figure~3).
We excluded very central points from the analysis. This galaxy has a bulge that contributes 10\% to the luminosity. The absence of central points leads to a somewhat underestimated value of the ratio $(M/L)_\mathrm{b}$ when this parameter is freely varied. A fixed value
of $(M/L)_\mathrm{b}=1.4(M/L)_\mathrm{d}$ compatible with the stellar population synthesis models (for example, \citealp{Querejeta+2015}) leads to a slight redistribution of the baryon mass between the bulge and disc but does not significantly affect the parameters of the dark halo. Although the observed rotation curve that we analyze does not support the result by \citet{Casertano_vanGorkom1991} about the falling rotation curve, all our model rotation curves that give the best solution show a downward trend.
\par
Regardless of the halo type, this galaxy 
does not have an increased content of baryon matter within four exponential scales of an outer disc. For a free value of $(M/L)_\mathrm{d}$, the ratio of the dark matter mass within $4h_2$ to the total mass of the baryon, including gas, turns out to be greater than 1.0. It should be noted that the outer disc, although extended, is low-mass\footnote{Its mass is almost 4 times less than the mass of the inner disc.} and it unlikely can affect the dynamics of the galaxy. If we consider the area within $R<4 h_1$, then the galaxy has undoubtedly reduced content of the dark matter, with $M_\mathrm{h}/M_\mathrm{baryon} \approx 1.0$ (Tables~\ref{tab:ISO_RCdec}-\ref{tab:Burkert_RCdec}). The same was noted in \citet{deBlok+2008}, where the disc component considered as a single component.

\begin{figure}
\centering
\includegraphics[width=0.23\textwidth]{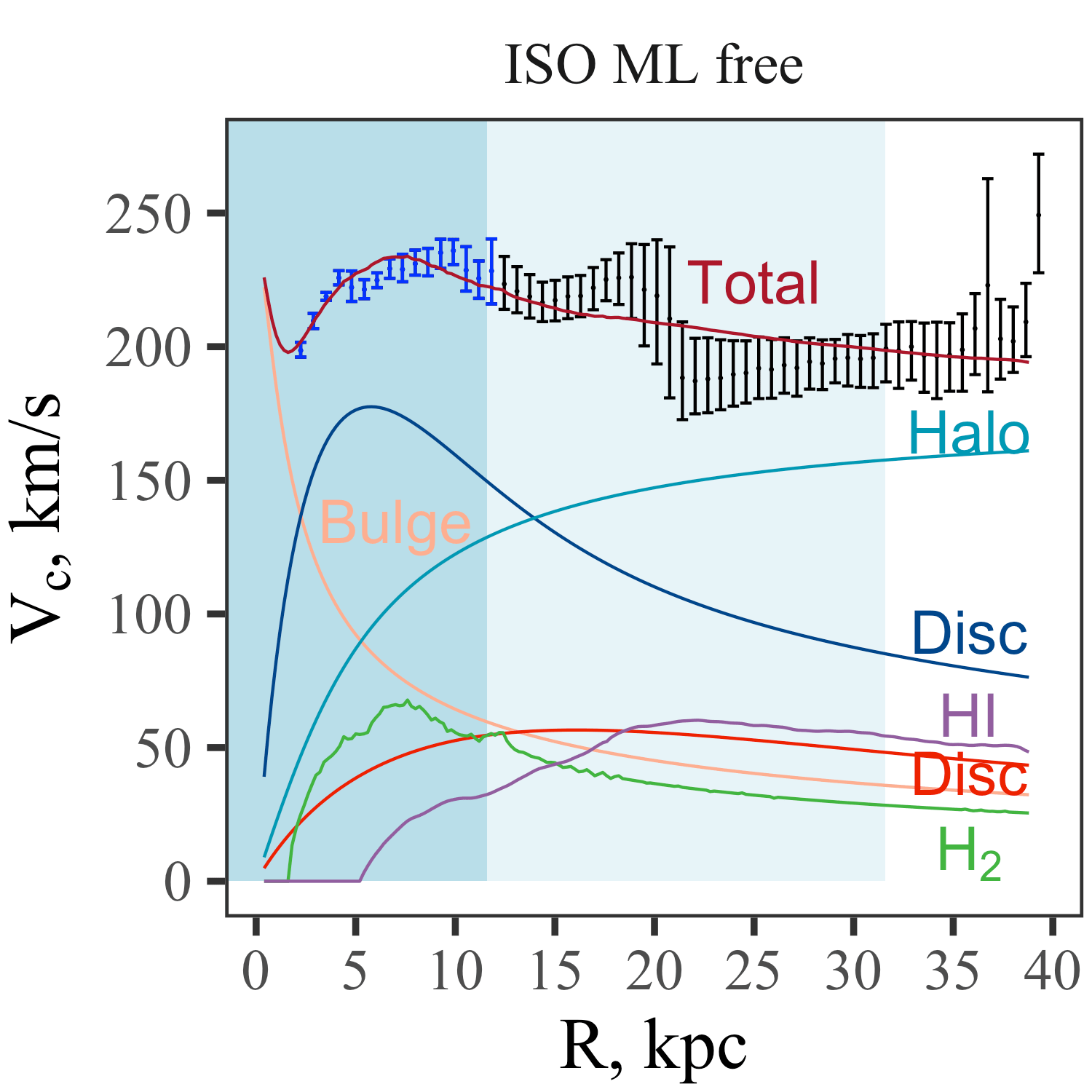}
\caption{Decomposition of the rotation curve~\textcolor{black} of NGC~3521 for the ISO halo with a free parameter $(M/L)_\mathrm{d}$ and $(M/L)_\mathrm{b}=1.4(M/L)_\mathrm{d}$. The vertical colored zones correspond to $R=4h$ of the inner and outer discs. The black points show data obtained by THINGS' data, the blue points  show data obtained by HERACLES data.}
\label{fig:N3521_ISO}
\end{figure}

\begin{figure}
\centering
\includegraphics[width=0.23\textwidth]{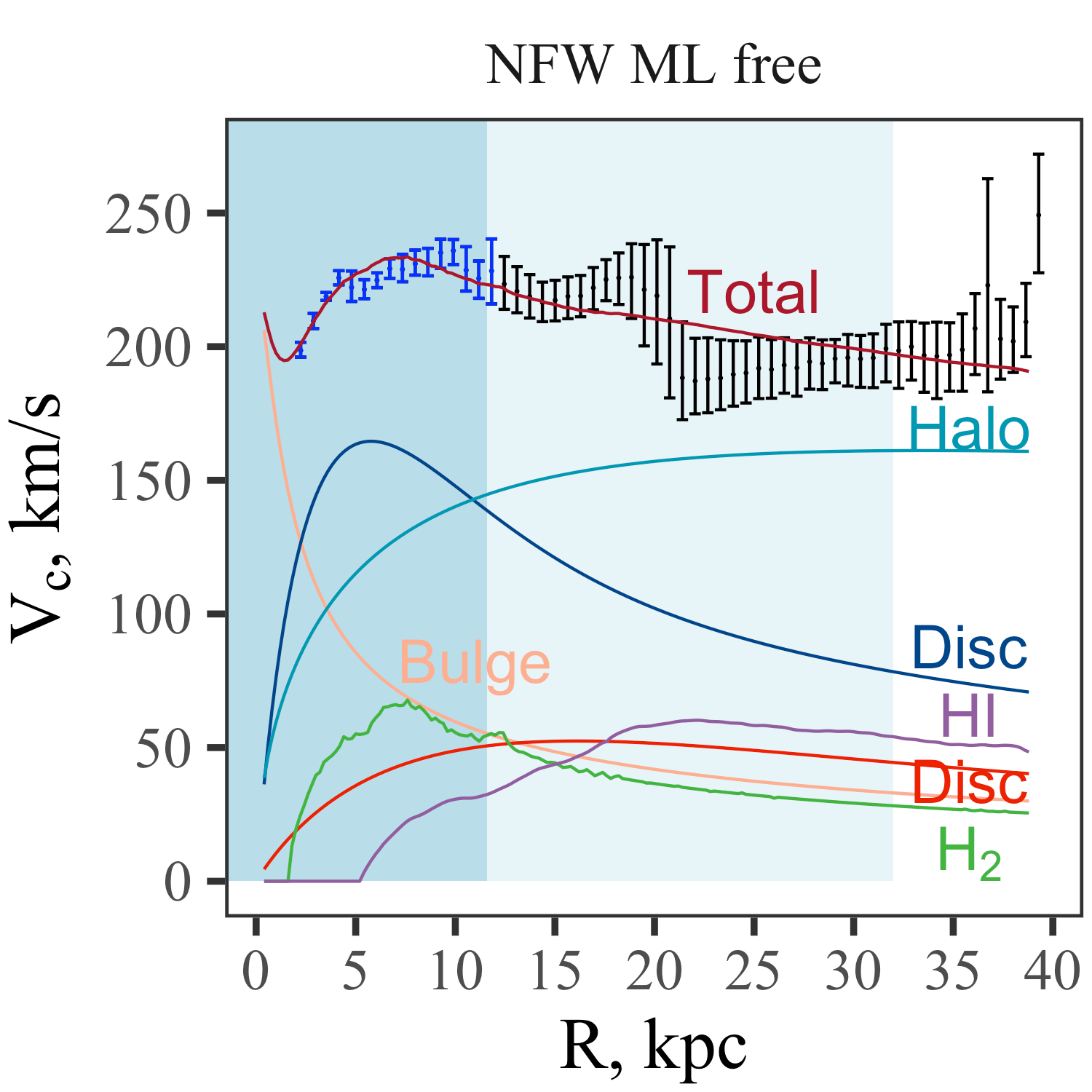}
\caption{Decomposition of the rotation curve of NGC~3521 for the NFW halo. Symbols, colored zones, $(M/L)_\mathrm{d}$ and $(M/L)_\mathrm{b}$ as in Fig.~\ref{fig:N3521_ISO}.}
\label{fig:N3521_NFW}
\end{figure}

\begin{figure}
\centering
\includegraphics[width=0.23\textwidth]
{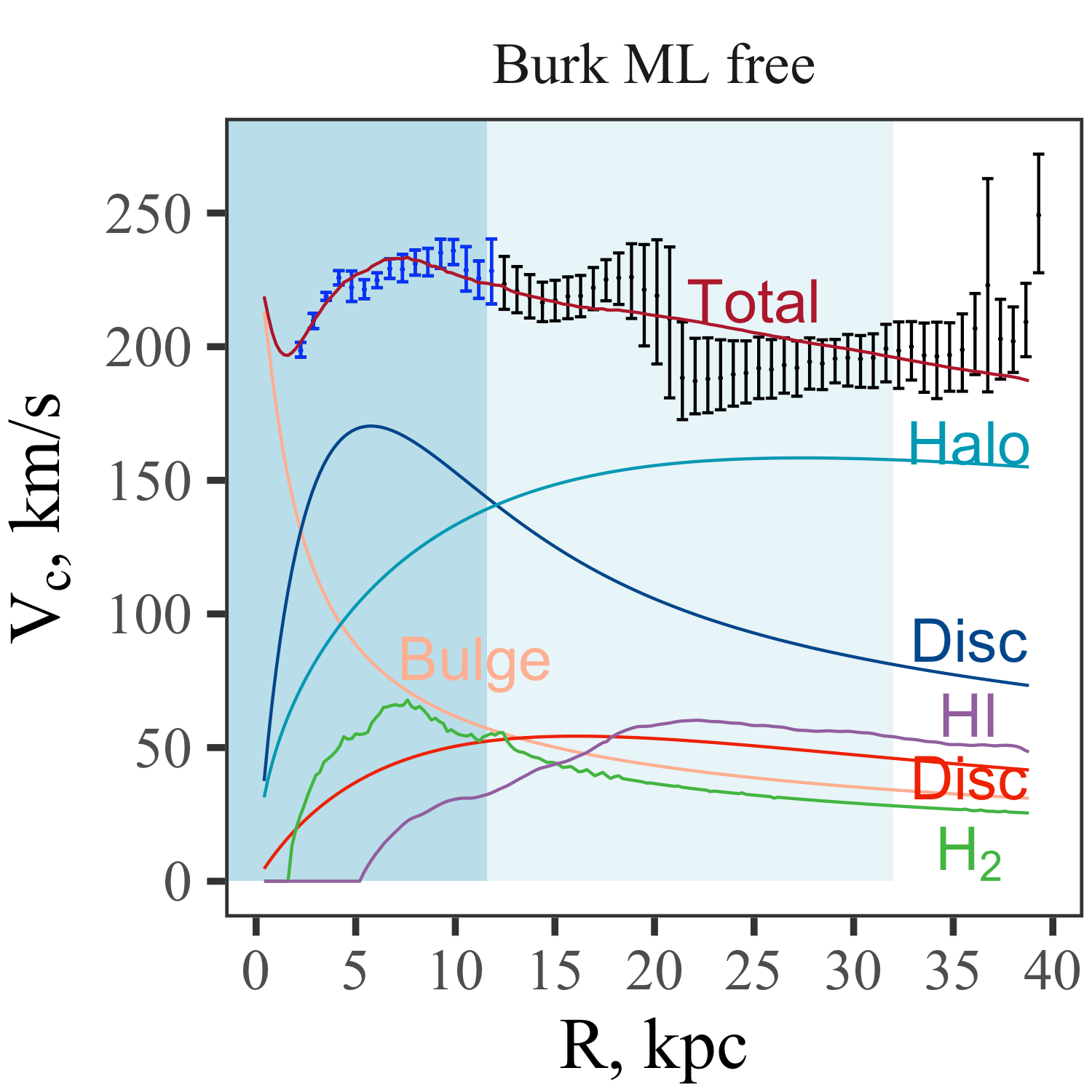}
\caption{Decomposition of the rotation curve of NGC~3521 for the Burkert halo. Symbols, colored zones, $(M/L)_\mathrm{d}$ and $(M/L)_\mathrm{b}$ as in Fig.~\ref{fig:N3521_ISO}.}
\label{fig:N3521_Burk}
\end{figure}

\subsection{NGC~5055}
NGC~5055 (M63, The Sunflower Galaxy) is a disc galaxy in the constellation Canes Venatici. It has a flocculent structure, no visible bar, and its photometric profile shows two discs. 
\par
The rotation curve of NGC~5055 extends to 900~arcsec ($\sim 39$~kpc) and has a maximal rotation speed of $V_{\mathrm{max}}\sim 210$~km/s at $\sim13$~kpc. It shows a minor fall by $\Delta V = 20$~km/s. 
A stronger fall down in the rotation curve was observed in the data by \citet{deBlok+2008} but \citet{DiTeodoro_Peek2021}, who took into account the warp of this galaxy, did not confirm the falling character of the rotation curve. Nevertheless, the results of the decomposition with both the isothermal halo \citep{deBlok+2008,Frank+2016} and the NFW halo \citep{ManceraPina+2022} hint at the fact that the mass of the stellar disc within the optical radius of the galaxy exceeds the mass of the dark matter.
In the region $R<900$~arcsec, our rotation curve is in good agreement with the rotation curve by \citet{DiTeodoro_Peek2021}, although other observational data were used in that work. 
\par
The results of the rotation curve decomposition for the three types of halo (ISO, NFW, Burkert) are given in Tables~\ref{tab:ISO_RCdec}-\ref{tab:Burkert_RCdec}. When the parameter $(M/L)_\mathrm{d}$ is fixed at the value of $0.46$, regardless of the type of halo, the galaxy \textcolor{black}{has reduced content of the dark matter}
within four exponential scales of the more extended disc ($h_2=12.9$~kpc). The ratio of the mass of dark matter within four exponential scales of the disc to the baryonic mass was $\approx 0.6$ for all types of halos. The situation does not change significantly with the free value of $(M/L)_\mathrm{d}$. The total mass of baryonic matter (including gas) within four exponential scales of a larger disc is equal to the mass of the dark matter. The greatest dark to baryonic ratio was obtained for the NFW halo $\sim 1.2$. It should be noted that all six decomposition results are physical, and the parameters of the NFW\footnote{If the ratio $(M/L)_\mathrm{d}$ is considered as a free parameter, the concentration parameter of the NFW halo is slightly overestimated (see Section\ref{sec:scaling} and Fig~\ref{fig:cosm-NFW}).} for all models do not contradict cosmological simulations.

\begin{figure}
\centering
\includegraphics[width=0.23\textwidth]{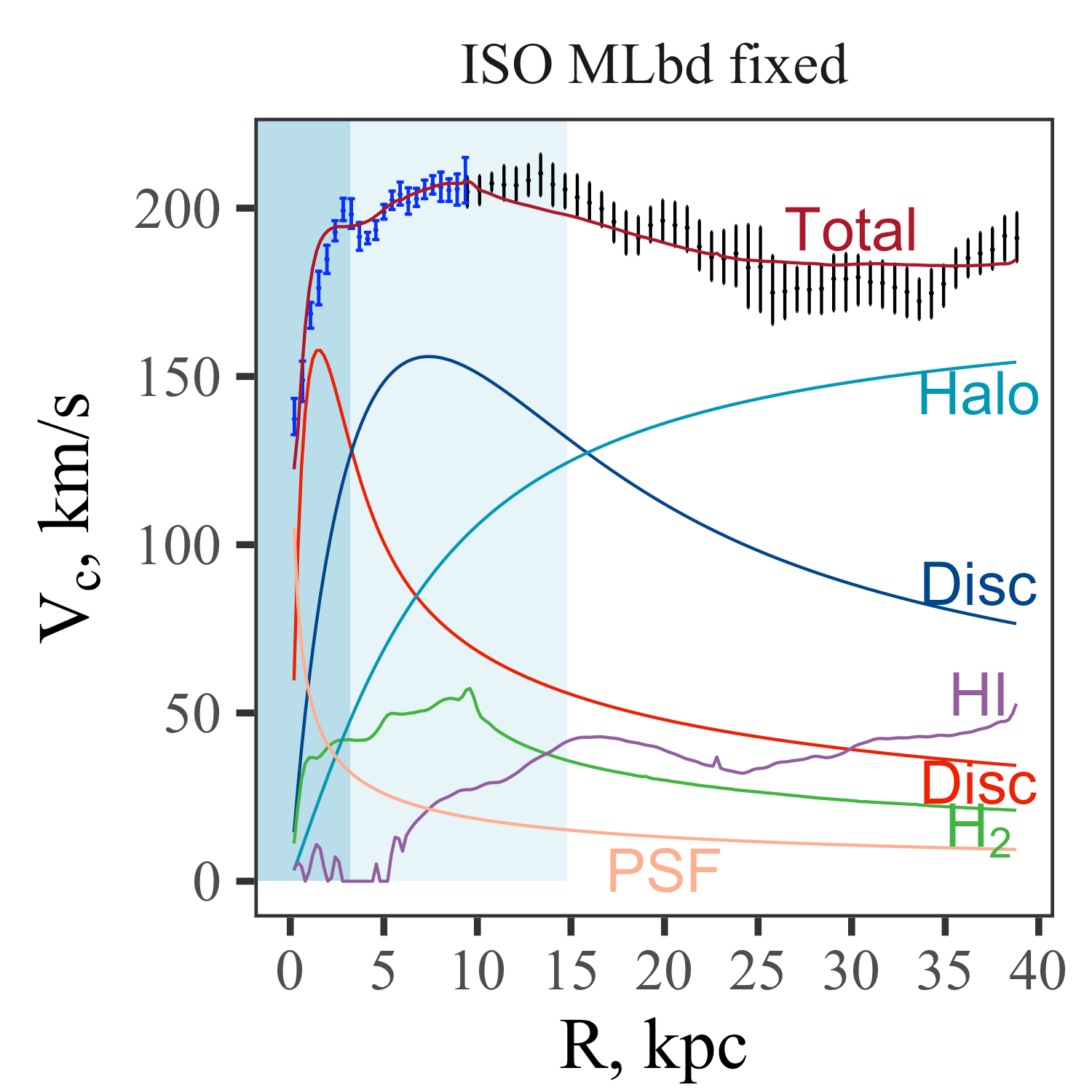}
\includegraphics[width=0.23\textwidth]{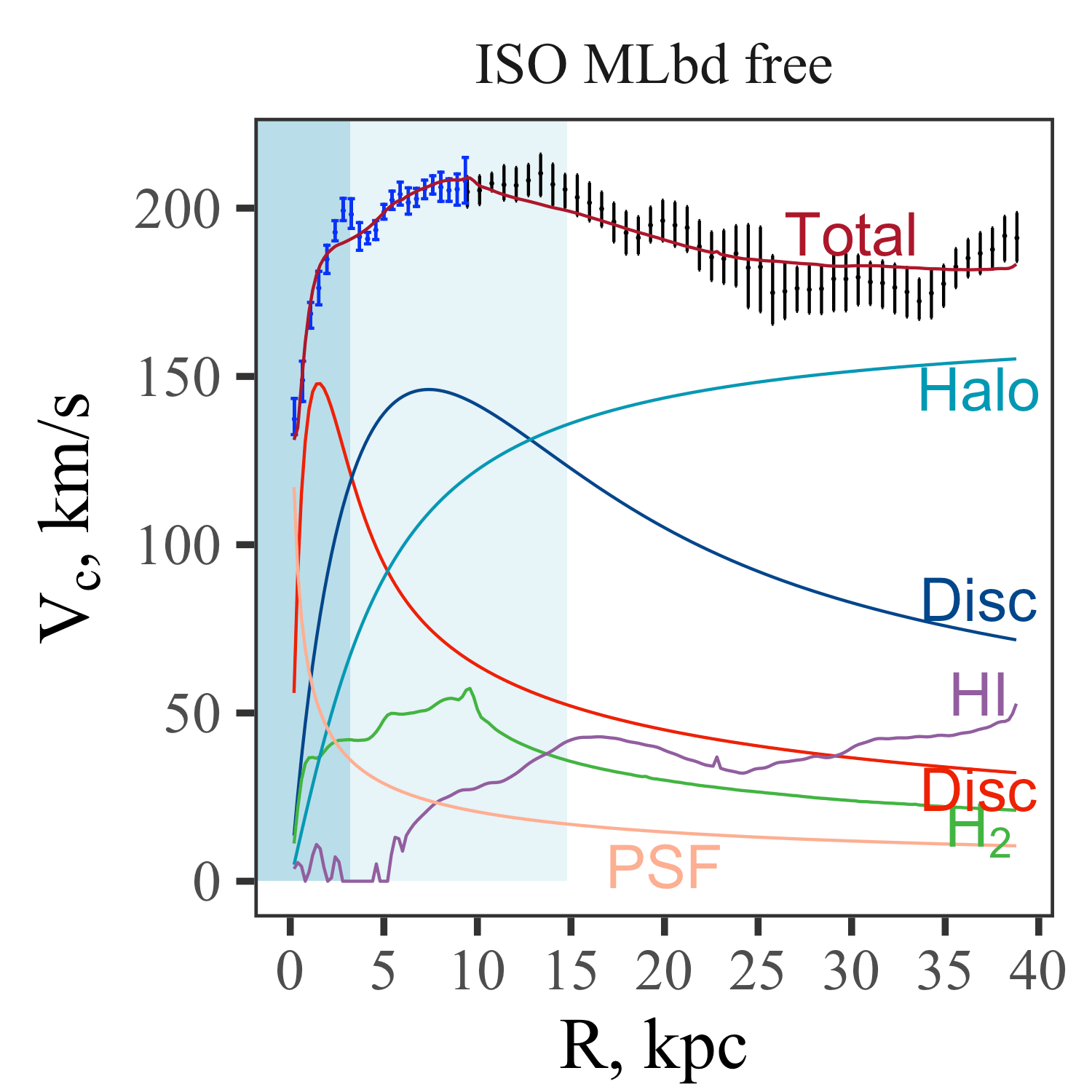}
\caption{Decomposition of the rotation curve~\textcolor{black}{(black and blue points)} of NGC~5055 for the ISO halo. {\it Left} --- for a fixed value $(M/L)_\mathrm{d}=0.46$ (free for a central unresolved core). {\it Right} --- with free values of $M/L$ for a disc  and for a central core. The colored zones correspond to $R=4h$ of the inner and outer discs.\textcolor{black}{The black points show data obtained by THINGS' data, the blue points  show data obtained by HERACLES data.}}
\label{fig:N5055_ISO}
\end{figure}

\begin{figure}
\centering
\includegraphics[width=0.23\textwidth]{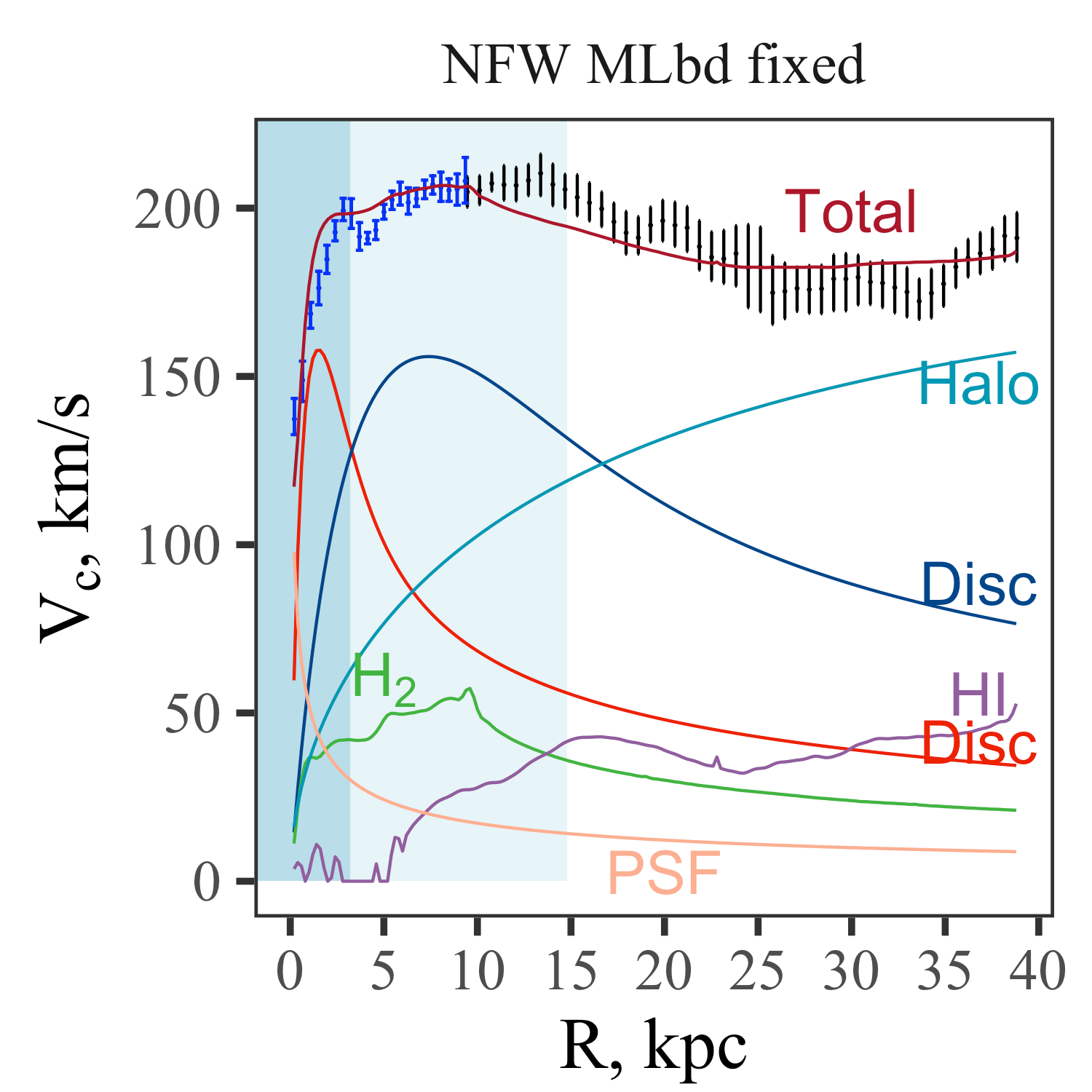}
\includegraphics[width=0.23\textwidth]{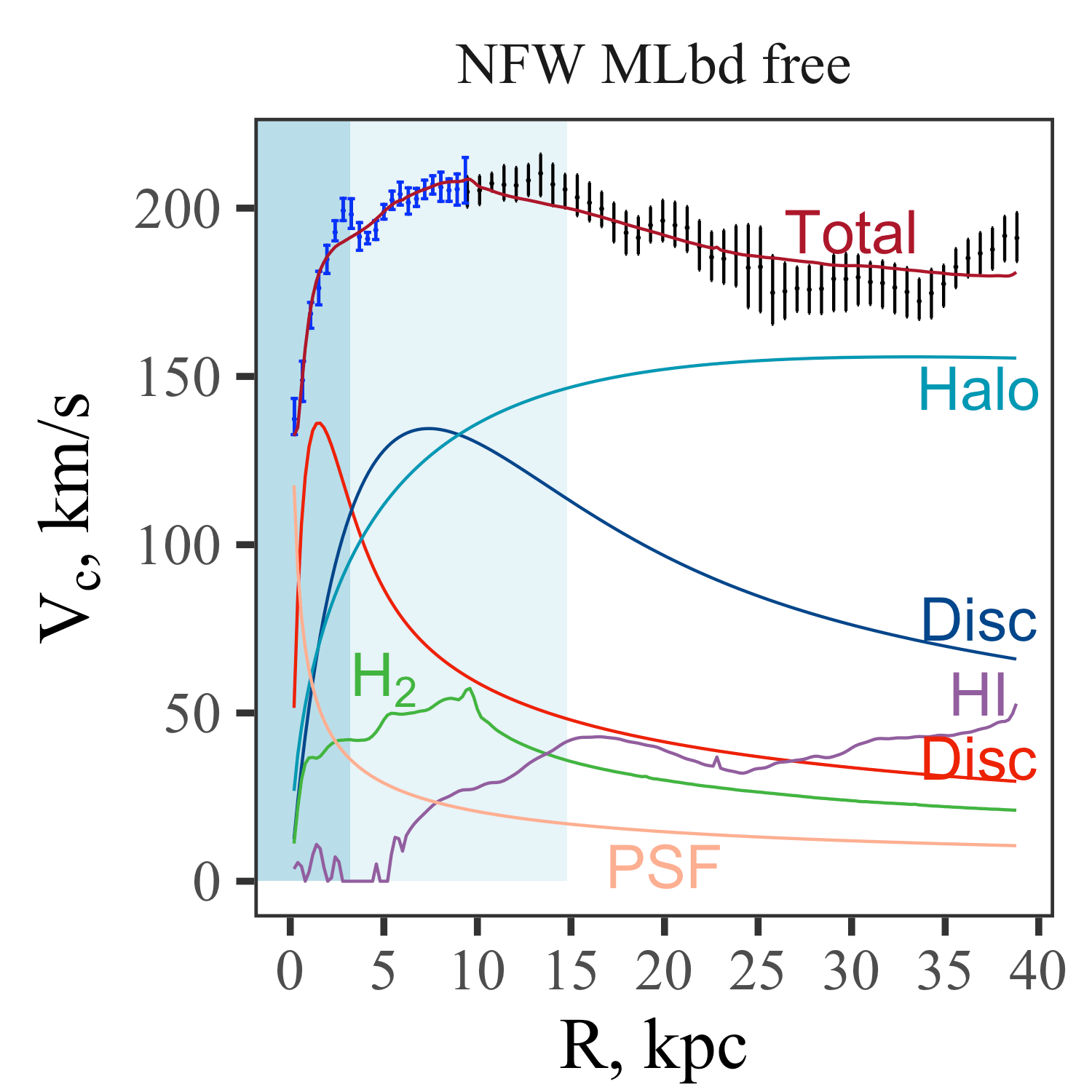}
\caption{Decomposition of the rotation curve of NGC~5055 for the NFW halo. Symbols, colored zones and $(M/L)_\mathrm{d}$ as in Fig.~\ref{fig:N5055_ISO}.}
\label{fig:N5055_NFW}
\end{figure}

\begin{figure}
\centering
\includegraphics[width=0.23\textwidth]{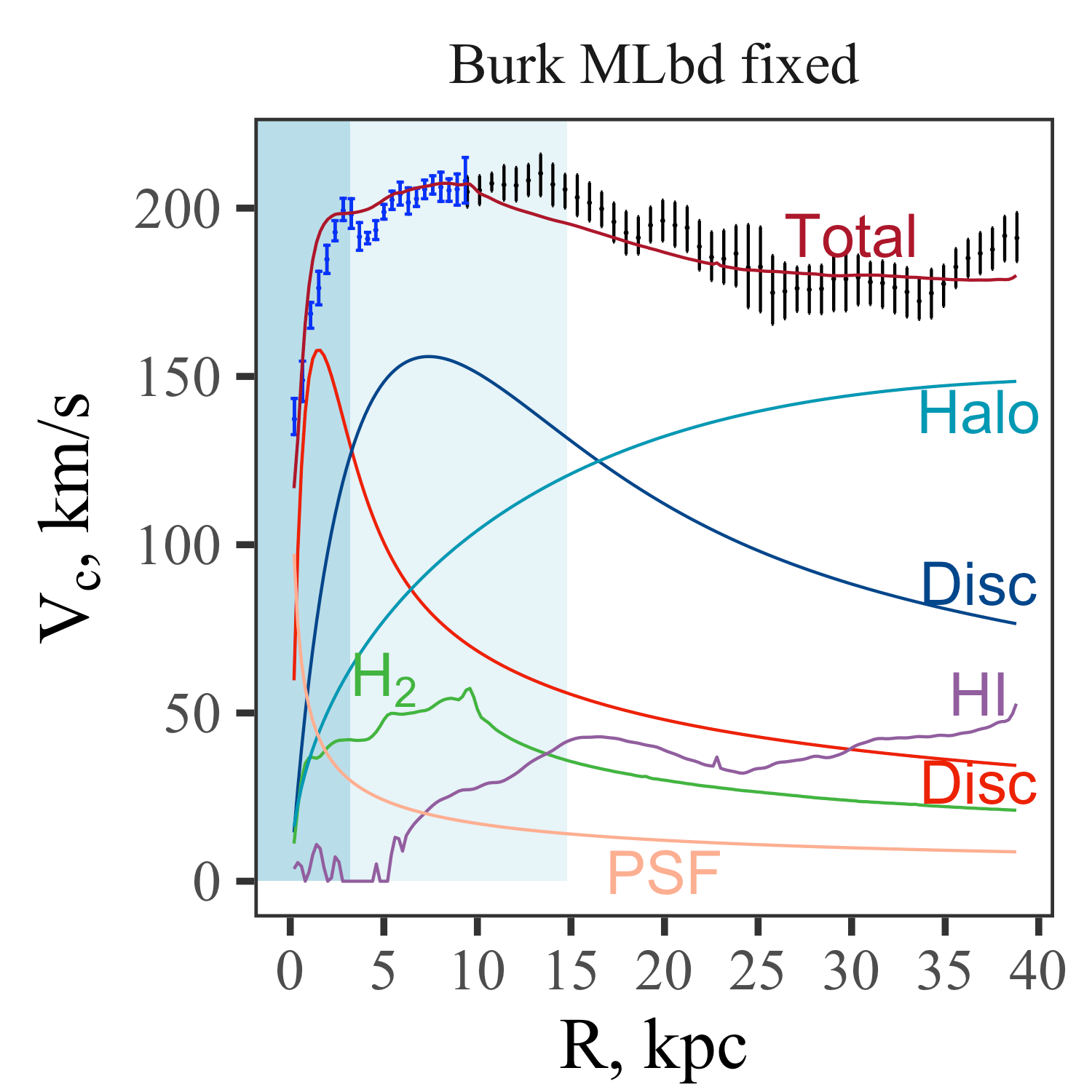}
\includegraphics[width=0.23\textwidth]{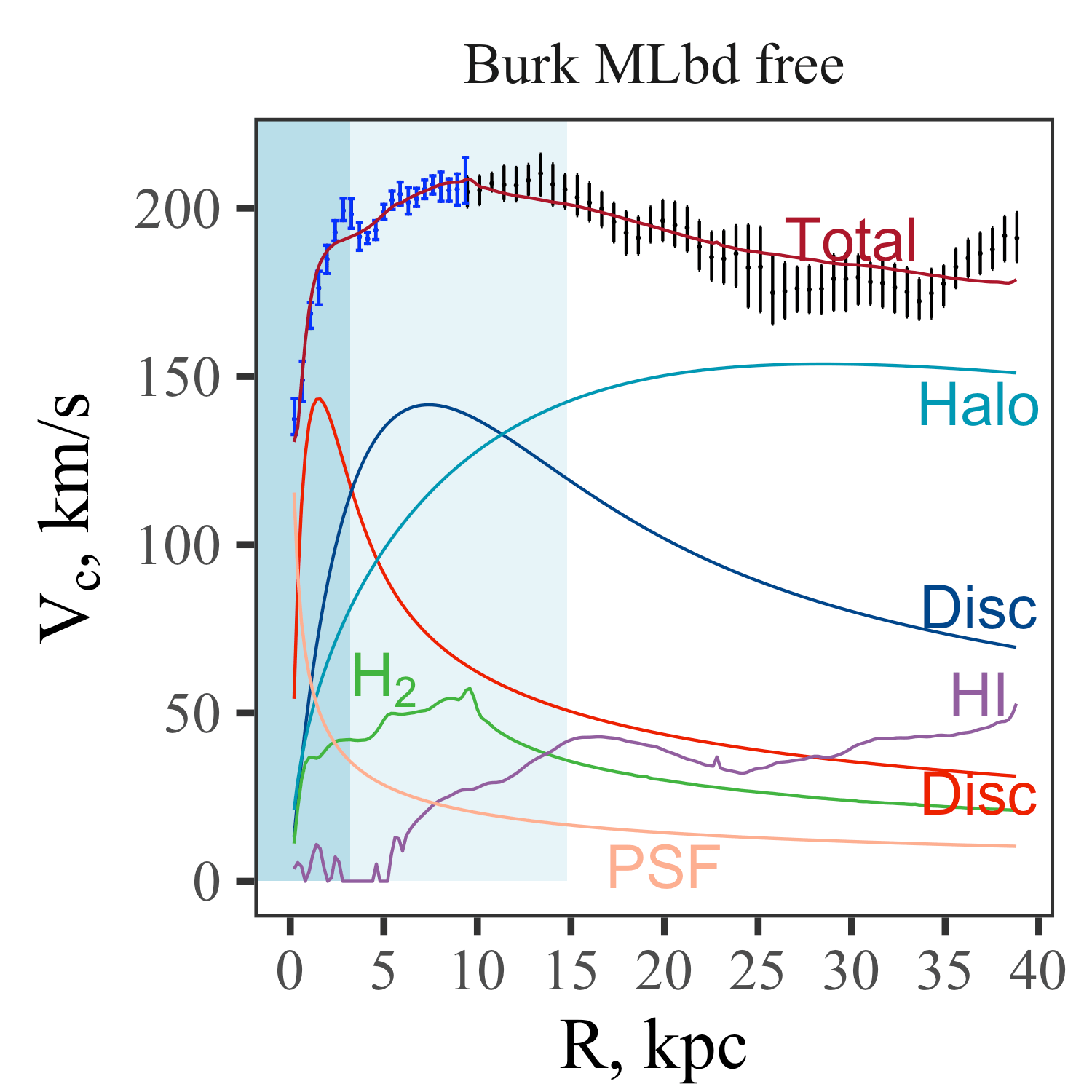}
\caption{\textcolor{black}{Decomposition of the rotation curve of NGC~5055 for the Burkert halo. Symbols, colored zones and $(M/L)_\mathrm{d}$ as in Fig.~\ref{fig:N5055_ISO}.}}
\label{fig:N5055_Burk}
\end{figure}

\subsection{NGC~7331}
NGC~7331 --- a spiral barless galaxy in the constellation Pegasus. 
\par
This galaxy has not clearly declining rotation (see \citet{deBlok+2008,DiTeodoro_Peek2021} and the Appendix~\ref{sec:appendix}). However, the decomposition of the rotation curve of this galaxy \citep{deBlok+2008,Frank+2016,Katz+2014}, although not very successful, suggests that \textcolor{black}{the disc rotation curve dominates within optical radius of the galaxy}. The absence of a bar and the flocculent structure were also the basis for its analysis in this work. It does not enter in S$^4$G that is why we performed our own photometric decomposition\footnote{The photometric model used by \citet{deBlok+2008}, in which the central exponential disc is singled out, and the rest of the galaxy is treated as an outer disc, although the profile of this component is far from exponential, seems to us too simplified.} (Table~\ref{tab:phot_decomposition}).
\par
{For a fixed ratio $(M/L)_\mathrm{d}$, the modelled rotation curve does not go beyond the observed curve only for values at the lower boundary of this ratio for the Kroupa IMF (0.47). However, in this case the ratio $(M/L)_\mathrm{b}$ is too small. The problem here is the same as with the galaxy NGC~3521. 
NGC~7331 has a high inclination and the bulge contribution to the rotation curve in the central region does not fit well into the observational data, with the inclination effects being significant even for the HERACLES data we used in the central regions. 
As for the galaxy NGC~3521, we did not take into account the most central points of the rotation curve, since they do not reflect the contribution of the central bulge.}
If the ratio $(M/L)_\mathrm{d}$ varies, then the best solutions are obtained even with lower values of $(M/L)_\mathrm{d} < 0.47$ if we set the ratio $(M/L)_\mathrm{b} = 1.4(M/L)_\mathrm{d}$. For all halo models, the galaxy appears to \textcolor{black}{have a reduce content of the dark matter} within four exponential scales of the massive inner disc.

\begin{figure}
\centering
\includegraphics[width=0.23\textwidth]{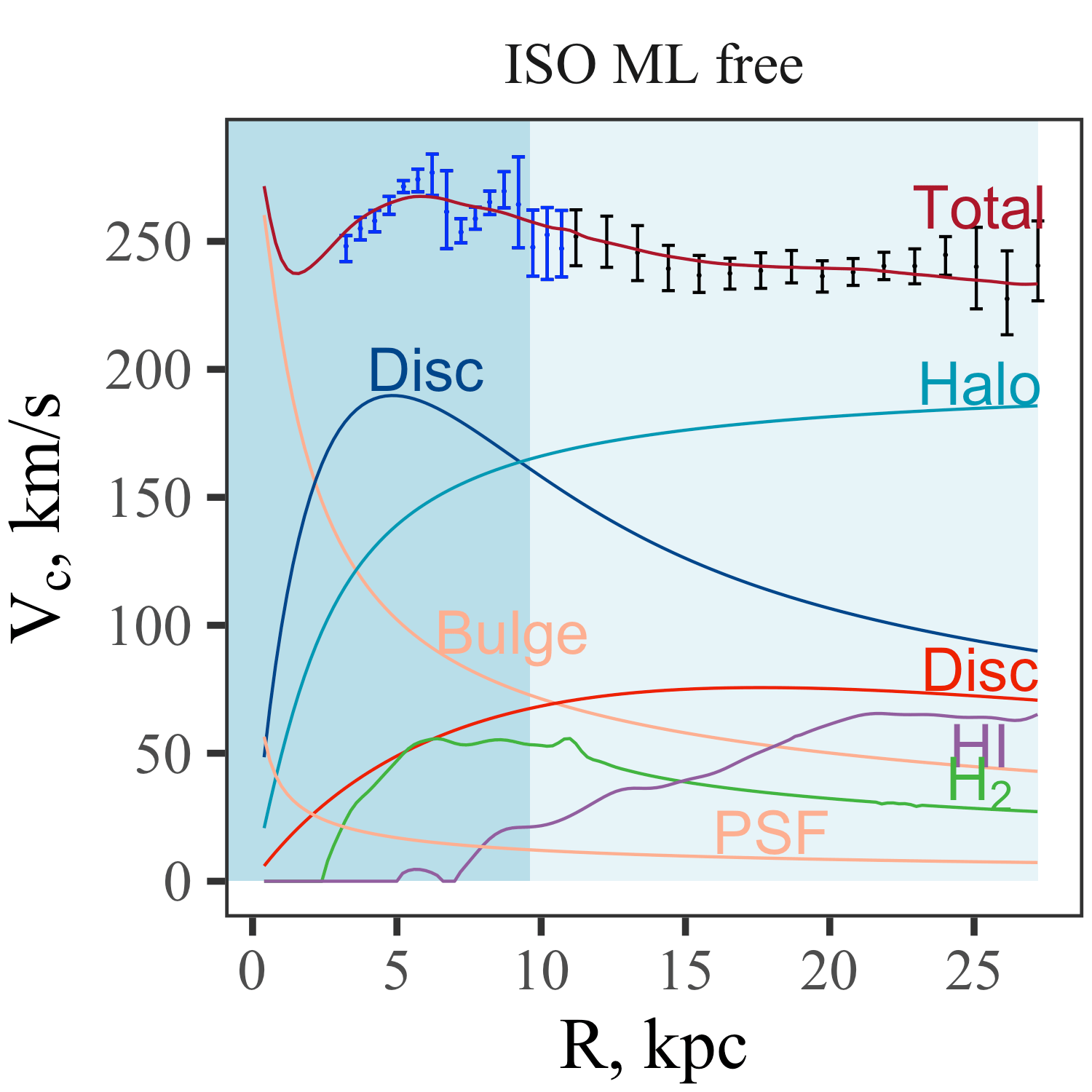}
\caption{Decomposition of the rotation curve~(black and blue points) of NGC~7331 for the ISO halo with a free parameter $(M/L)_\mathrm{d}$ and $(M/L)_\mathrm{b}=1.4(M/L)_\mathrm{d}$. The vertical colored zones correspond to $R=4h$ of the inner and outer discs. The black points show data obtained by THINGS' data, the blue points  show data obtained by HERACLES data.}
\label{fig:N7331_ISO}
\end{figure}

\begin{figure}
\centering
\includegraphics[width=0.23\textwidth]{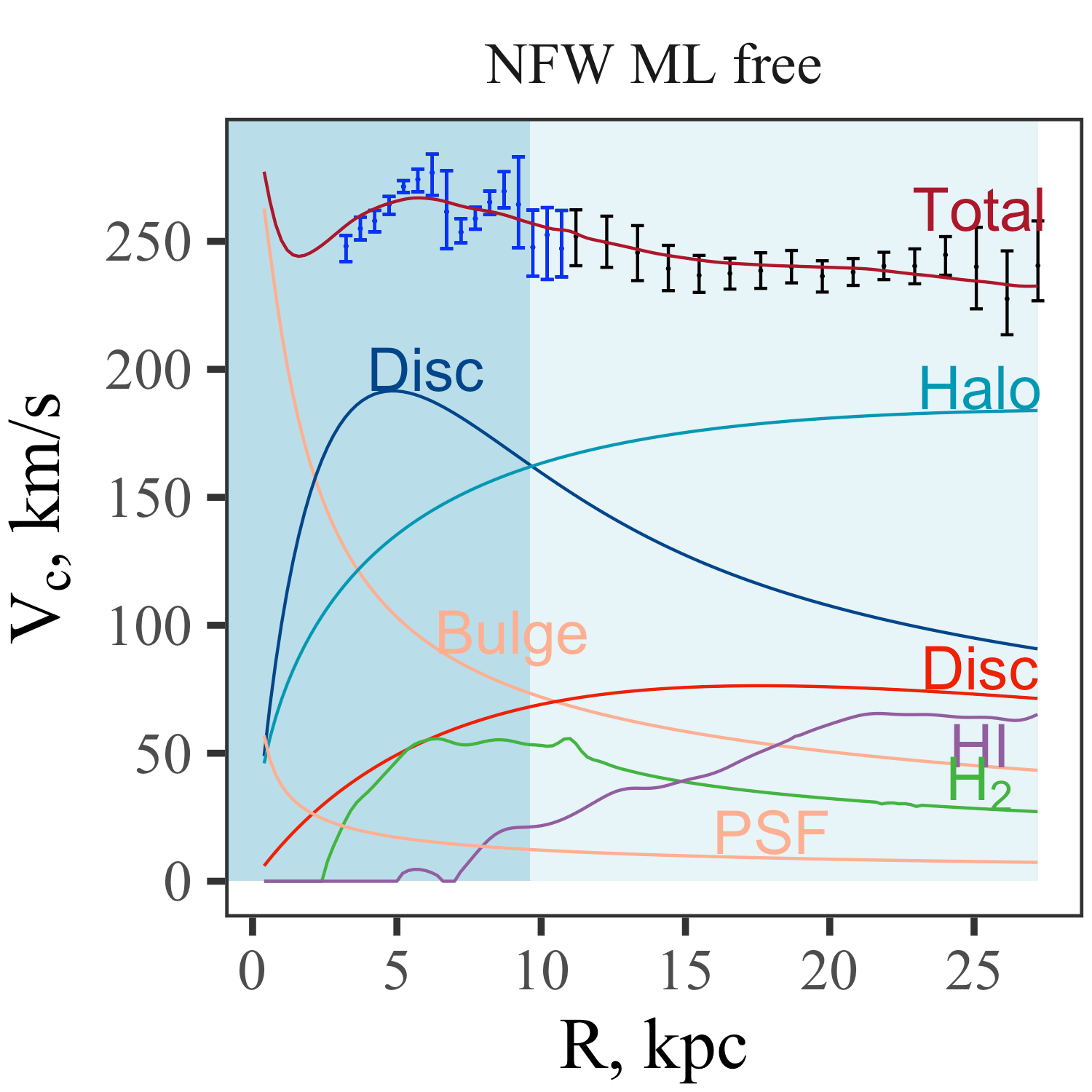}
\caption{\textcolor{black}{Decomposition of the rotation curve of NGC~7331 for the NFW halo. Symbols, colored zones, $(M/L)_\mathrm{d}$ and $(M/L)_\mathrm{b}$ as in Fig.~\ref{fig:N7331_ISO}.}}
\label{fig:N7331_NFW}
\end{figure}

\begin{figure}
\centering
\includegraphics[width=0.23\textwidth]{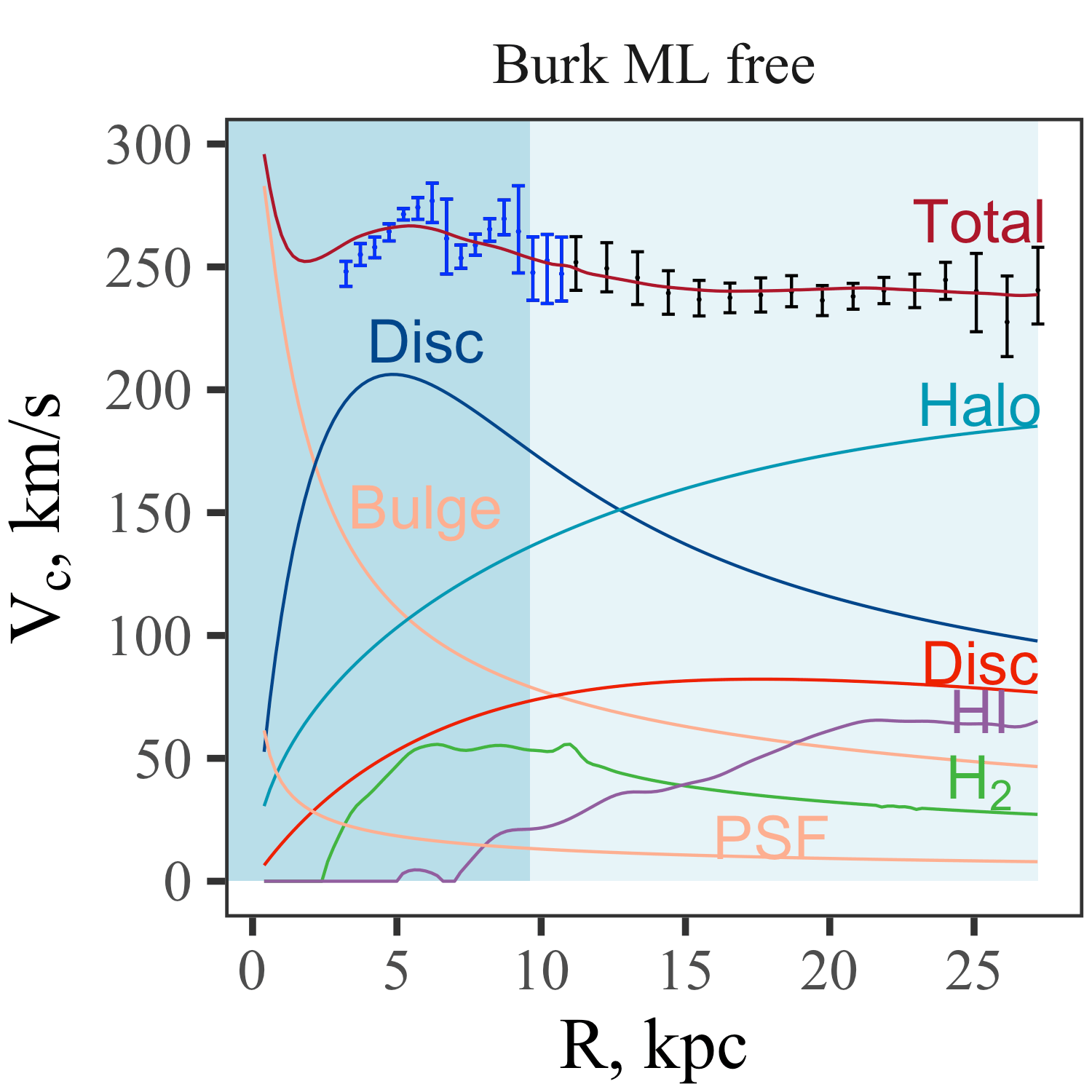}
\caption{\textcolor{black}{Decomposition of the rotation curve of NGC~7331 for the Burkert halo. Symbols, colored zones, $(M/L)_\mathrm{d}$ and $(M/L)_\mathrm{b}$ as in Fig.~\ref{fig:N7331_ISO}.}}
\label{fig:N7331_Burk}
\end{figure}

\section{Scaling relations for the dark halo parameters}
\label{sec:scaling}

\begin{figure}
\centering
\includegraphics[width=0.23\textwidth]{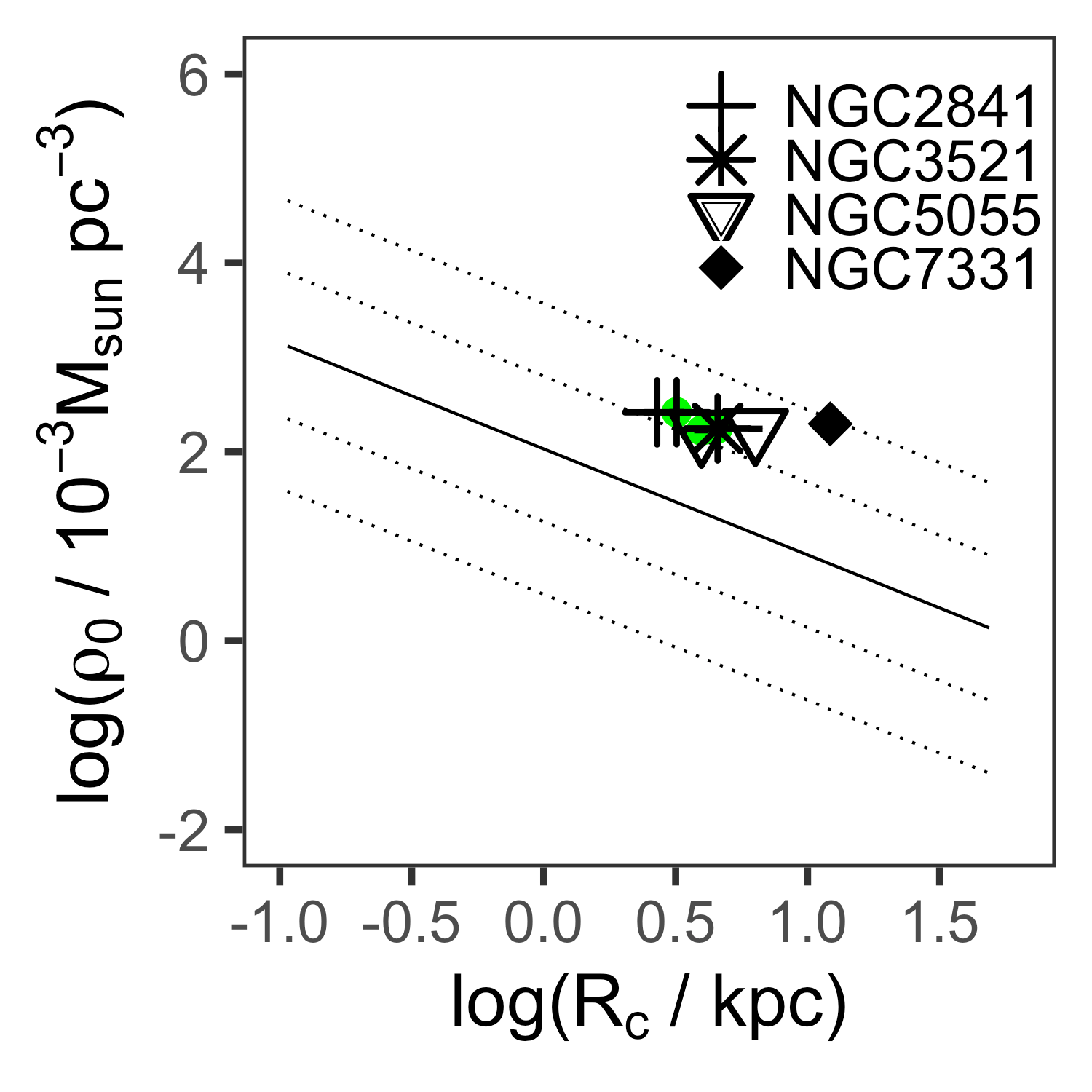}
\includegraphics[width=0.23\textwidth]{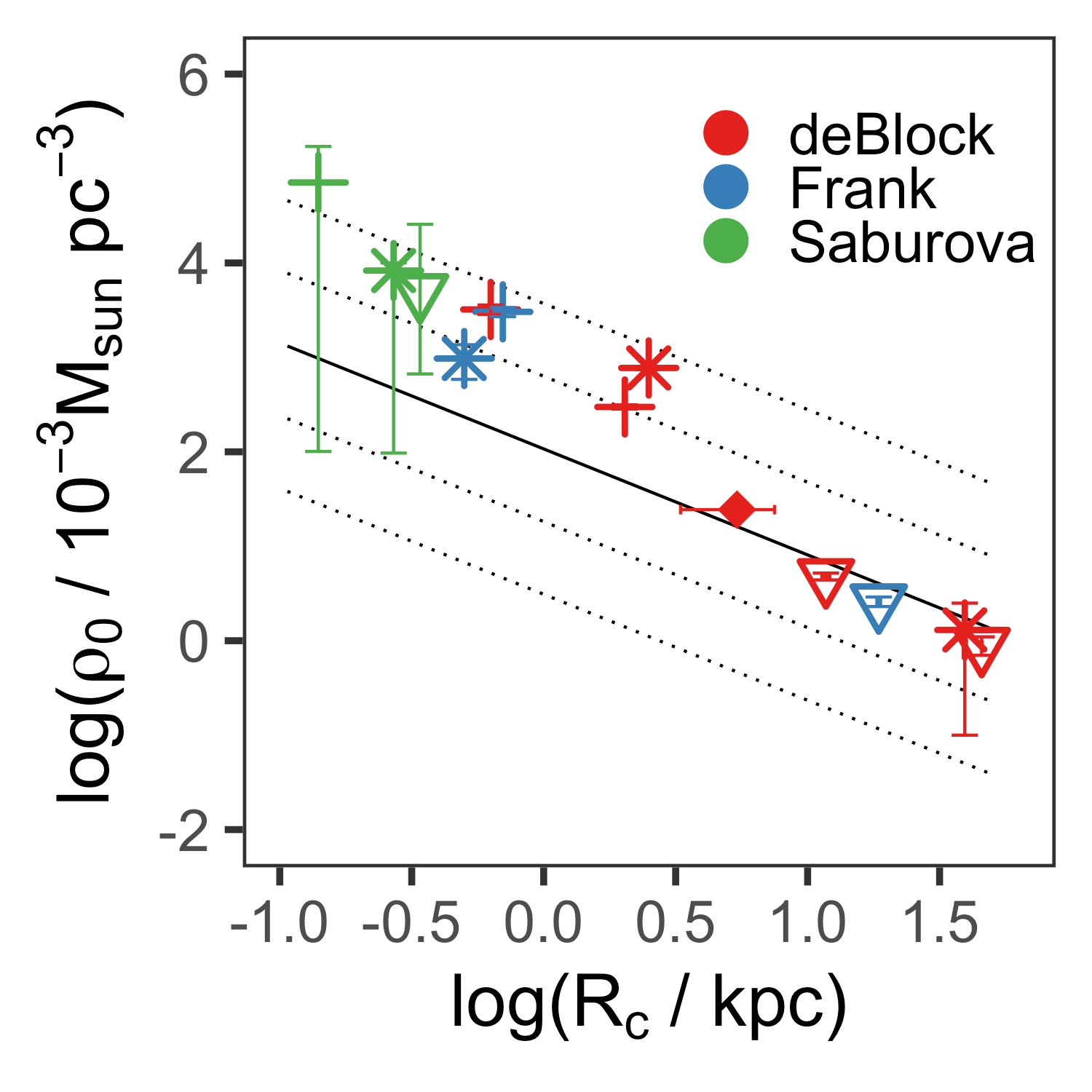}
\caption{Fit halo parameters for the ISO model; $\log(\rho_0/10^{-3}\,M_{\sun}\,\mathrm{pc}^{-3})$ vs. $\log(R_\mathrm{c}/\mathrm{kpc})$. The thick dotted line is the relation from \citet{Kormendy_Freeman2004}, two thin dotted lines indicate the $1\sigma$ and $2\sigma$ scatter. {\it Left}: the parameters from the present work (the results for the free value of the ratio $(M/L)_\mathrm{d}$ are marked with green circles). {\it Right}: the parameters from \citet{deBlok+2008}, \citet{Katz+2014}, \citet{Frank+2016}, \citet{Saburova+2016}. The  shapes correspond to the names of the galaxies. 
}
\label{fig:cosm-ISO}
\end{figure}

Fig.~\ref{fig:cosm-ISO} demonstrates the Kormendy--Freeman relationship $\log(\rho_0)$ vs. $\log(R_\mathrm{c})$ from \citet{Kormendy_Freeman2004} for the ISO dark halo parameters. Two thin dotted lines show the $1\sigma$ and $2\sigma$ scatter of this relation. Our fit dark halo parameters for all models from Table~\ref{tab:ISO_RCdec} are shown on the left\footnote{The parameter $\rho_0$ was calculated from $R_\mathrm{c}$ and $v_{\infty}$ via $\rho_0=v^2_{\infty}/4 \pi G R^2_\mathrm{c}$.}. The right plot presents the results of other authors. The parameters of the halo obtained in this work follow the expected relation and practically do not go beyond the $2\sigma$ line.

\begin{figure}
\centering

\includegraphics[width=0.23\textwidth]{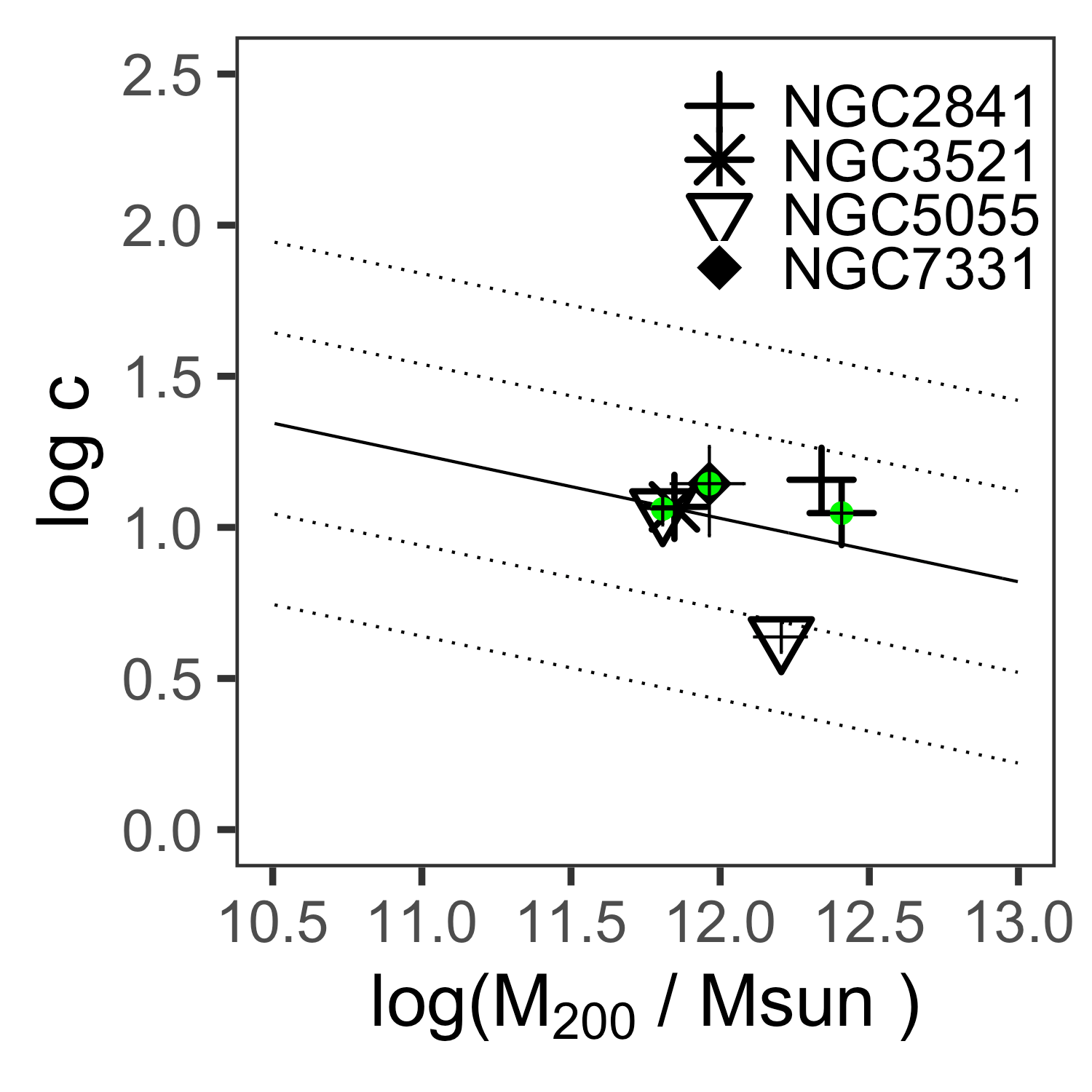}
\includegraphics[width=0.23\textwidth]{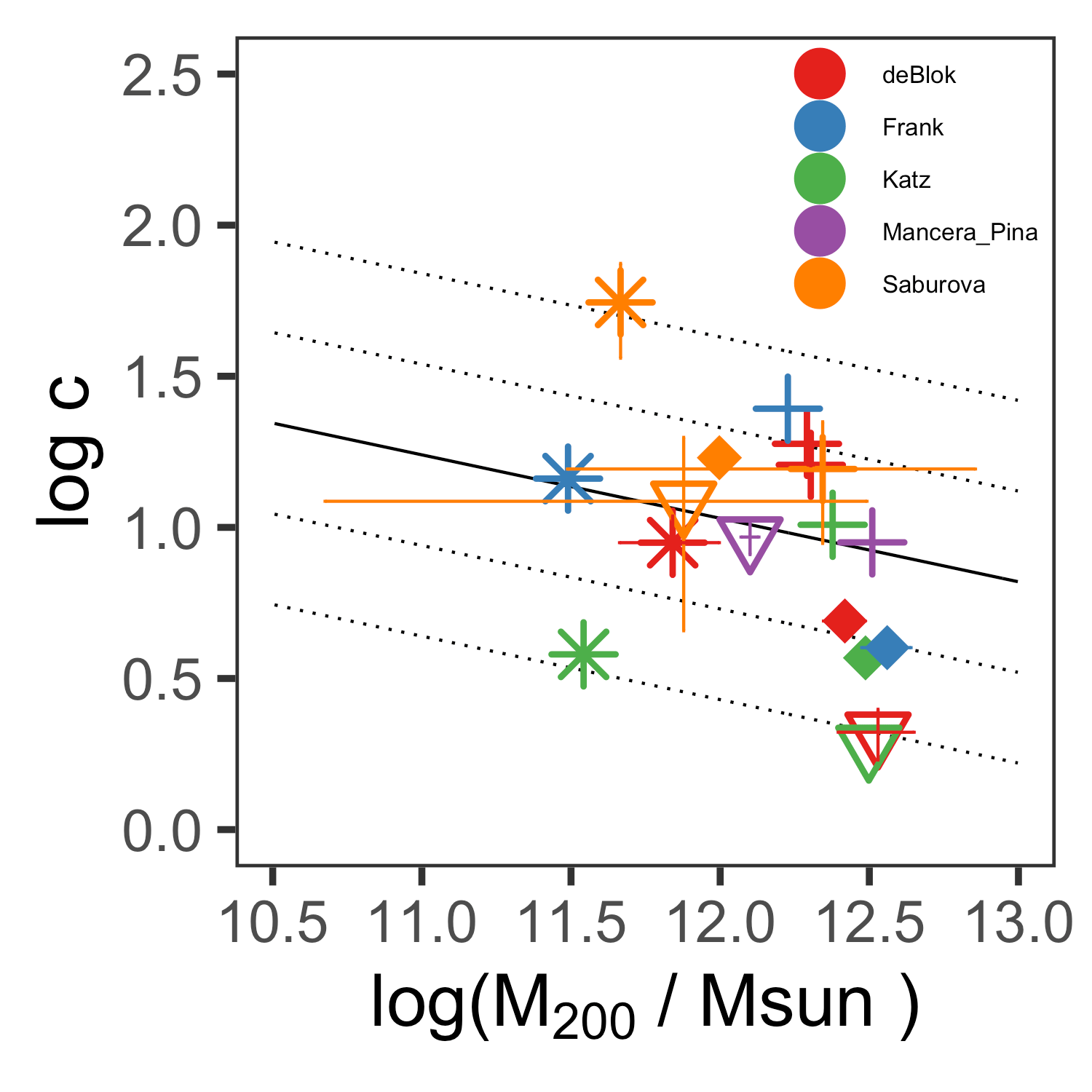}
\caption{Fit halo parameters for the NFW model; $\log(c)$ vs. $\log(M_{200}/(M_\mathrm{sun})$. The straight line is the relation from \citet{Macci+2008}, the dotted lines indicate the $1\sigma$ and $2\sigma$ scatter. {\it Left}: the parameters from the present work (the results for the {free} value of the ratio $(M/L)_\mathrm{d}$ are marked with green empty circles). 
{\it Right}: the parameters from \citet{deBlok+2008}, \citet{Katz+2014} , \citet{Frank+2016}, \citet{Saburova+2016}, \citet{ManceraPina+2022}. }

\label{fig:cosm-NFW}
\end{figure}

The fit parameters $c$ and $V_{200}$ for the NFW model are not completely independent but related. Their values are determined by the assumed cosmology. We plot (Fig.~\ref{fig:cosm-NFW}, {\it left} plot) fit parameters corresponding to all models that produce the lowest $\chi^2$ for each galaxy in order to make sure that our results are cosmologically justified. 
A straight thick line in Fig.~\ref{fig:cosm-NFW} demonstrates the expected $c$--$M_{200}$ relation from LCDM cosmology \citep{Macci+2008}. Two thin dotted lines show the $1\sigma$ and $2\sigma$ scatter of this relation. The parameters of the NFW halo from \citet{deBlok+2008,Katz+2014,Frank+2016,Saburova+2016, ManceraPina+2022} are shown in the {\it right} plot of Fig.~\ref{fig:cosm-NFW}. The parameters derived in the present work follow well the relation.

\begin{figure}
\centering
\includegraphics[width=0.23\textwidth]{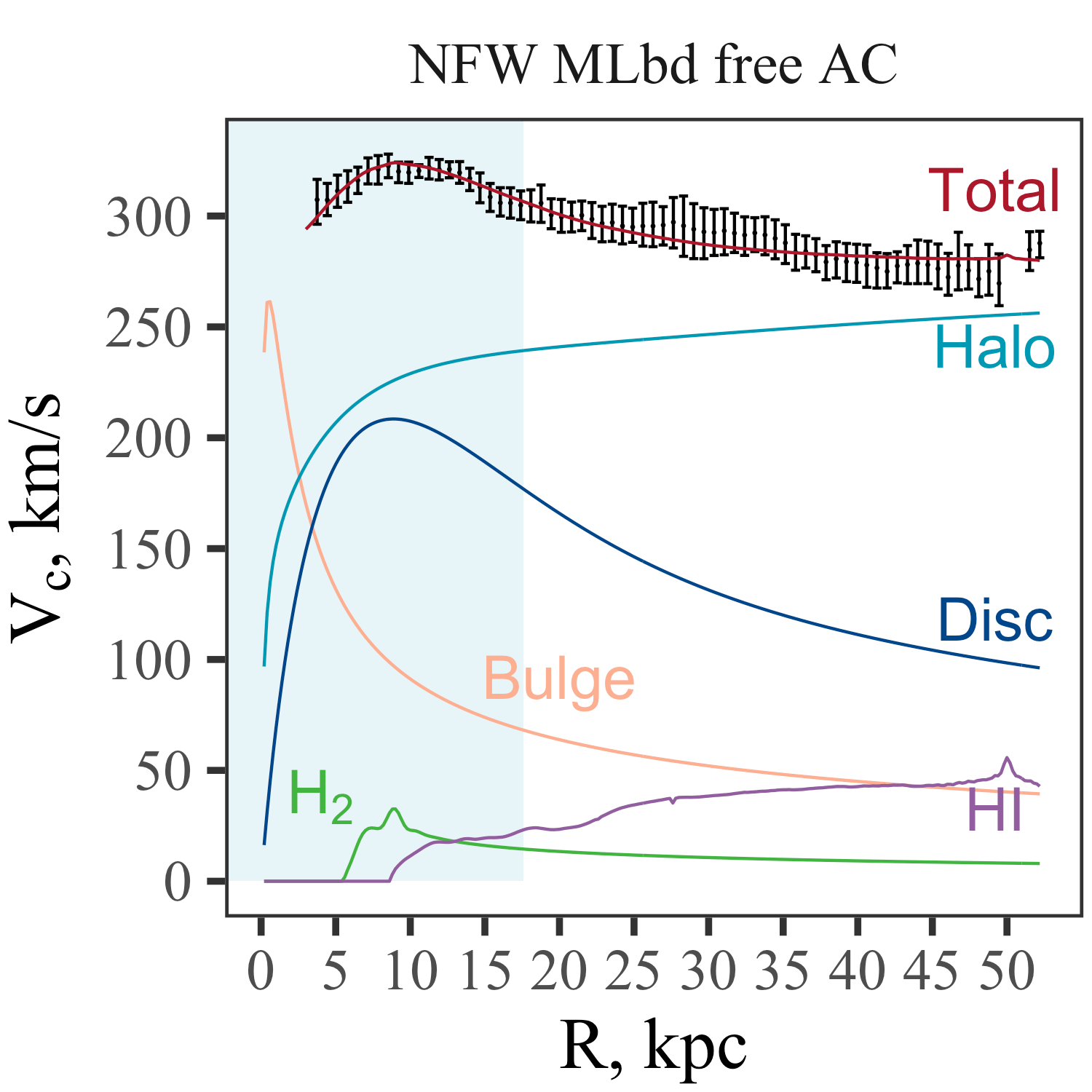}
\includegraphics[width=0.23\textwidth]{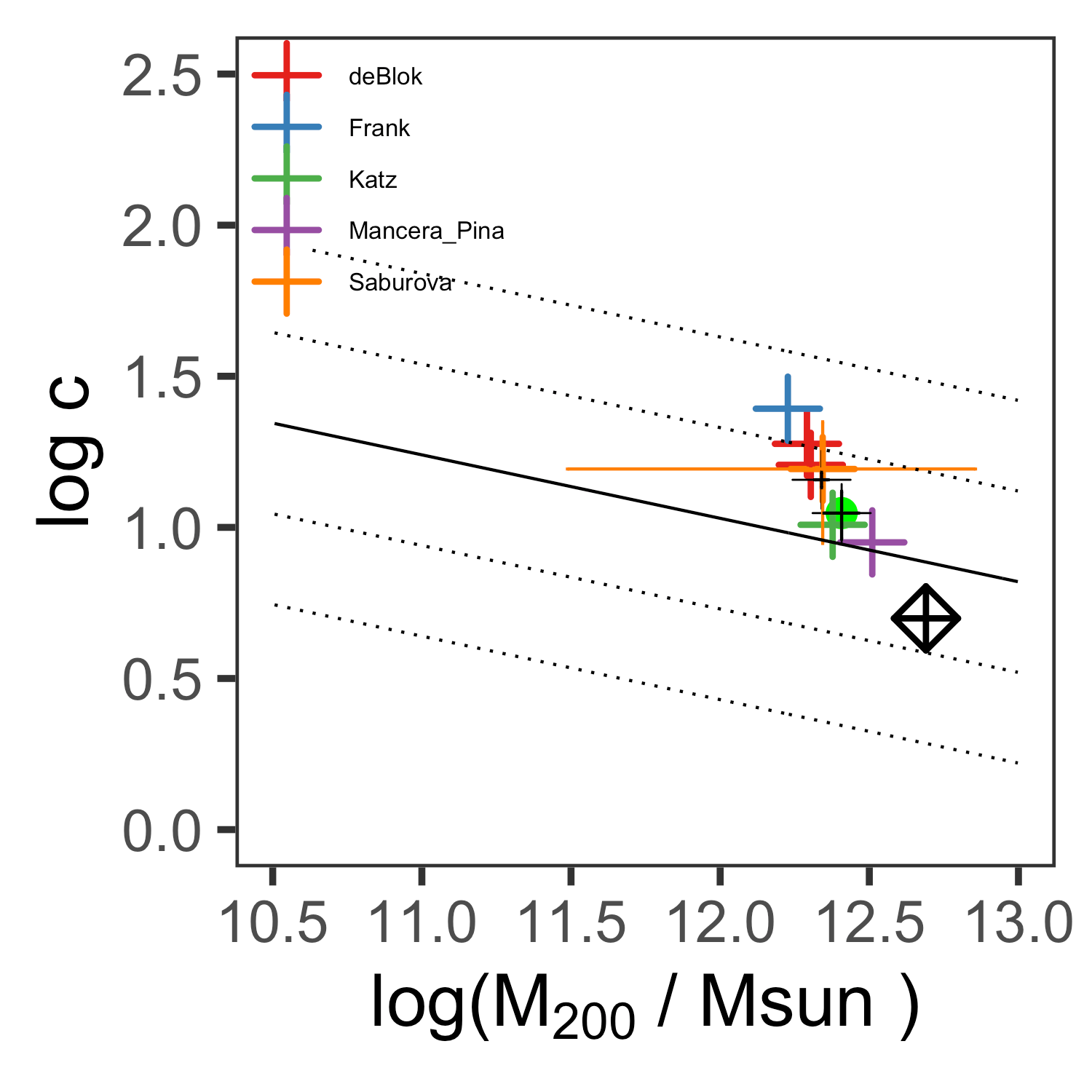}
\caption{The results of the rotation curve fit for NGC~2841. {\it Left}: mass model with the compressed NFW dark halo and a diet Salpeter IMF (fixed value $(M/L)_{\mathrm{d}}=0.68$). {\it Right}: comparison of fit NFW halo parameters with parameters taken from other works~(legend correspond to the Fig~\ref{fig:cosm-NFW}). Results of this work are shown by different colour, black points are represent values for this work. Green circle shows the models with free $(M/L)$. Diamond symbol represent the AC case.}

\label{fig:cosm-AC}
\end{figure}

Although the parameters of the NFW halo are in good agreement with the scaling relations resulting from cosmological simulations, we have tested the effect of adiabatic contraction (AC) of the halo. We have applied this procedure to the galaxy NGC~2841~(Section~\ref{sec:AC}). The fit parameters of the NFW halo for this galaxy, as they were before contraction, are also given in Table~\ref{tab:NFW_RCdec} and are plotted in Fig.~\ref{fig:cosm-AC} that demonstrates the scaling relation $c$--$V_{200}$ \citep{Macci+2008}. We want to emphasize that in the new model, the contribution of the dark halo within the optical radius of the galaxy becomes large, but we have considered the extreme case of contraction according to \citet{Blumenthal+1986} recipe. In \citet{Katz+2014}, who used a more gentle compression procedure, halo contraction did not significantly change the contributions of the disc and halo to the rotation curve.

\par
Decomposition of the rotation curve of NGC~2841 with compressed halo is shown in Fig.~\ref{fig:cosm-AC} ({\it left} plot). Fig.~\ref{fig:cosm-AC} ({\it right} plot) presents $c$ vs. ${M}_{200}$ for NGC~2841 according to the present work and works of other authors. The black circles indicates the model without AC with a free ratio $(M/L)_{\mathrm{d}}$. Fit halo without AC with a fixed value of $(M/L)_{\mathrm{d}}=0.68$ corresponds to the black diamond with an asterisk. The AC model with a fixed ratio  $(M/L)_{\mathrm{d}}=0.68$ is indicated by a green diamond with a cross\footnote{The primordial NFW halo parameters are plotted.}. Accounting for AC increased the contribution of the dark matter halo inside $4h$, but did not change the tracking of the cosmological dependence between $c$ and $M_{200}$.

\section{Discussion}
\label{sec:discussion}
The selected galaxies have a very simple structure. Their bulges, if any, are small and have color indices that differ from the disc, which makes it possible to set physically reasonable conditions for the ratio $(M/L)_\mathrm{b}$. We impose numerous and fairly clear restrictions on the parameters of the rotation curve decomposition ($(M/L)_\mathrm{d}$ from the IR photometry, free or fixed, three different models for the halo). Using the Monte Carlo method, we made sure that the decomposition parameters converge to the same best solution (see Appendix~\ref{sec:appendix2}). We also make sure that, despite the variety of these restrictions, the contribution of dark matter within the optical radius of the galaxy for selected barless galaxies with flocculent morphology is less than the contribution of baryonic matter for all halo models, i.e., the total mass of baryonic matter (disc, bulge, gas) within the four disc scales is equal to or even higher than the mass of dark matter. This result is independent of the dark halo model and this is of crucial importance for the dynamics of galaxies.
\par
Galaxies with similar properties have recently been discovered at large redshifts \citep{Lang+2017,Genzel+2017,Genzel+2020}. \citet{Lang+2017} have shown that the rotation curve of these galaxies is consistent with a high mass fraction of baryons, relative to the total dark matter halo. \citet{Genzel+2017} concluded that two-thirds or more of the $z \geq 1.2$ of massive rotating discs are baryon-dominated within a few $R_\mathrm{eff}$. At lower redshifts ($z<1.2$) that fraction is less than one-third, and at $z\sim0$ the fraction of baryon-dominated, massive discs is less than 10\%. The galaxies at high redshifts have massive discs with powerful star formation, which is localized in giant star-forming clumps. Perhaps the stellar discs of these galaxies are going through a stage of violent instability. We in no way associate our flocculent galaxies with high redshift galaxies with a low content of dark matter, but among the latter there are a number of objects with small bulges and axisymmetric discs with many clumps lined up along fragments of flocculent spirals in which star formation is observed (for example, the galaxy EGS\_13035123, figure~2, \citealp{Genzel+2020}).
\par
If the galaxies we are studying are the objects that have experienced early violent instability, then the evolution of their discs should proceed differently compared to those that start from a marginally stable state. If, at the same time, they retained some of the gas, then they may be similar to the flocculent galaxies we are studying. Models of such galaxies should be built taking into account two circumstances: a reduced content of dark matter and small values of the Toomre parameter $Q$ \citep{Toomre1964}, which characterizes how unstable the disc is. We were motivated by the results of \citet{Saha_Cortesi2018}. The authors started their simulations from isolated dynamically cold disc ($Q<0.5$) and showed that such a disc ends up with a featureless structure, while passing trough the short stage of violent instability with fragmentation and formation of stellar clumps. In the next section, we consider similarly unstable models, varying the contribution of the dark matter and the disc initial properties.
\section{Dynamical models of galaxies with low contentent of dark matter}
\label{sec:nbody}
The existence of galaxies without bars and with reduced content of a dark matter within the optical radius is poorly explained in terms of galactic dynamics.
\par
In $N$-body simulations it is almost impossible to get a model without a bar. To prevent the formation of a bar in numerical simulations, special models are most often used: either with a very massive halo, or with a very compact and fairly massive bulge, or with a dynamically very hot disc ($Q>2.0$), or with an initially very thick disc. In real galaxies, such features are not always present.
\citet{Sellwood+2019} recently pondered a similar mystery of the galaxy M~33. For M~33, they constructed a mass model of all components based on an extended rotation curve with high space resolution. This rotation curve was exploited as the basis for the equilibrium $N$-body model. All simulations with model parameters slightly varying within uncertainties led to the formation of a bar. Meanwhile, the galaxy shows neither a bar nor even traces of any oval-like distortion.
\par
For galaxies NGC~2841, NGC~3521, NGC~5055, and NGC~7331, it is impossible to exclude the presence of a small oval-like distortion due to large inclination. An oval-like distortion, or a weak bar can be obtained in models with \mbox{$\mu_\mathrm{h} \equiv M_\mathrm{h}(R <4h)/M_\mathrm{d}\approx$ 0.5--1.0} starting from marginally stable conditions (\citealp{Athanassoula_Misiriotis2002}, figure~3, middle plot, model MD; \citealp{Smirnov_Sotnikova2018}, figure~8, left plot, $\mu_\mathrm{h} = 1.0$). If we compare the model with $\mu_\mathrm{h} = 1.5$ and the model with $\mu_\mathrm{h} = 1.0$ from \citet{Smirnov_Sotnikova2019} (figure~1 there, upper left plot) we can notice that the amplitude of a bar $A_2$ decreases when going from $\mu_\mathrm{h} = 1.5$ to $\mu_\mathrm{h} = 1.0$. \citet{Athanassoula2003} explains the difference in amplitude by the smaller number of halo particles at the co-rotation. These particles effectively absorb the angular momentum of the resonant particles that make up the bar. It is the efficient exchange of angular momentum that drives the bar evolution and determines its strength and length. More massive halo and more angular momentum exchange lead to stronger bars.
\par
To reduce the influence of dark matter particles on the growth of the bar, we built several additional to  \citealp{Smirnov_Sotnikova2018,Smirnov_Sotnikova2019} models with $\mu_\mathrm{h} = 0.5$. 

\subsection{Model setup}
\label{sec:setup}
We do not consider here the models with the gaseous component. All models contain only stars dark matter.
Each model initially consists of an exponential disc ($h$ is the radial scale length) isothermal in the vertical direction ($z_0$ is the vertical scale height) embedded in a live dark matter halo. The halo is modelled by a truncated sphere with the density profile close to the Navarro-Frenk-White (NFW) profile \citep{NFW1996,NFW1997}. One model possesses a small classical bulge of a \cite{Hernquist1990} profile with total bulge mass $M_\mathrm{b}$ and the scale parameter $r_\mathrm{b}$. For some models, we also vary the Toomre parameter of a disc $Q$ at $R=2\,h$ from 0.5 to 2.0. The details about model parameters are given in Table~\ref{tab:models_pars}.
\par
For each of the models, we assume that the disc scale length is $h=3.5$ kpc and the disc mass is $M_\mathrm{d}=5 \cdot 10^{10} M_\mathrm{\sun}$. {Then the time unit will be $t_\mathrm{u}\approx 14$ Myr.} 
We use 4$kk$ particles for the disc, 4.5$kk$ for the halo, and 0.4$kk$ for the bulge. The evolution of the model was followed up to 8~Gyr.
\par
The whole procedure of model setup according to the specified initial conditions is described in details in~\citet{Smirnov_Sotnikova2018}. A brief description is as follows. An $N$-body representation of each model was prepared via {\tt{mkgalaxy}} code of~\cite{McMillan_Dehnen2007}. This code is a part of {\tt{NEMO}} project~\citep{Teuben_1995}. The evolution of the models was calculated via {\tt{gyrfalcON}} integrator~\citep{Dehnen2002}.

\begin{table}
\centering
\caption{Parameters of models}
\begin{tabular}{|c|c|c|c|c|}
\hline
$\mu_\mathrm{h}$ & 
$z_0 / h$ & 
$Q$ & 
$M_\mathrm{b}$ & 
$r_\mathrm{b}$ \\
\hline 
\hline
 1.5 & 0.05 & 1.2 & 0 & -- \\
 \hline
 1.5 & 0.05 & 0.5 & 0 & -- \\
\hline
 0.5 & 0.05 & 1.2 & 0 & -- \\
 \hline
 0.5 & 0.20 & 1.2 & 0 & -- \\
\hline
 0.5 & 0.05 & 1.2 & 0.1 & 0.1 \\
 \hline
 0.5 & 0.05 & 2.0 & 0 & -- \\
 \hline
 0.5 & 0.05 & 0.5 & 0 & -- \\
\hline
\multicolumn{5}{p{0.35\textwidth}}
{\footnotesize{\textit{Notes}: each column represents parameters of the models, one model on one line. $\mu_\mathrm{h}=M_\mathrm{h}(R < 4\, h)$ is the mass of the halo in units of the disc mass $M_\mathrm{d}$ within a sphere with radius $R=4\,h$, where $h$ is the scale length of the disc, $z_\mathrm{d} \equiv z_0/h$ is the initial ratio of the disc scale height to the disc scale length. $Q$ is the Toomre parameter at $R=2\,h$. 
$M_\mathrm{b}$ and $r_\mathrm{b}$ are the total mass and the scale length of the bulge, respectively.
}}
\end{tabular}
\label{tab:models_pars}
\end{table} 

\subsection{Model evolution and final morphology}

\begin{figure}
\centering
\includegraphics[width=0.47\textwidth]{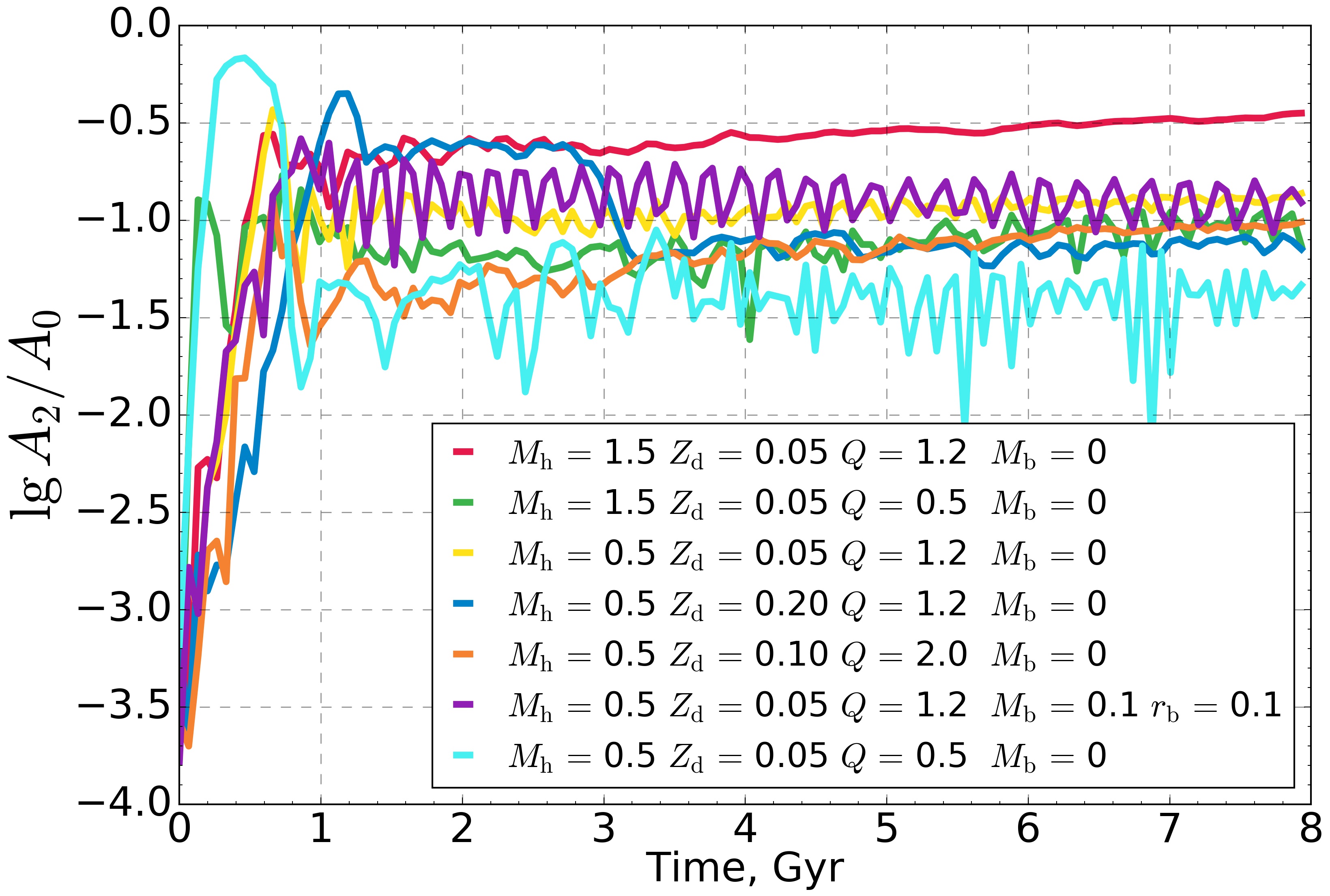}
\caption{The logarithm of the bar amplitude for all models from Table~\ref{tab:models_pars}.}
\label{fig:a2}
\end{figure}

\begin{figure*}
\centering
\includegraphics[width=0.95\textwidth]{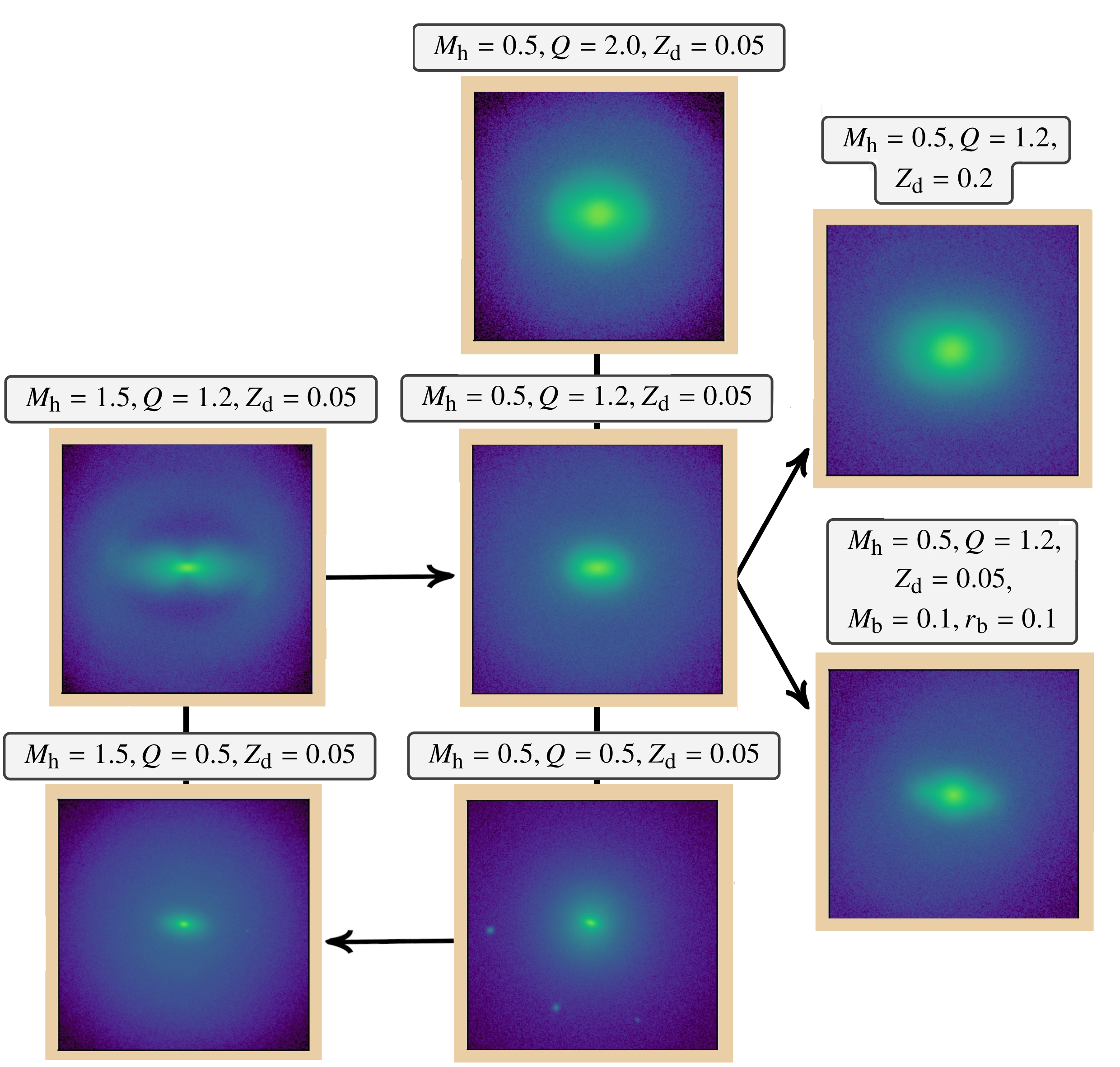}
\caption{The {face-on surface density plots of the stellar component for the} models with different initial parameters (see legends) at $t \approx 8$~Gyr. The size of each square is 5x5 in units of the disc initial scale length, i.e. 17.5x17.5 kpc.}
\label{fig:collage}
\end{figure*}
\par
Fig.~\ref{fig:a2} demonstrates the difference in the bar amplitude $A_2$ between several models. Let us compare, for example, two marginally stable and initially thin models ($Q=1.2$, $z_\mathrm{d}=0.05$), but with different dark matter contribution, with $\mu_\mathrm{h}=1.5$ (dark red line) and $\mu_\mathrm{h}=0.5$ (yellow line). The bar at time after 2~Gyr in the model with a reduced dark halo is substantially weaker.
\par
In Fig.~\ref{fig:collage} one can see the difference in the morphology between these two models with $\mu_\mathrm{h}=1.5$ (left middle plot) and $\mu_\mathrm{h}=0.5$ (at the middle). For both models $Q=1.2$, $z_\mathrm{d}=0.05$. The bar for $\mu_\mathrm{h}=0.5$ is weak and resembles a large scale oval distortion.
\par
We consider the model with $Q=1.2$, $z_\mathrm{d}=0.05$ and $\mu_\mathrm{h}=0.5$ as a base model. We varied several parameters for this model keeping $\mu_\mathrm{h}=0.5$. An initially thicker model ($Q=1.2$, $Z_\mathrm{d}=0.2$) leads to an even weaker bar (Fig.~\ref{fig:a2}, blue line) but the model itself looks like a lenticular galaxy at a face-on view (Fig.~\ref{fig:collage}, upper right plot). The dynamically hotter model\footnote{We increased the initial thickness of a disc because it is impossible to build up an equilibrium model which is hot in-plane and cold in the vertical direction} ($Q=2.0$, $Z_\mathrm{d}=0.1$) ends up with a faint oval-like perturbation (Fig.~\ref{fig:a2}, dark orange line), but the final structure also resembles a lenticular galaxy (Fig.~\ref{fig:collage}, upper middle plot). The addition of a small and compact bulge ($M_\mathrm{b}=0.1$, $r_\mathrm{b}=0.1$) to the base model diminishes the bar amplitude $A_2$ (Fig.~\ref{fig:a2}, blue violet line) but leads to a clear barlens morphology (Fig.~\ref{fig:collage}, bottom right plot).
\par
Recently \citet{Saha_Cortesi2018} have shown that an isolated dynamically cold disc ($Q < 0.5$) settled into rotational equilibrium passes trough the short stage of violent instability with fragmentation and formation of stellar clumps. After that it evolves passively and ends up with a featureless structure. The final galaxy models (with $Q=0.17$ and $Q=0.42$) resemble the morphology of the present-day S0 galaxies with two almost exponential {sub-structures}, the inner S{\'e}rsic-like bulge (with $n=1.3$) and the extended exponential disc.
\par
The model considered by \citet{Saha_Cortesi2018} has $\mu_\mathrm{h} \approx 1.0$ (according to the rotation curve, figure~1, upper plot). We modified our base model with $\mu_\mathrm{h} = 0.5$ and started simulations from $Q=0.5$. As a result, we obtained a featurless at late stages of its evolution (Fig.~\ref{fig:collage}, bottom middle plot). The model is almost barless, the amplitude $A_2$ is very small (Fig.~\ref{fig:a2}, cyan line).
\par
How can one be sure that the same picture is not observed for $\mu = 1.5$? We constructed an additional model with $\mu_\mathrm{h} = 1.5$, $Q=0.5$ and $z_\mathrm{d}=0.05$. The model clearly exhibits a bar (Fig.~\ref{fig:a2}, green line; Fig.~\ref{fig:collage}, bottom left plot). Thus, there are two crucial factors to get a barless galaxy. First, a small amount of the dark matter $\mu_\mathrm{h} < 1.0$ leads to a weak bar for initial marginally stable disc. Secondly, a weak bar can be completely destroyed by violent instability ($Q<0.5$).
\par
Our simulations are very similar to those of \citet{Saha_Cortesi2018}. \citet{Saha_Cortesi2018} used models starting from highly unstable conditions and showed that the final models resemble S0 galaxies. On the one hand, this is consistent with Genzel et al. (\citeyear{Genzel+2020}) suggestion that the $z\sim0$ ATLAS-3D passive, early type galaxies with large bulges (ETGs, \citealp{Cappellari_etal2012,Cappellari_etal2013}) may be descendants of the baryon-dominated $z\sim2$ galaxies. 
\par
\begin{figure}
\centering
\includegraphics[width=0.47\textwidth]{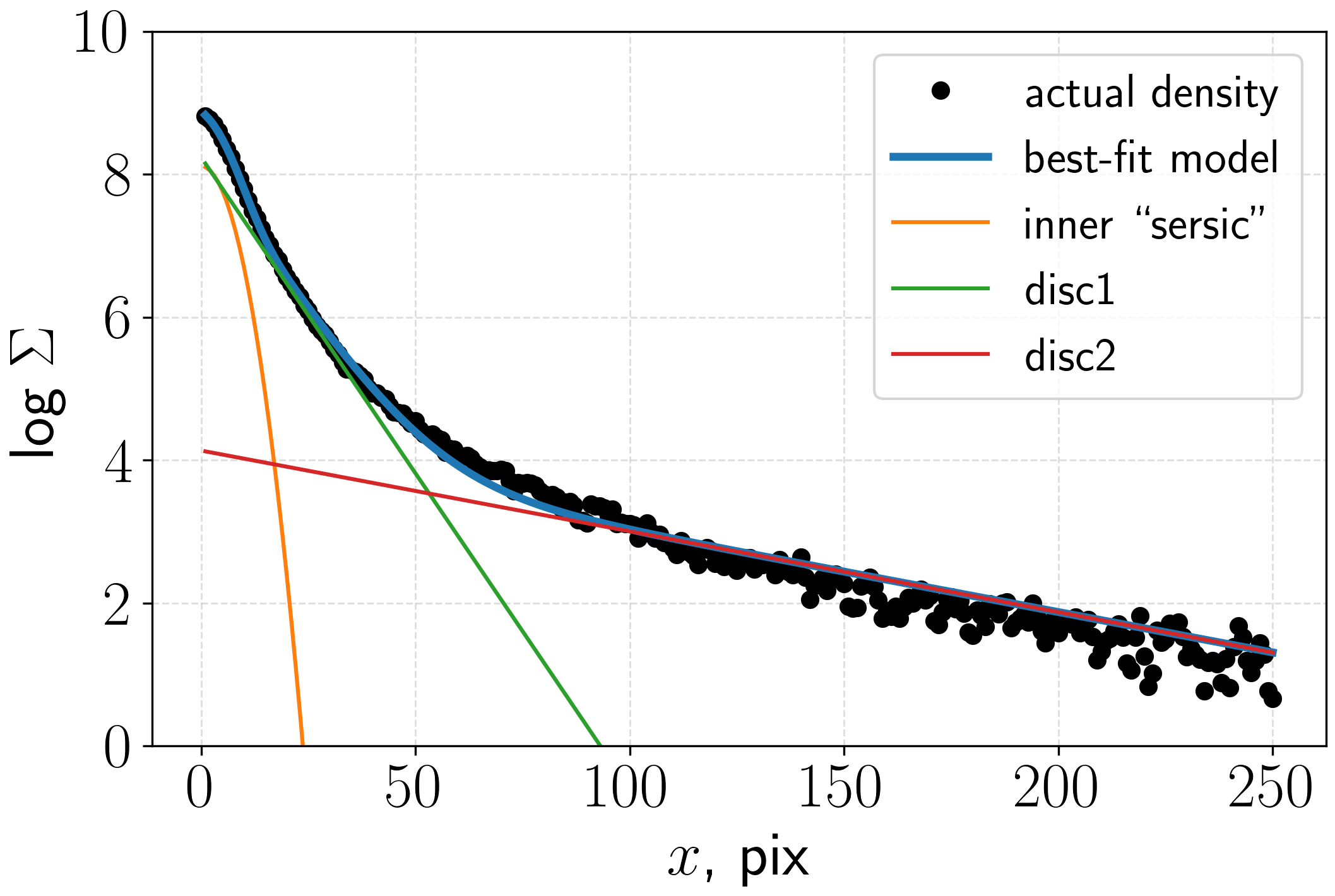}
\caption{{Three-component} decomposition of the final galaxy model. Surface density radial profile and {a central S{\' e}rsic} and two linear fits of the profile for the model with $\mu_\mathrm{h}=0.5$, $Q=1.2$ and $z_\mathrm{d}=0.05$ at $t \approx 8$~Gyr.}
\label{fig:cut}
\end{figure}
\begin{figure}
\centering
\includegraphics[width=0.47\textwidth]{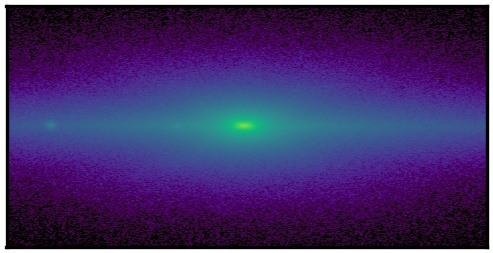}
\includegraphics[width=0.47\textwidth]{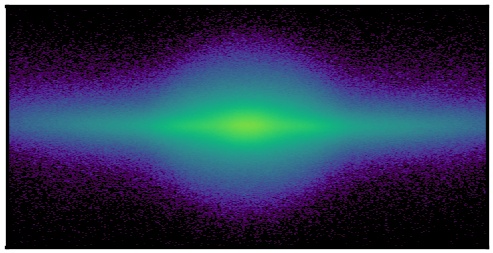}
\caption{The edge-on view of two models with $\mu_\mathrm{h}=0.5$ at $t \approx 8$~Gyr. The size of each rectangular is 5x2.5 in units of the disc initial scale length. {\it Top}: $Q=0.5$, $z_\mathrm{d}=0.05$. {\it Bottom}: $Q=2.0$, $z_\mathrm{d}=0.1$.}
\label{fig:edgeon}
\end{figure}

Actually, {our final model demonstrates three components. The central component is of ``sersic'' type. Two outer components have exponential profiles} (Fig.~\ref{fig:cut}), similar to the results of decomposition by \citet{Saha_Cortesi2018}. The inner {component can be treated as an S{\'e}rsic-like bulge (with $n \approx 0.5$)}. The contribution of the inner structure is $B/T=0.34$. The bulge need not have $n=4$, but the bulge is a spheroidal system. \citet{Saha_Cortesi2018} does not provide an edge-on view for their models with $Q<0.5$ but the contribution of the inner {sub-structure} to the total mass is large ($B/T=0.35$). To distinguish between a S{\'e}rsic-like bulge (with $n \approx 0.5$) and an inner near exponential disc, we provide an edge-on snapshot of our model with initial parameters $\mu_\mathrm{h} = 0.5$, $Q=0.5$ and $z_\mathrm{d}=0.05$ at the late stage of its evolution (Fig.~\ref{fig:edgeon}, upper plot). Our model is definitely flat and has rather two discs as NGC~3521 and NGC~5055 (Table~\ref{tab:phot_decomposition}) {and a central component with slightly elliptical isophotes}. It should be noted that the galaxy NGC~3521 also has a {classical} bulge in its structure (Table~\ref{tab:phot_decomposition}), but in our model there was no bulge from the very beginning. As for the galaxy NGC~7331, it shows small traces of the presence of an inner {sub-structure} (Fig.~\ref{fig:N7331_mu}) that is different from the main disc. Moreover, the galaxy is visible at a very large inclination; nevertheless, no signs of a noticeable 3D bulge in its structure are observed (Fig.~\ref{fig:flocc}). At the same time, an initially ``hot'' model ($\mu_\mathrm{h}=0.5$, $z_\mathrm{d}=0.1$, $Q=2.0$) eventually comes to an almost barless state with a large bulge (Fig.~\ref{fig:edgeon}, bottom plot). Such models are more reminiscent of the $z\sim0$ ATLAS-3D passive, early type galaxies with large bulges than initially very unstable and very ``cold'' models.
\par
The main difference between the structure of our model (Fig.~\ref{fig:collage}, bottom middle plot; Fig.~\ref{fig:edgeon}, upper plot) and the structure of NGC~2841, NGC~3521, NGC~5055 and NGC~7331 galaxies is the absence of flocculent spirals in our model. But we carried out simulations without gas. As a result we got featureless models (without bar and without spirals). However, NGC~2841, NGC~3521, NGC~5055 and NGC~7331 have a lot of gas. They are active in starformation. The addition of gas in our model can lead to the flocculent structure. In simulations by \citet{Saha_Cortesi2018} and in our simulations the model passes trought the stage of violent instability and breaks up into clamps (see, for example, \citealp{Saha_Cortesi2018}, figure~2). Some clumps are preserved even at the stage of passive evolution. If galaxies have stored a lot of gas, then star formation concentrated in clumps throughout the disc will lead to the appearance of multiple scraps of spirals due to differential rotation.

\par
\begin{figure}
\centering
\includegraphics[width=0.47\textwidth]{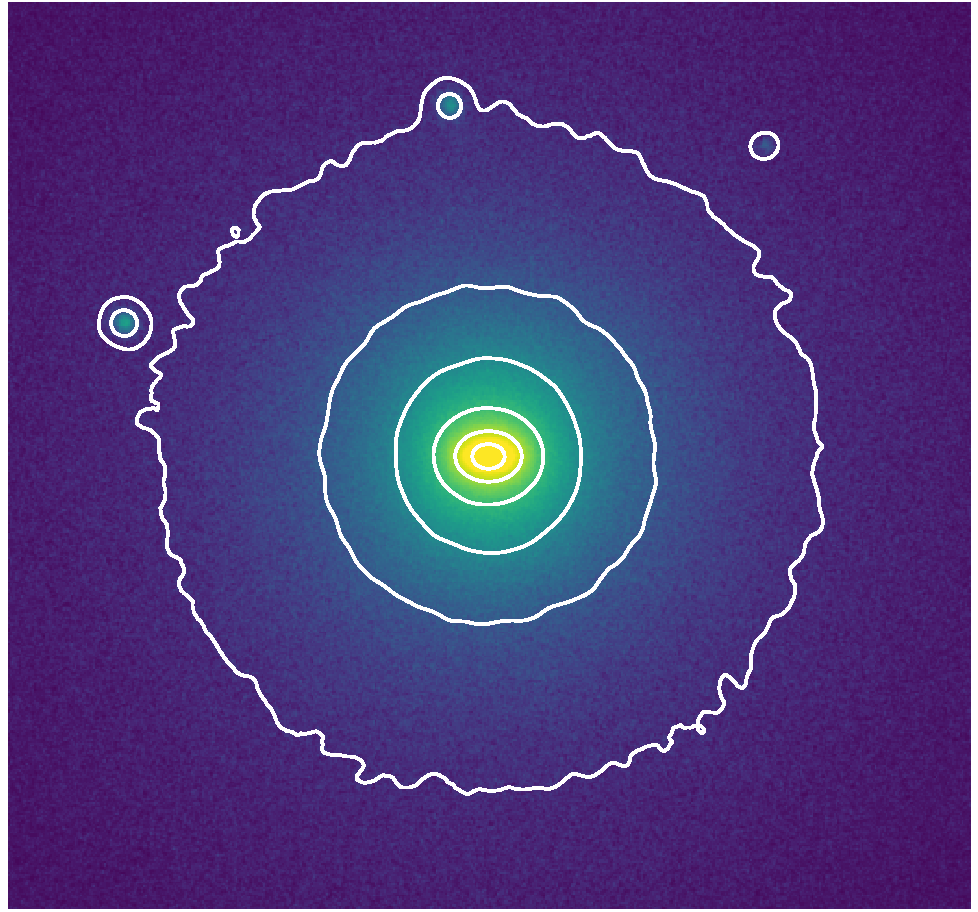}
\includegraphics[width=0.47\textwidth]{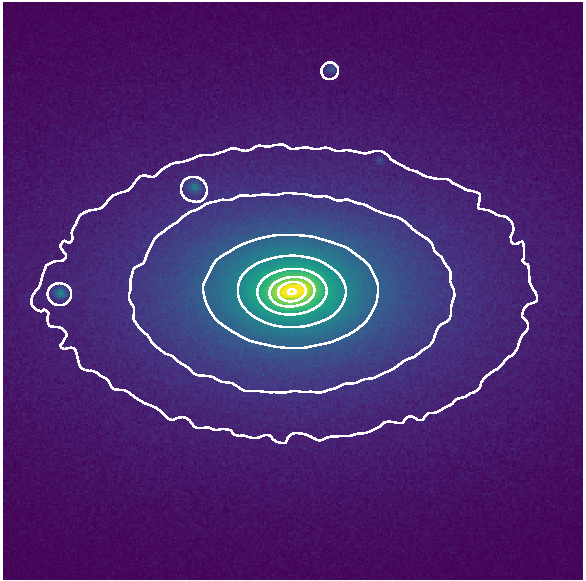}
\caption{Snapshots of the model with $\mu_\mathrm{h}=0.5$, $Q=1.2$ and $z_\mathrm{d}=0.05$ at $t \approx 8$~Gyr with superimposed isophotes. {\it Top}: the face-on view. {\it Bottom}: the snapshot was rotated in azimuth by $21.4^\mathrm{o}$ and further inclined by $56.2^\mathrm{o}$. The size of each square is 5x5 in units of the disc initial scale length.}
\label{fig:isophotes}
\end{figure}

Fig.~\ref{fig:isophotes} shows an image of a model galaxy ($\mu_\mathrm{h}=0.5$, $z_\mathrm{d}=0.05$, $Q=0.5$) at $t=600$ ($\approx 8$~Gyr) with superimposed isophotes. At the top, the model is shown face-on; isophotes in the centre outline a very subtle oval-like distortion. At the bottom, an image is rotated at an angle with P.A. and inclination taken from \citet{Laurikainen_Salo2017} for NGC~5055. It can be seen that with this position of the galaxy, very tiny methods of photometric analysis are needed to reveal the presence of an oval-like distortion.
\par
Perhaps the galaxies under consideration are objects that have survived violent instability from large $z$. At high $z$, they can be observed as objects studied by \citet{Lang+2017,Genzel+2017,Genzel+2020}. 

\section{Conclusions}
\label{sec:conclusions}

We analyzed the THINGS' and HERACLES' kinematic data of four barless galaxies with flocculent structure (NGC~2841, NGC~3512, NGC~5055, NGC~7331) and constructed their rotation curves. 
We decomposed the extracted rotation curves taking into account S$^4$G photometric models of galaxies, modern $M/L$ calibrations, the contribution of the atomic and molecular gas, reliable distance to galaxies from Cepheids, and several models for the dark halo. 
\par
As a result of the analysis, it was found that the mass of the baryonic matter in galaxies studied, within four exponential scales of the disc, is not less than the mass of dark matter, and sometimes even exceeds it. 
\par
In an attempt to understand the phenomenon of galaxies without bars, we have constructed several $N$-body models of a galaxy with $M_{\mathrm{h}} = (0.5-1.0) M_{\mathrm{baryon}}$. 
It turned out that the bar is an inevitable consequence of the evolution of such models, although the forming bar is rather faint. In models that evolve from marginally stable states, even in the case of a light halo, the formation of an oval-like distortion cannot be completely avoided. 
\par
Dynamically hot and/or fairly thick models also end up with a very weak bar, although the entire model resembles rather a lenticular galaxy. The addition of a small compact bulge does not lead to the destruction of the bar, but to the formation of a barlens. However, in galaxies NGC~2841, NGC~3521, NGC~5055, NGC~7331 even a weak bar is not observed. This creates a dynamic puzzle. 
\par
We see the following solution to this problem. Such galaxies in the past could have gone through a stage of violent instability, which could lead to the formation of a featureless disc. The decisive factor for the formation of a disc without structural features is the low mass of the dark halo. Otherwise, a small bar might still survive. 
However, only a reduced content of the dark halo is not enough. For the formation of a galaxy without features in its structure, a start from strongly unstable conditions is also necessary. In this case, the bar does not survive. Flocculent spirals in such galaxies can be explained by the presence of gas. It is the gas that can lead to the flocculent structure at the stage of violent instability, when many clumps are formed. If they preserved at the stage of passive evolution and galaxies have stored a lot of gas, then star formation concentrated in clumps throughout the disc could lead to the appearance of multiple scraps of spirals due to differential rotation.

\section*{Data availability}
The data underlying this article will be shared on reasonable request to the corresponding author.



\section*{Acknowledgements}
We thank the referee for the comments that helped to improve the presentation of the results. We thank the anonymous referee for his/her review and appreciate the comments. The authors thank THINGS’ and HERACLES team for available data. This research made use $^{\mathrm{3D}}${\tt BBarolo}.  The authors express gratitude for the grant of the Russian Foundation for Basic Researches number 19-02-00249. 

\bibliographystyle{mnras}
\bibliography{barless_galaxies}

\appendix
\section{Data, rotation curves and surface densities}
\label{sec:appendix}
In this appendix, for all investigated galaxies, we plot the results of 3D tilted-ring modeling of THINGS and HERACLES data (by $^{\mathrm{3D}}${\tt BBarolo}): rotation curves, inclination and PA profiles, the velocity fields and the best-field model, the P-V diagrams for major and minor axes.

\begin{figure}
\centering
\includegraphics[width=0.47\textwidth]{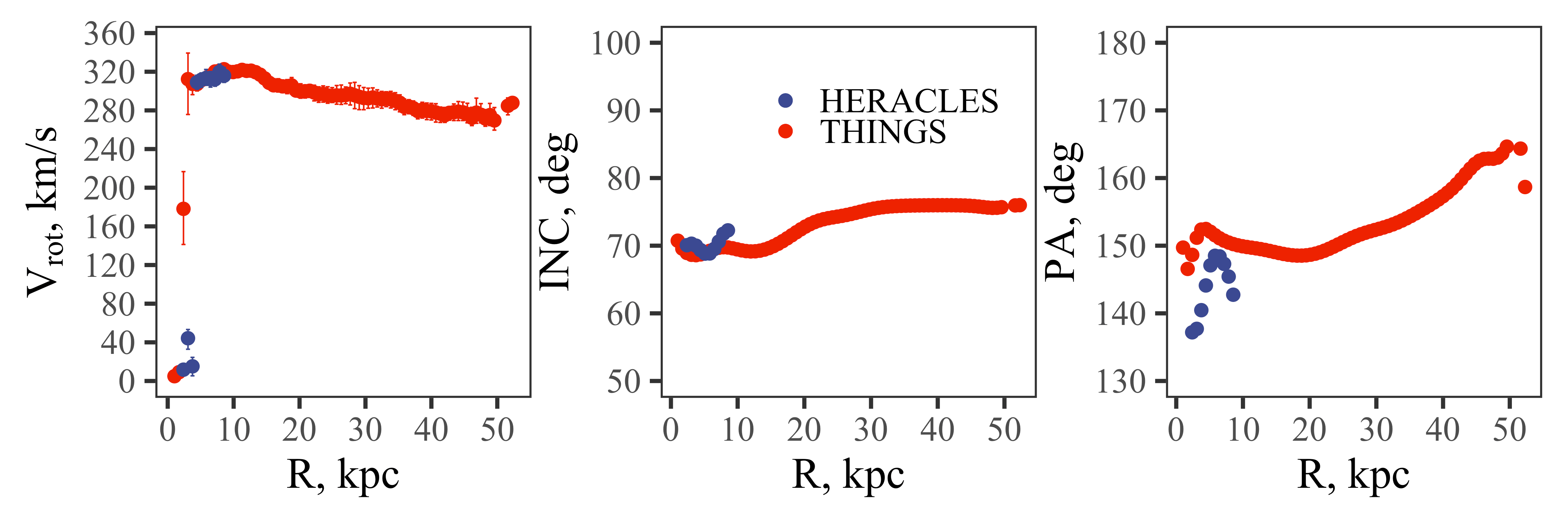}
\includegraphics[width=0.23\textwidth]{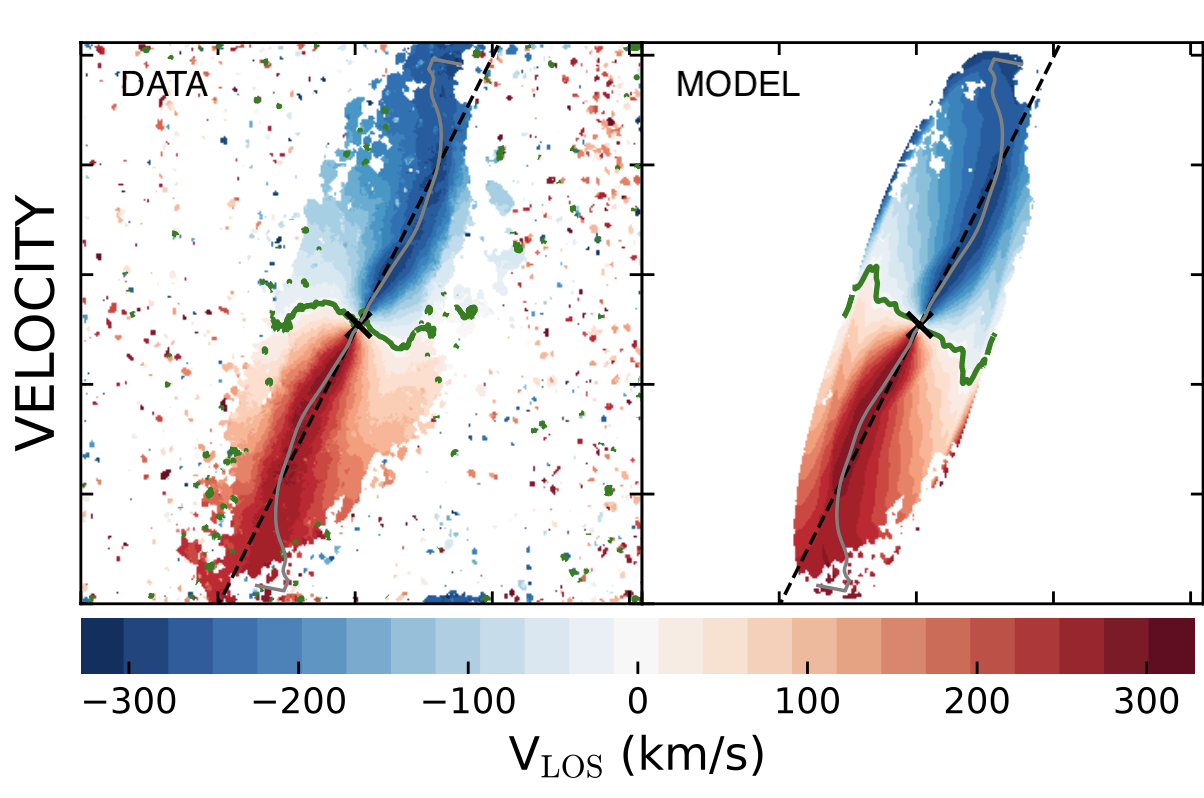}
\includegraphics[width=0.23\textwidth]{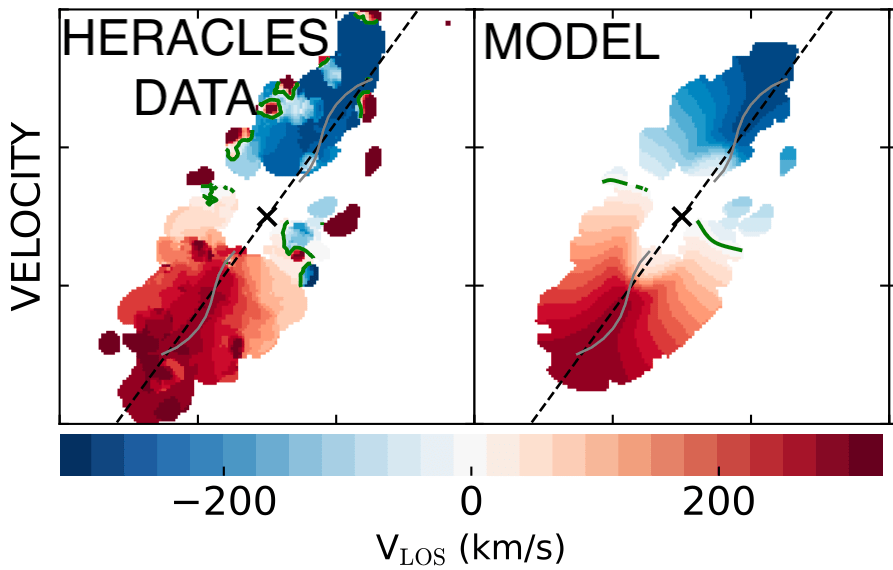}
\includegraphics[width=0.23\textwidth]{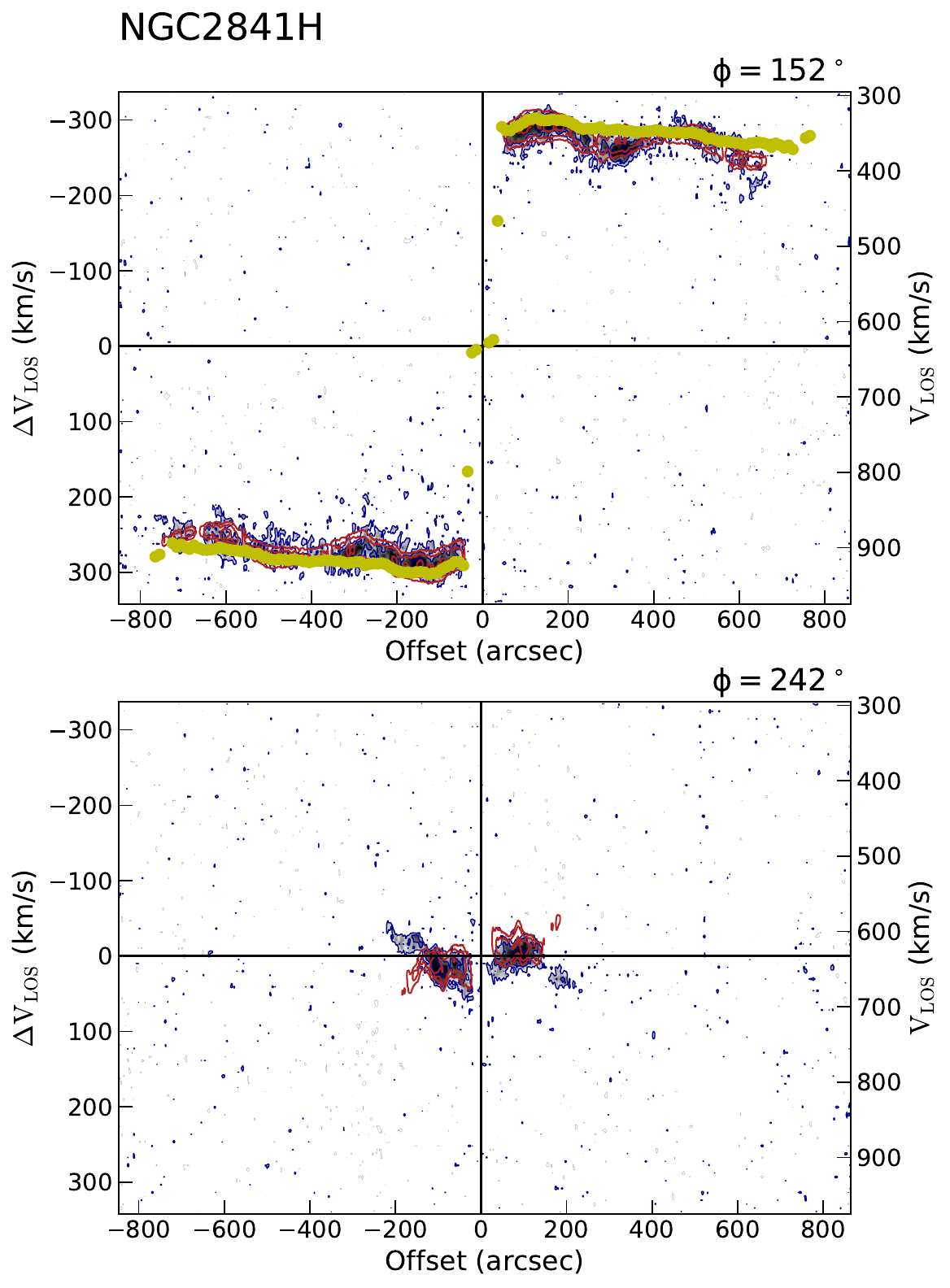}
\includegraphics[width=0.23\textwidth]{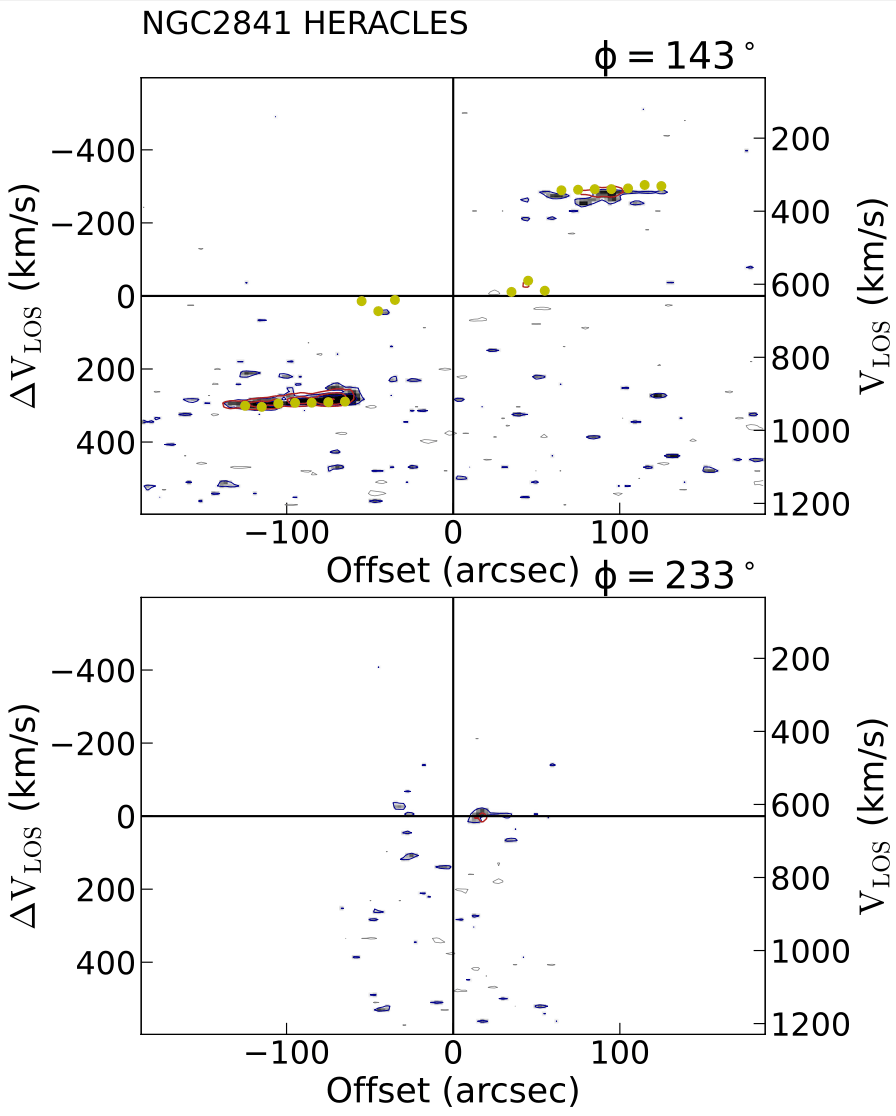}
\caption{Summary panel for NGC~2841. The top panels show RC, inclination and position angle for each ring for THINGS (red dots) and HERACLES (blue dots) data. The middle panels show the velocity map of the data and for fitted model both for THINGS~(left plots) and HERACLES~(right plots). The bottom panels show the position-velocity diagrams both for THINGS~(left plots) and HERACLES~(right plots) and for the major (upper plots) and minor (bottom plots) axes.}
\label{fig:NGC2841-cube}
\end{figure}

\begin{figure}
\centering
\includegraphics[width=0.47\textwidth]{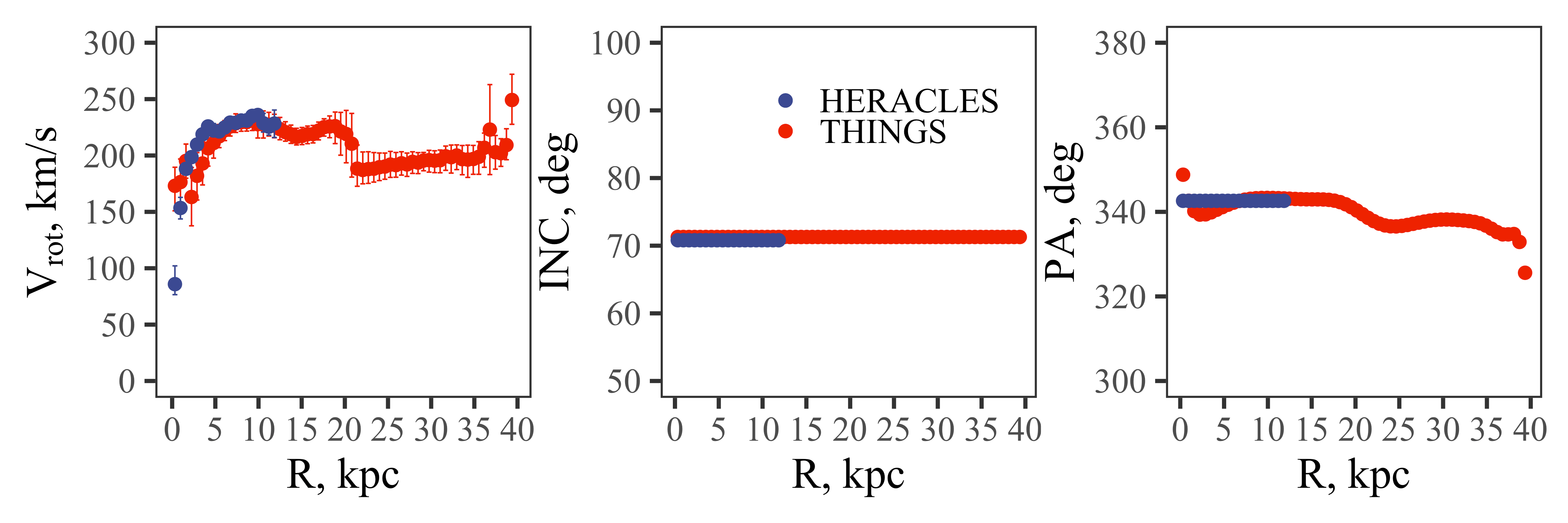}
\includegraphics[width=0.23\textwidth]{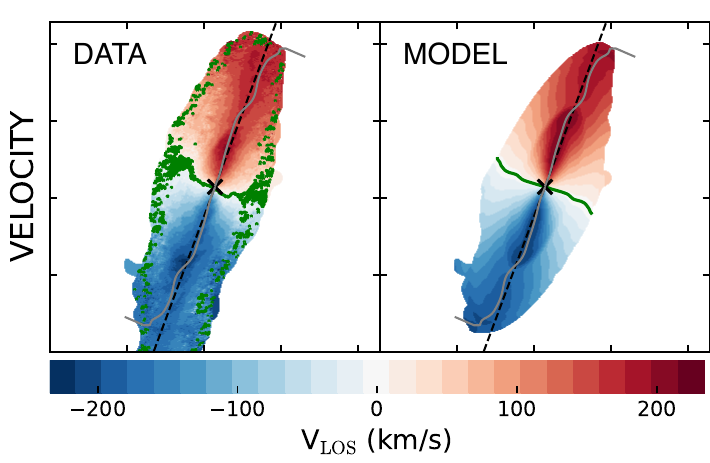}
\includegraphics[width=0.23\textwidth]{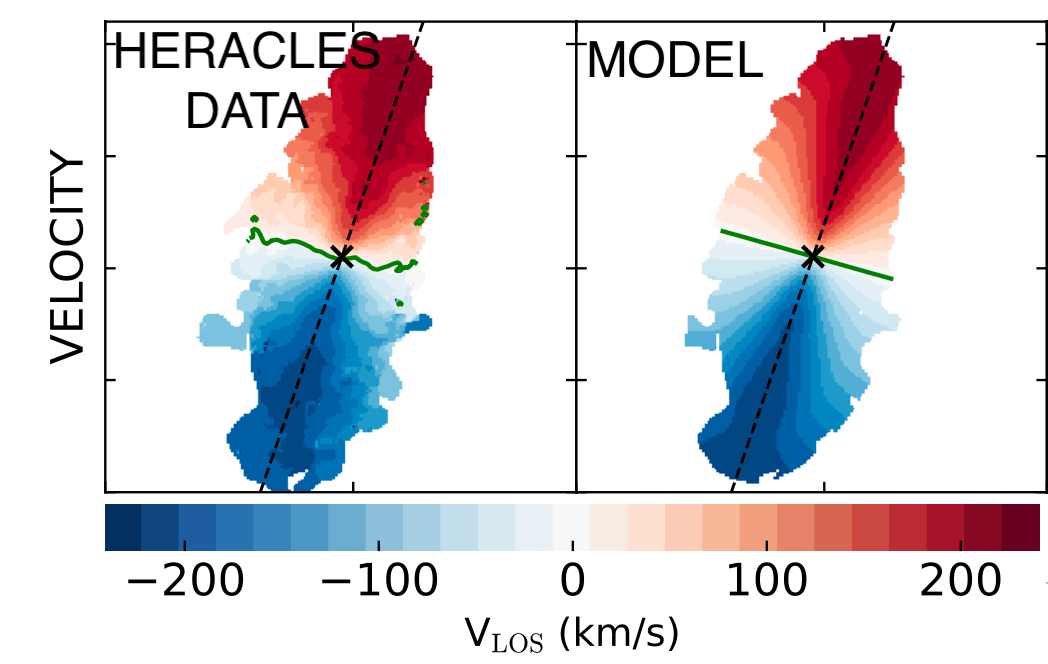}
\includegraphics[width=0.23\textwidth]{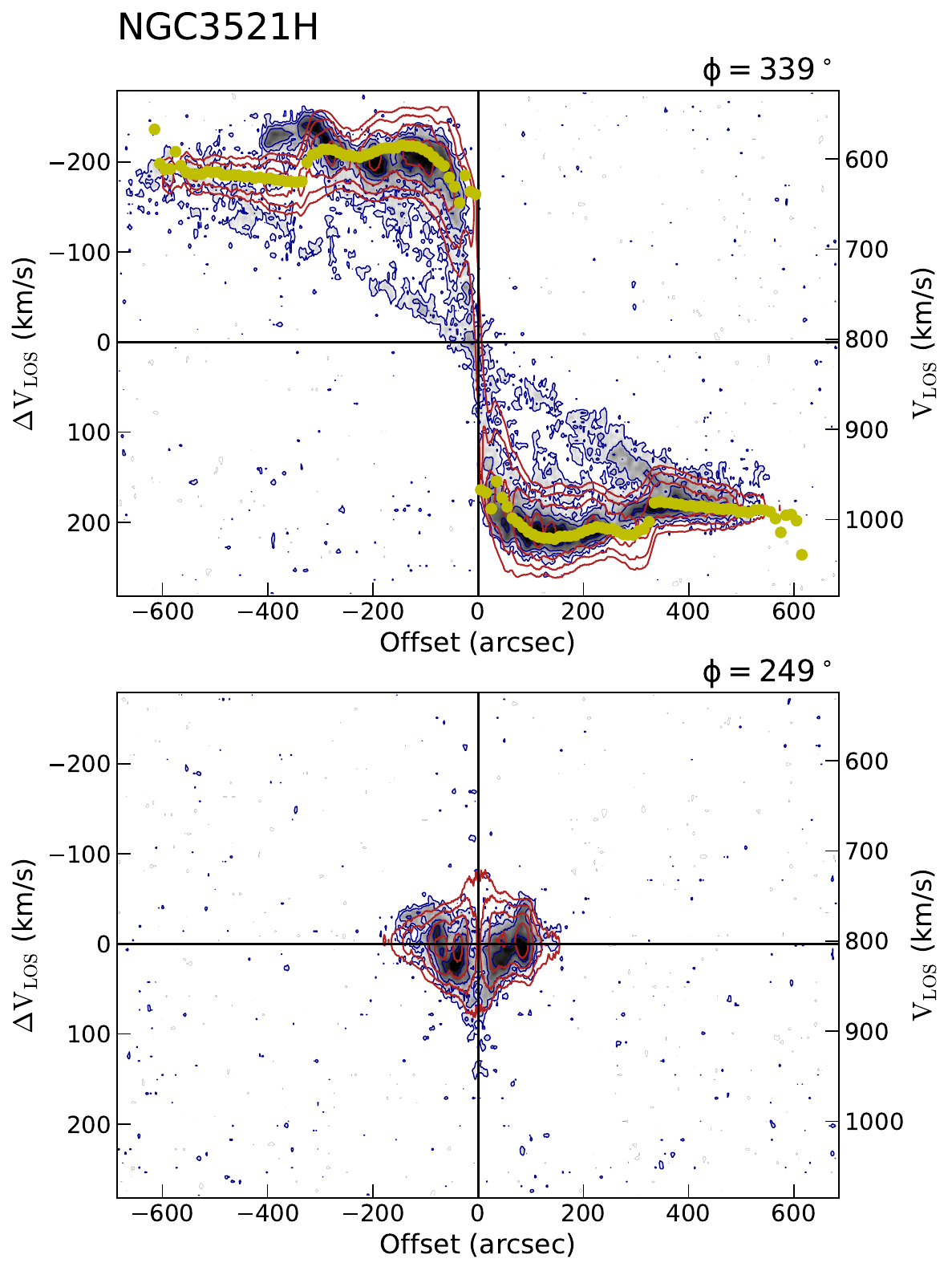}
\includegraphics[width=0.23\textwidth]{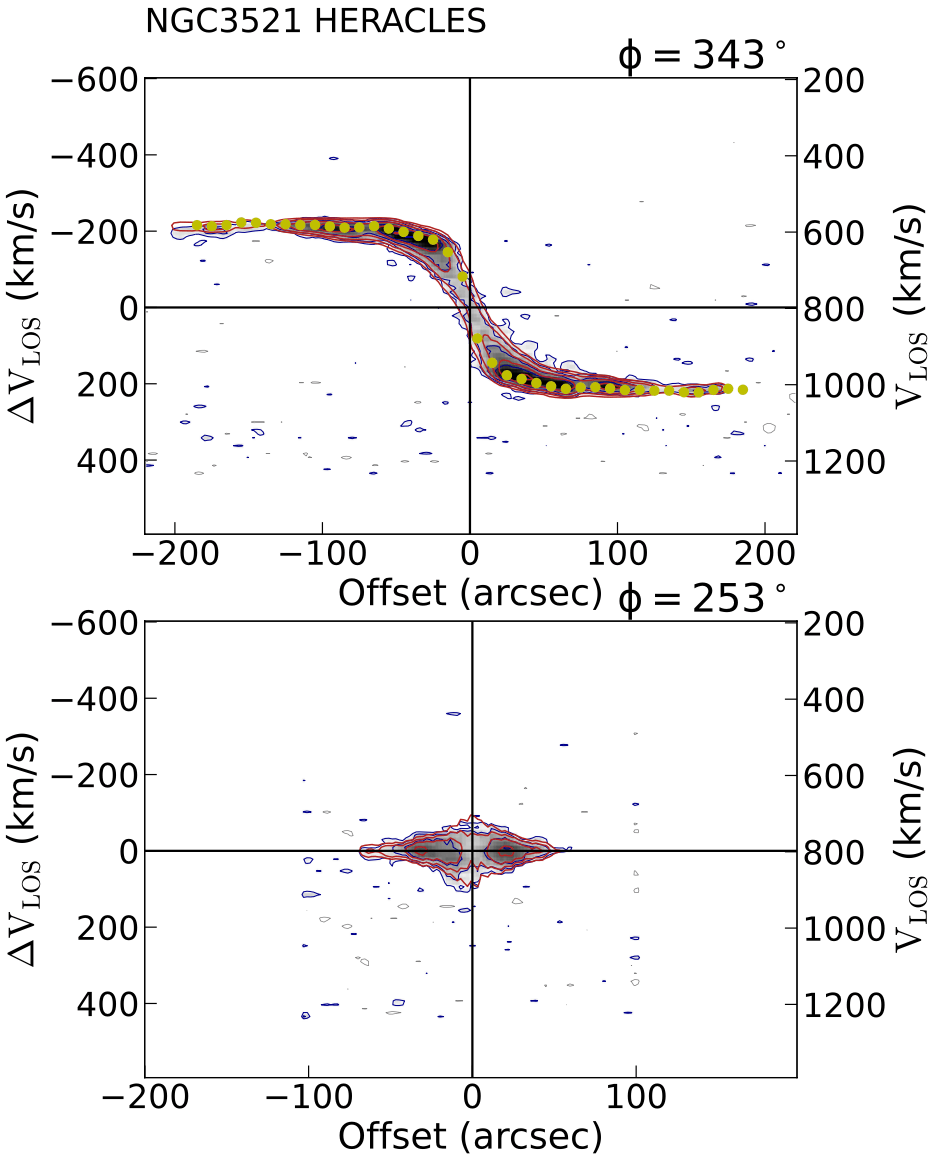}
\caption{Summary panel for NGC~3521. The same as in Fig.~\ref{fig:NGC2841-cube}.}
\label{fig:NGC3521-cube}
\end{figure}

\begin{figure}
\centering
\includegraphics[width=0.47\textwidth]{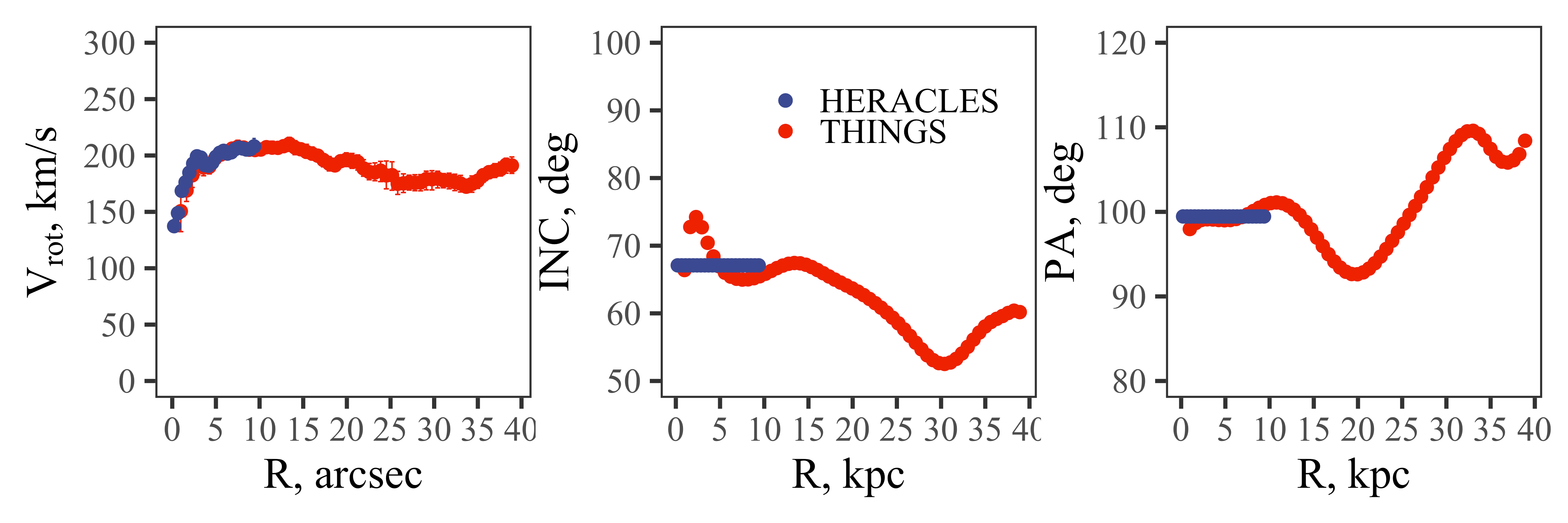}
\includegraphics[width=0.23\textwidth]{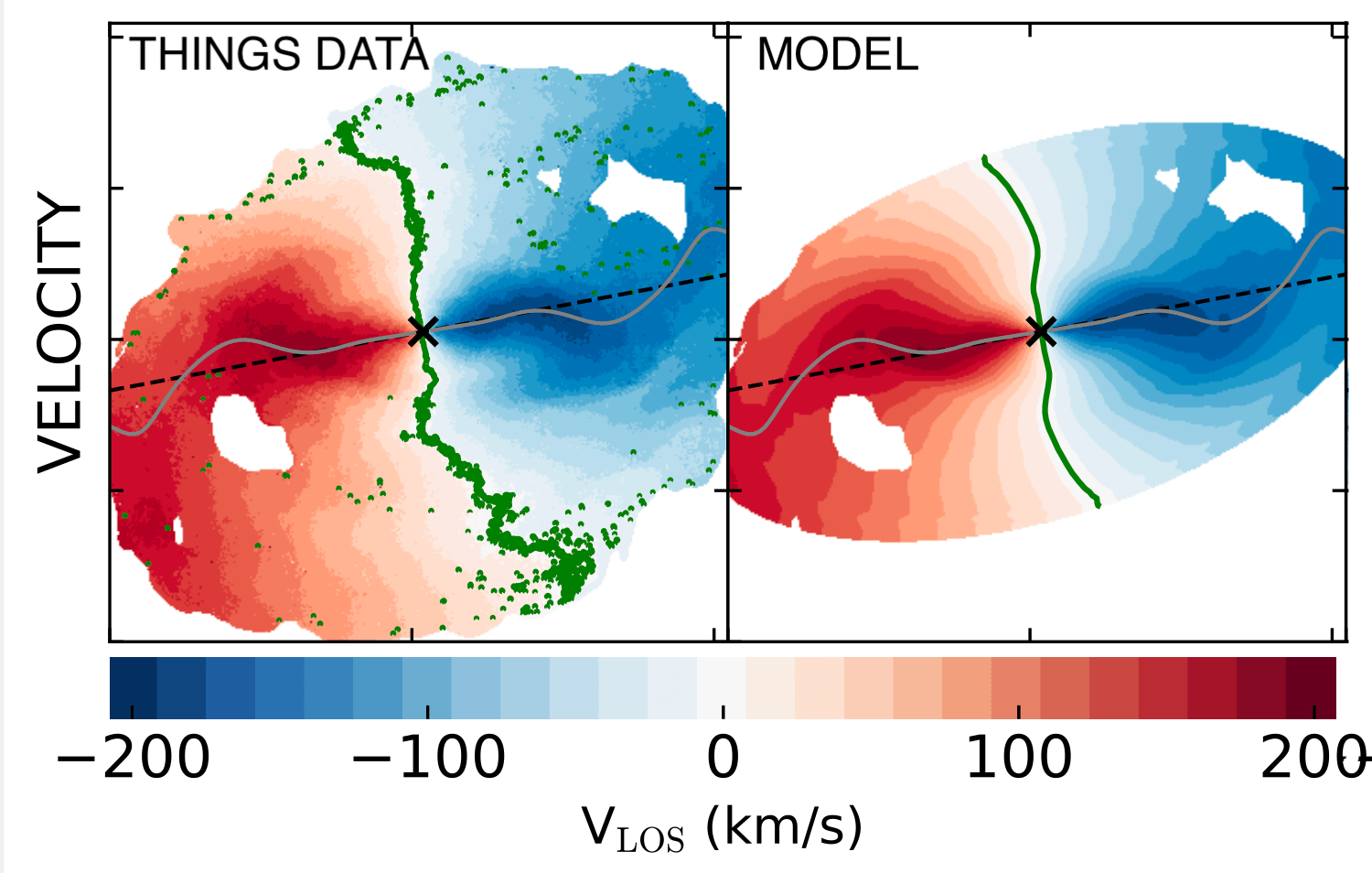}
\includegraphics[width=0.22\textwidth]{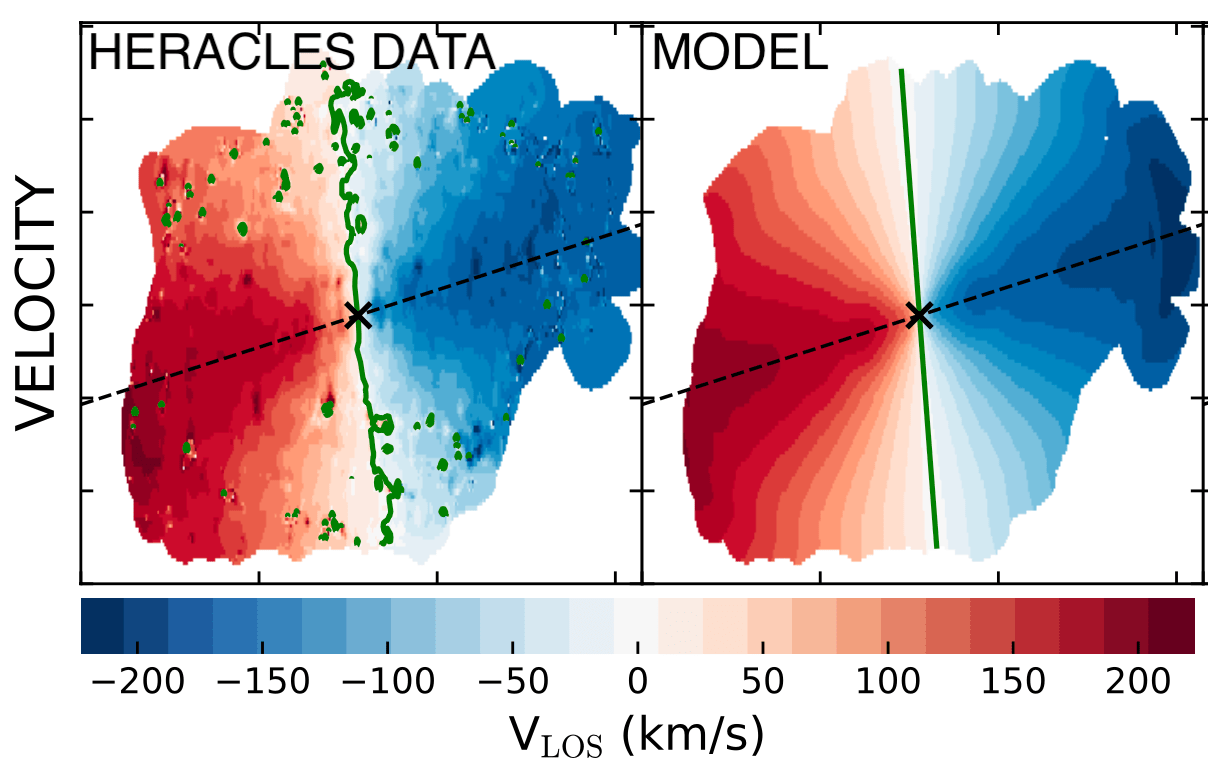}
\includegraphics[width=0.23\textwidth]{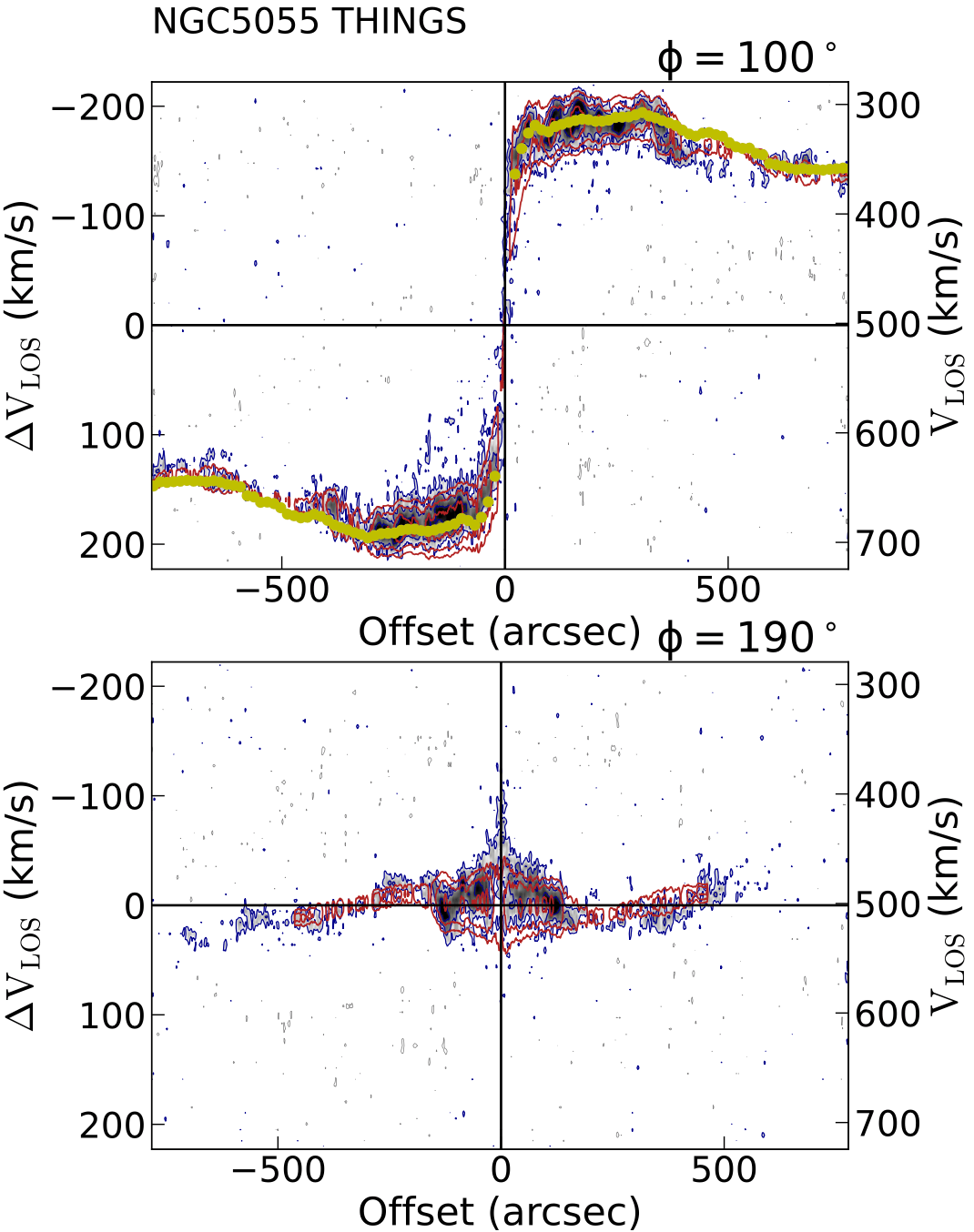}
\includegraphics[width=0.23\textwidth]{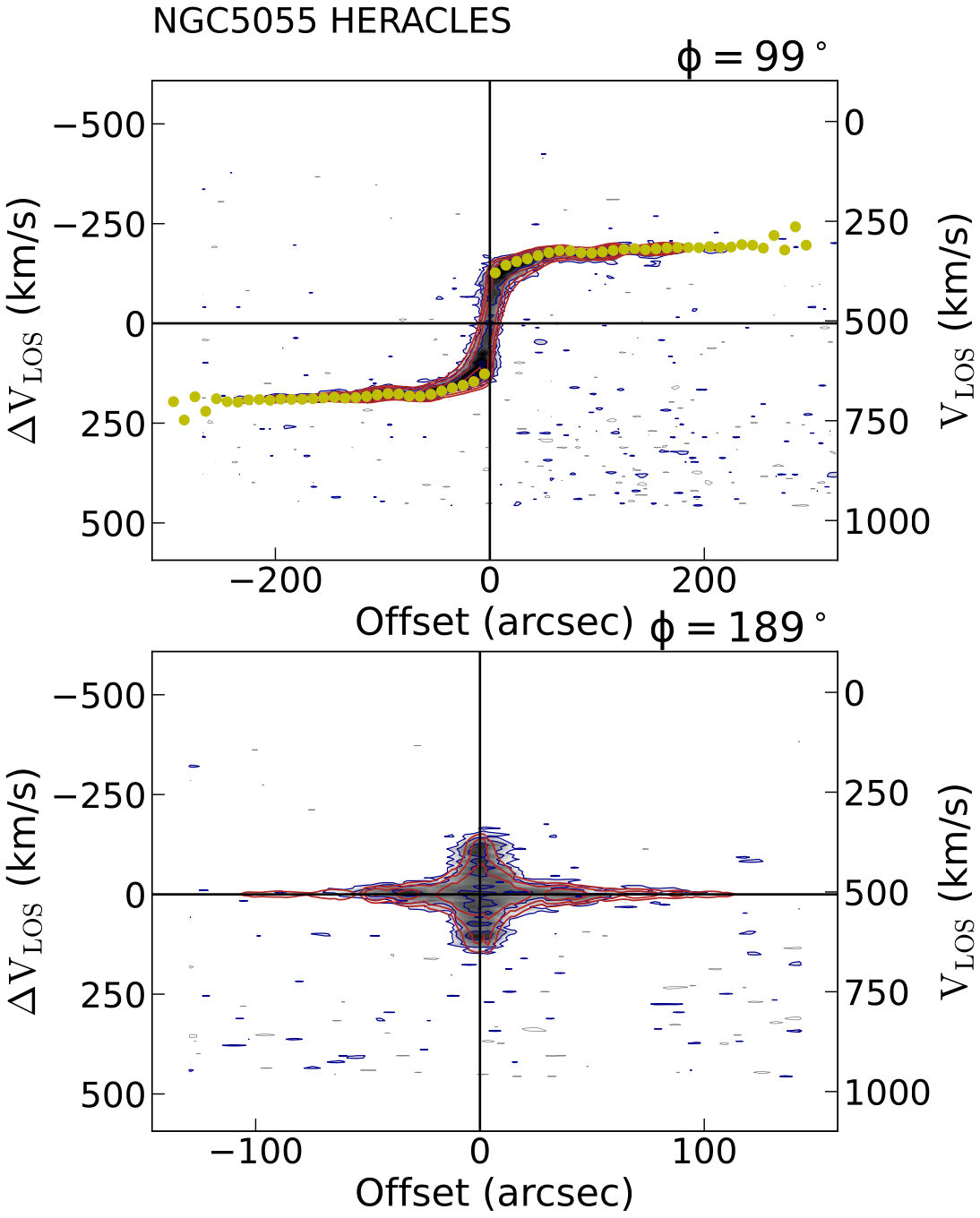}
\caption{Summary panel for NGC~5055. The same as in Fig.~\ref{fig:NGC2841-cube}.}
\label{fig:NGC5055-cube}
\end{figure}

\begin{figure}
\centering
\includegraphics[width=0.47\textwidth]{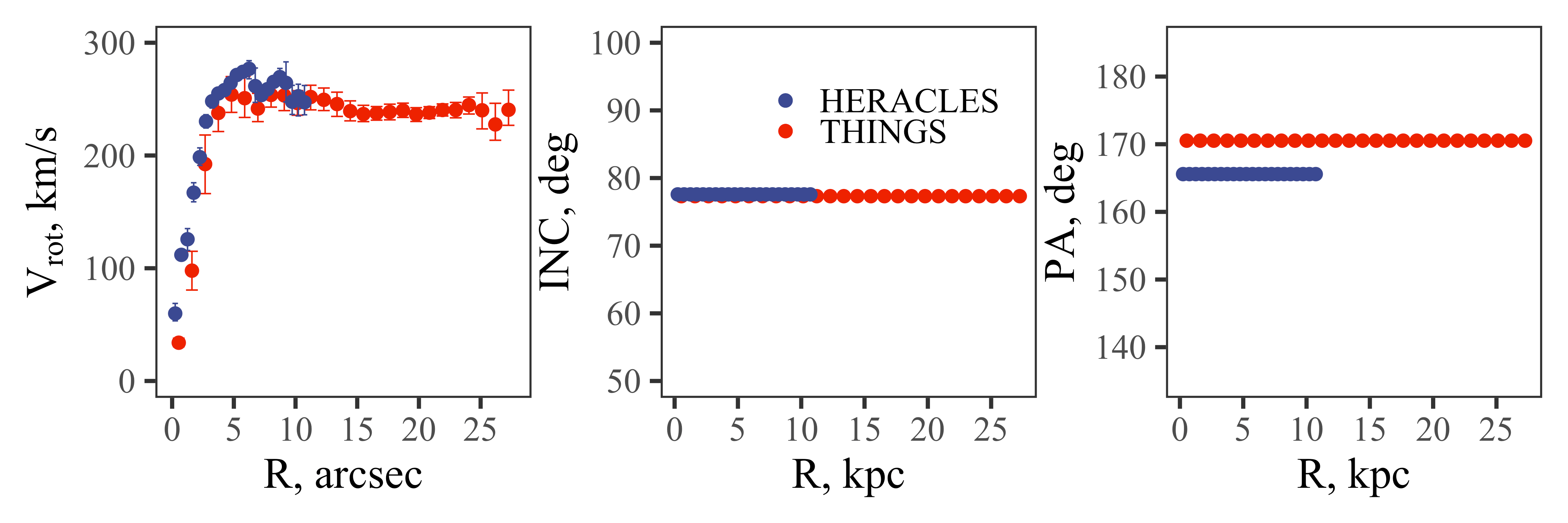}
\includegraphics[width=0.22\textwidth]{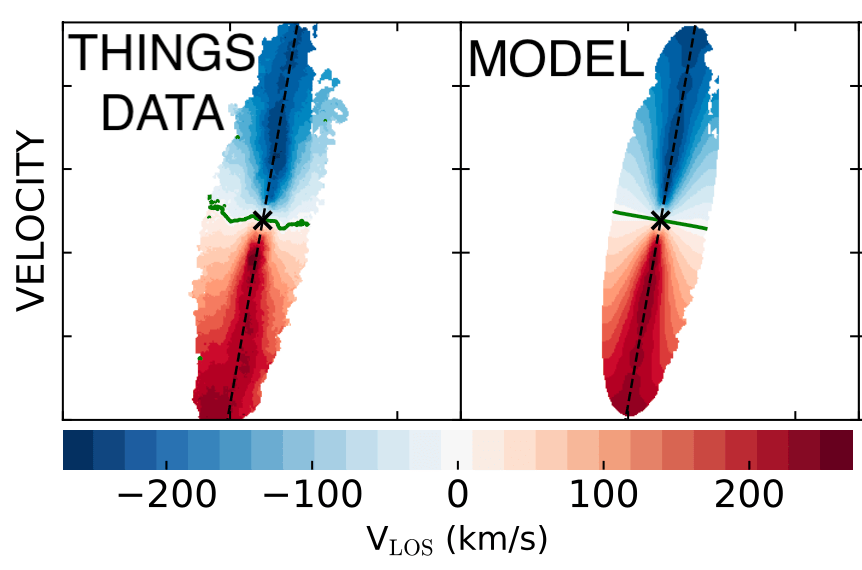}
\includegraphics[width=0.23\textwidth]{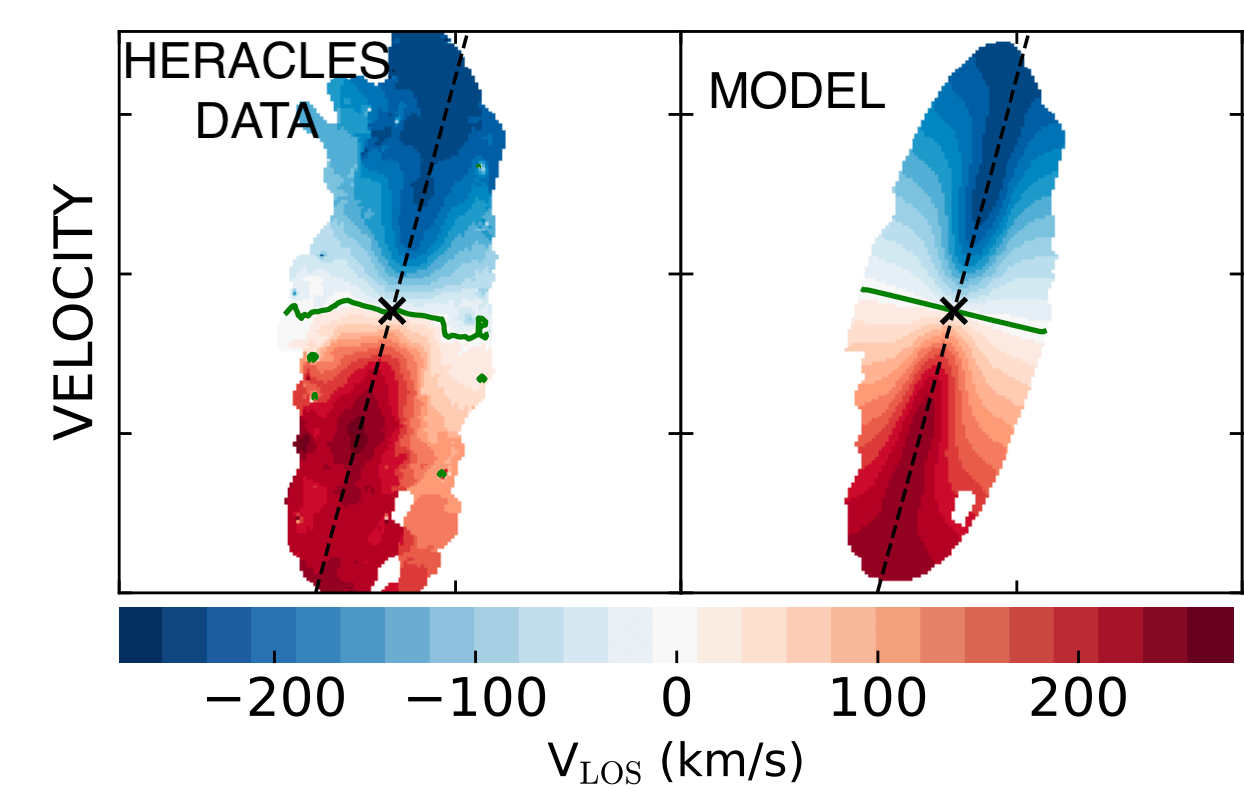}
\includegraphics[width=0.23\textwidth]{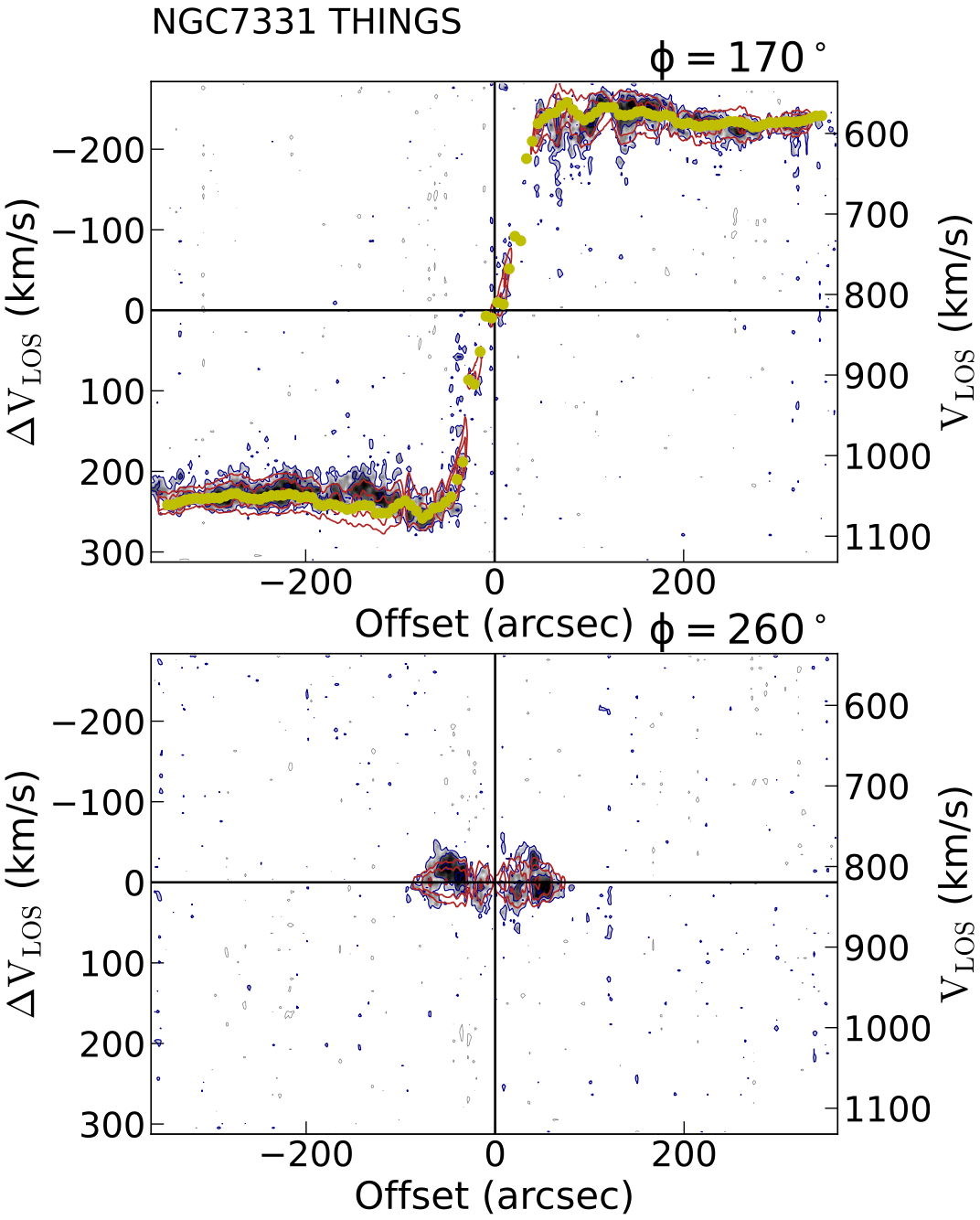}
\includegraphics[width=0.23\textwidth]{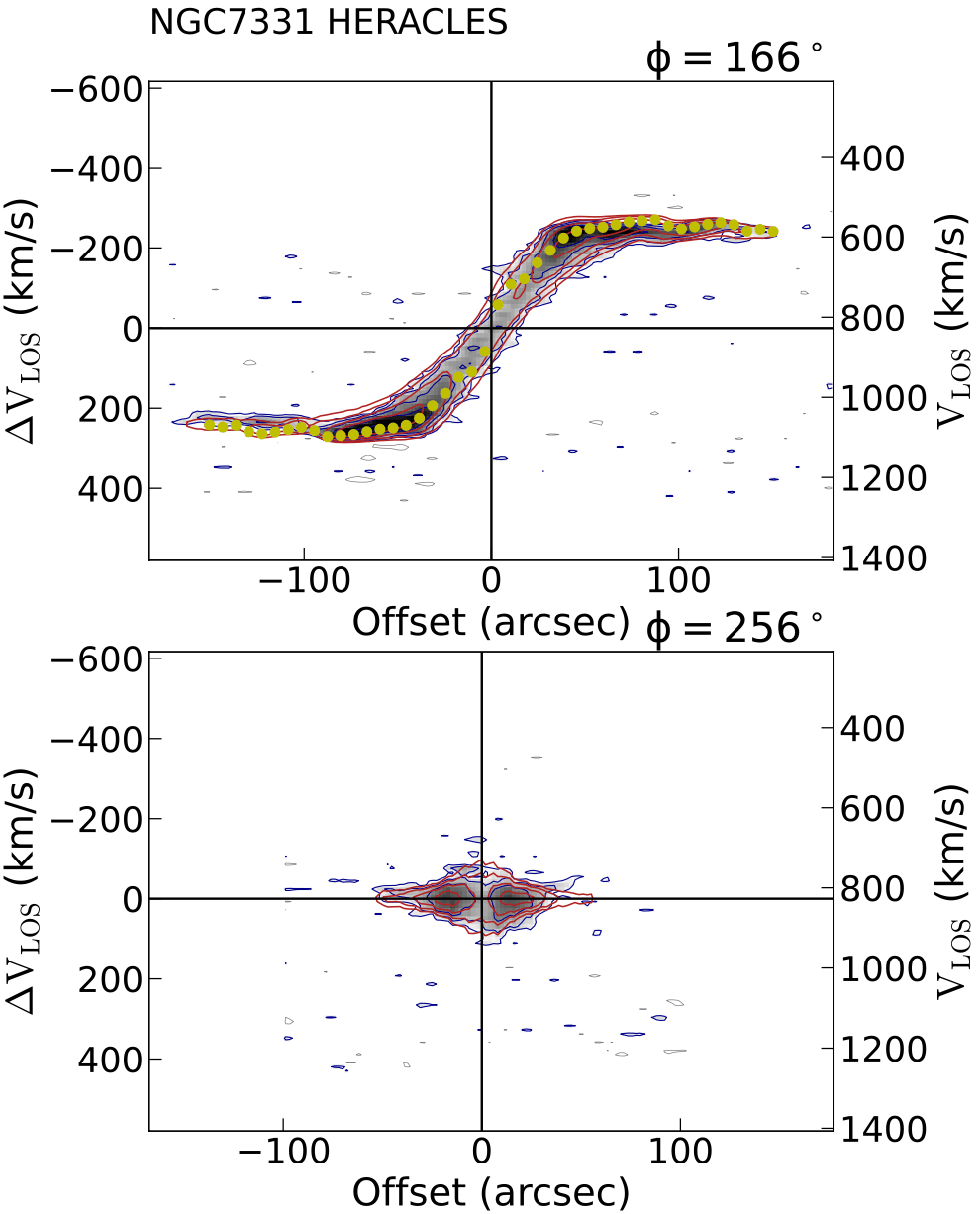}
\caption{Summary panel for NGC~7331. The same as in Fig.~\ref{fig:NGC2841-cube}.}
\label{fig:NGC7331-cube}
\end{figure}

\section{Monte Carlo simulations}
\label{sec:appendix2}

\begin{figure}
\centering
\includegraphics[width=0.48\textwidth]{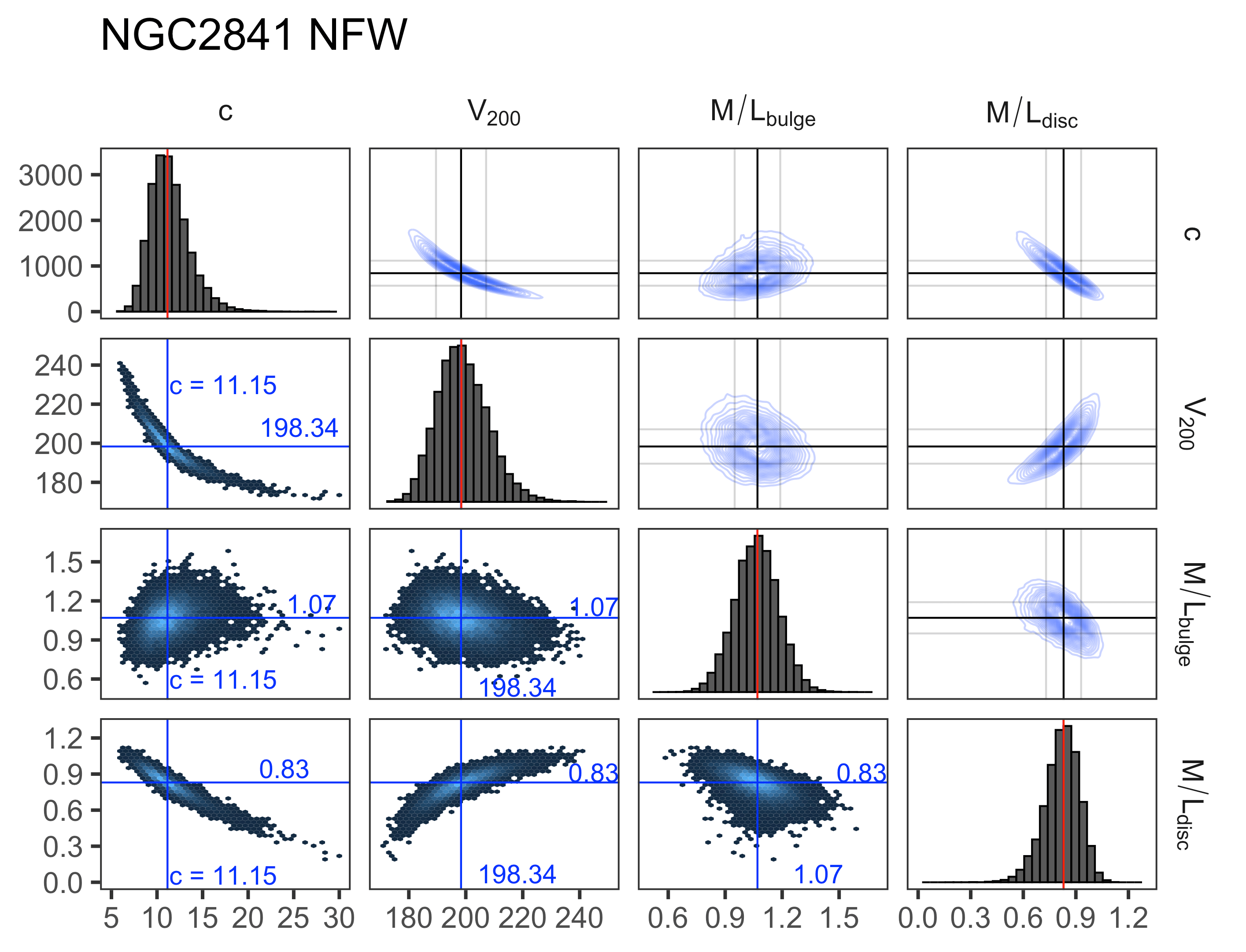}
\caption{Convergence of solutions for 20.000 random realizations of the rotation curve for NGC~2841. NFW halo, $(M/L)_\mathrm{d}$ and $(M/L)_\mathrm{b}$ are free. Blue lines indicate our best solution, red lines are locations of medians of the distributions.}
\label{fig:2841_mc}
\end{figure}

\begin{figure}
\centering
\includegraphics[width=0.48\textwidth]{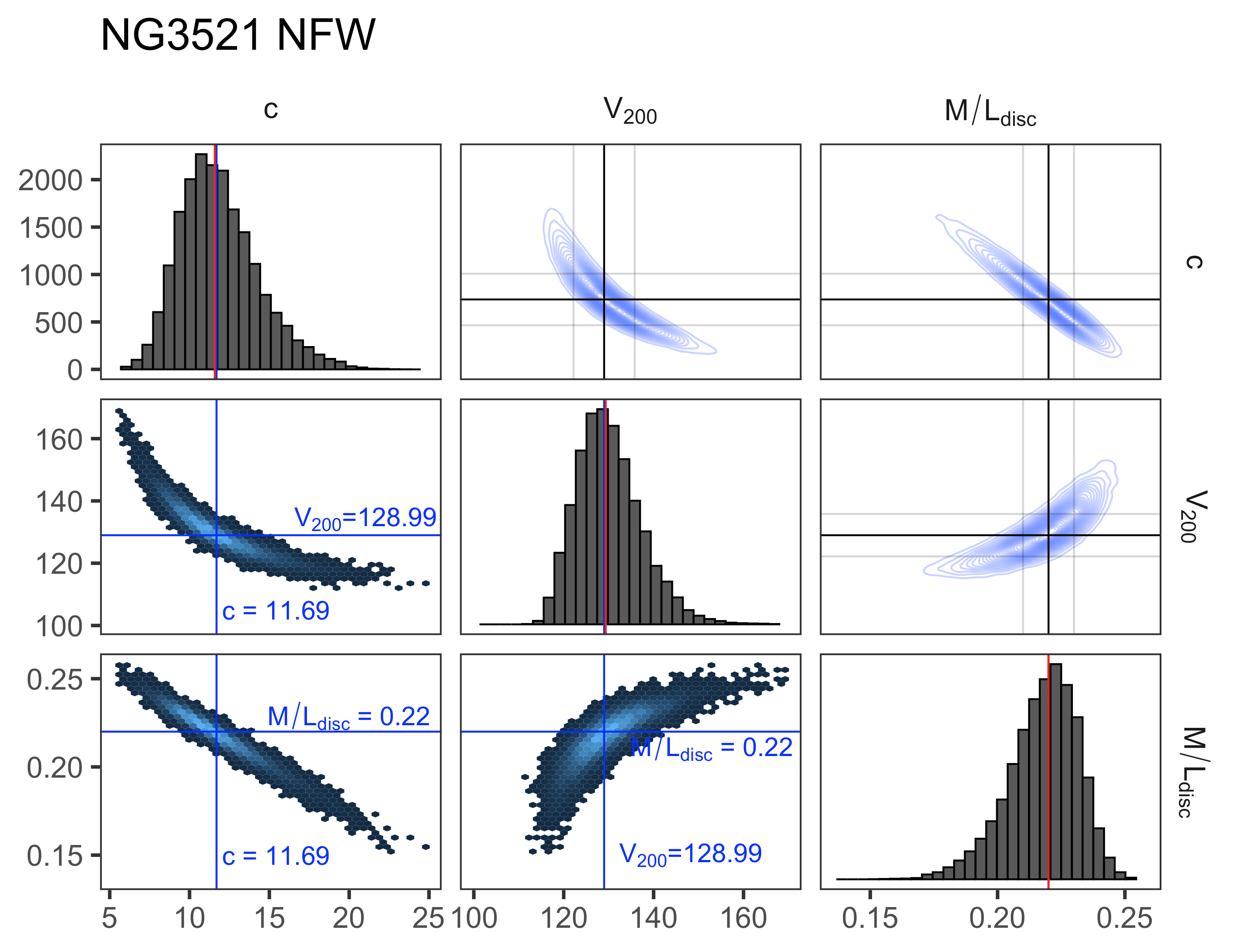}
\caption{The same as in Fig.~\ref{fig:2841_mc} but for NGC~3521.}
\label{fig:3521_mc}
\end{figure}

\begin{figure}
\centering
\includegraphics[width=0.48\textwidth]{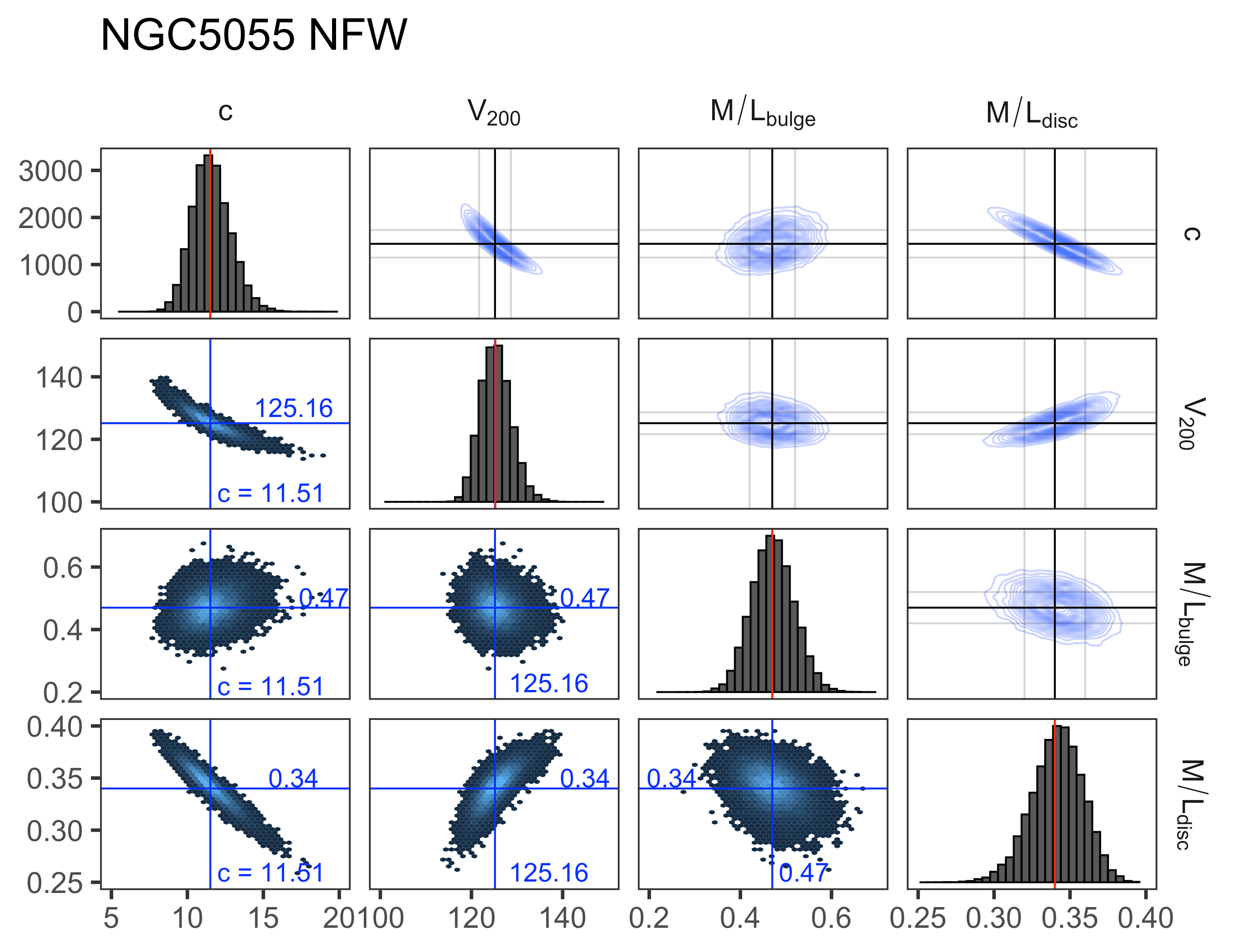}
\caption{The same as in Fig.~\ref{fig:2841_mc} but for NGC~5055.}
\label{fig:5055_mc}
\end{figure}

\begin{figure}
\centering
\includegraphics[width=0.48\textwidth]{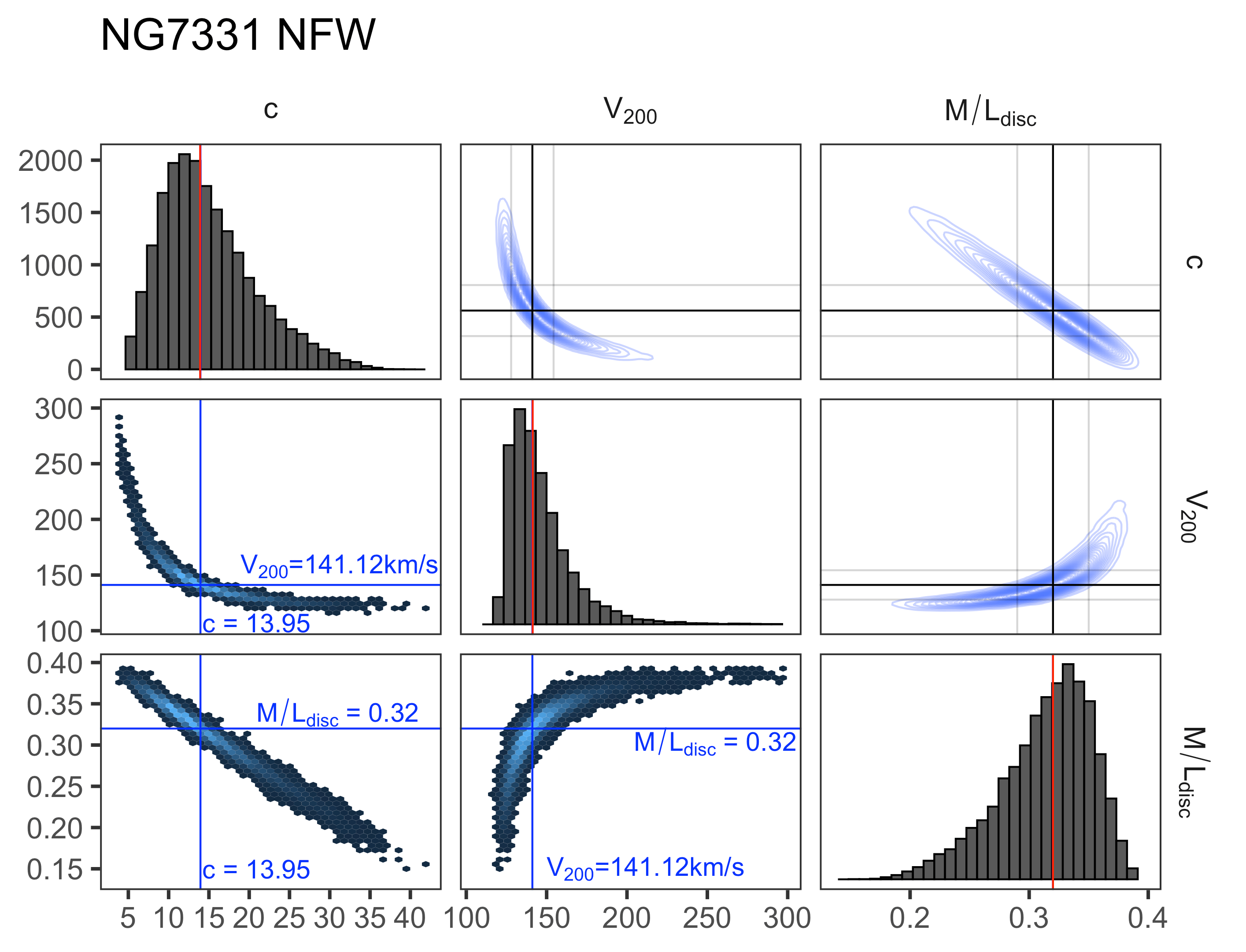}
\caption{The same as in Fig.~\ref{fig:2841_mc} but for NGC~7331.}
\label{fig:7331_mc}
\end{figure}

In this appendix, for all investigated galaxies, we plot (Figs.~\ref{fig:2841_mc}--\ref{fig:7331_mc}) the results of Monte Carlo realization of the rotation curves and parameters derived from the subsequent decomposition. Here only parameters for the NFW halo and free $(M/L)_\mathrm{d}$ are presented.

\label{lastpage}
\end{document}